\documentclass[twocolumn,tighten]{aastex63}
\usepackage{amssymb,mathtools}
\usepackage{graphicx}

\begin{document}

\newcommand{\afluxa}{450~$\mu$m\ }
\newcommand{\afluxb}{850~$\mu$m\ }
\newcommand{\afluxar}{450~$\mu$m}
\newcommand{\afluxbr}{850~$\mu$m}
\newcommand{\lfir}{$L_{8-1000~\mu{\rm m}}$}

\newcommand{\almafluxratio}{$f_{870~\mu{\rm m}}/f_{450~\mu{\rm m}}$}
\newcommand{\smmfluxratio}{$f_{850~\mu{\rm m}}/f_{450~\mu{\rm m}}$}
\newcommand{\firfluxratio}{$f_{250~\mu{\rm m}}/f_{850~\mu{\rm m}}$}
\newcommand{\radfluxratio}{$f_{850~\mu{\rm m}}/f_{20~{\rm cm}}$}
\newcommand{\flipradfluxratio}{$f_{20~{\rm cm}}/f_{850~\mu{\rm m}}$}
\newcommand{\flipsmmfluxratio}{$f_{450~\mu{\rm m}}/f_{850~\mu{\rm m}}$}

\title{A Submillimeter Perspective on the GOODS Fields (SUPER GOODS).~V. Deep 450$~{\micron}$ Imaging}

\author[0000-0002-3306-1606]{A.~J.~Barger}
\affiliation{Department of Astronomy, University of Wisconsin-Madison, 475 N. Charter Street, Madison, WI 53706, USA}
\affiliation{Department of Physics and Astronomy, University of Hawaii, 2505 Correa Road, Honolulu, HI 96822, USA}
\affiliation{Institute for Astronomy, University of Hawaii, 2680 Woodlawn Drive, Honolulu, HI 96822, USA}

\author[0000-0002-6319-1575]{L.~L.~Cowie}
\affiliation{Institute for Astronomy, University of Hawaii, 2680 Woodlawn Drive, Honolulu, HI 96822, USA}

\author{A.~H.~Blair}
\affiliation{Department of Astronomy, University of Wisconsin-Madison, 475 N. Charter Street, Madison, WI 53706, USA}

\author[0000-0002-1706-7370]{L.~H.~Jones}
\affiliation{Department of Astronomy, University of Wisconsin-Madison, 475 N. Charter Street, Madison, WI 53706, USA}
\affiliation{Space Telescope Science Institute, 3700 San Martin Drive, Baltimore, MD 21218, USA}

%-----------------------------------------------------------------------------
%   Abstract
%-----------------------------------------------------------------------------
\begin{abstract}
\noindent We present deep SCUBA-2 \afluxa imaging 
of the two GOODS fields, achieving a 
central rms of 1.14~mJy for the GOODS-N and 1.86~mJy for the GOODS-S.
For each field, we give a catalog of  $>4\sigma$ detections 
(79 and 16 sources, respectively).
We construct the \afluxa number counts, finding excellent
agreement with others from the literature.
We match the \afluxa sources to
20~cm data (both fields) and ALMA 870~$\mu$m data (GOODS-S) 
to gauge the accuracy of the \afluxa positions.
We use the extensive
redshift information available on the fields to test how well
redshift can be estimated from simple flux ratios (\afluxar/\afluxb
and 20~cm/\afluxbr), finding tight correlations. 
We provide a catalog
of candidate high-redshift submillimeter galaxies.
We look for evolution in dust temperature with redshift by fitting 
the spectral energy distributions of the sources, but we do not find any
significant redshift evolution after accounting for the far-infrared luminosity.
We do not find evidence for the \afluxa selection picking out warmer sources
than an \afluxb selection.
We find that a \afluxa selected sample only adds low-redshift
($z<1.5$) galaxies beyond an \afluxb sample. 
\end{abstract}


%-----------------------------------------------------------------------------
\section{Introduction}
%-----------------------------------------------------------------------------
Submillimeter galaxies (SMGs) are some of the most powerfully star-forming galaxies 
in the universe. The use of deep imaging with the Submillimeter Common-User Bolometer 
Array (SCUBA; Holland et al.\ 1999) on the 15~m 
James Clerk Maxwell Telescope (JCMT) brought these distant, dusty, and ultraluminous 
galaxies into view for the first time
(e.g., Smail et al.\ 1997; Barger et al.\ 1998; Hughes et al.\ 1998; Eales et al.\ 1999).  
Their star formation rates (SFRs) are in excess of $500\,{\rm M}_\odot$~yr$^{-1}$.
They are significant contributors to the 
star formation history out to redshifts of at least 5
(e.g., Barger et al.\ 2000, 2012, 2014; Chapman et al.\ 2005; Wardlow et al.\ 2011;
Casey et al.\ 2013; Swinbank et al.\ 2014; Cowie et al.\ 2017) and hence
critical for understanding galaxy evolution.

SCUBA was replaced by the second-generation camera SCUBA-2 (Holland et al.\ 2013), 
which vastly improved ground-based submillimeter astronomy by achieving a field-of-view 
of 50~arcmin$^2$, 16 times larger than its predecessor. It is
the preeminent instrument
for obtaining the wide-field maps required to identify large numbers of SMGs. 
There is currently no other submillimeter instrument that can match its mapping speed.
SCUBA-2 observes simultaneously at \afluxa and \afluxbr,
but useful \afluxa observations are possible only 
during the small fraction ($\sim15$\%) of nights on Mauna Kea when 
the precipitable water vapor is exceptionally low.

The natural limit of single-dish submillimeter observations is the depth at which
confusion, which refers to the blending of sources or where the noise is dominated
by unresolved contributions from fainter sources, becomes important.
At \afluxbr, the JCMT has a resolution of $14''$ FWHM, which allows the construction 
of very large and uniform samples of SMGs brighter than the
$\sim1.65$~mJy ($4\sigma$) confusion limit (Cowie et al.\ 2017).
However, it also results in source blending and unresolved galaxies.
Targeted follow-up observations with submillimeter interferometers, such as the 
Atacama Large Millimeter/submillimeter Array (ALMA),
are able to provide accurate positions, to resolve the sources, and to determine
multiplicity, but their small fields-of-view mean interferometers are too costly
for the initial selection of large samples of SMGs.

An alternative way to get higher resolution is to go to shorter wavelengths.
At \afluxar, the JCMT has a resolution of $7\farcs5$ FWHM. (Note that this is 
a considerably higher resolution than the $\sim35''$ FWHM of {\em Herschel\/} 
at 500~$\mu$m.)
These higher resolution data (i) substantially reduce the confusion limit,
thereby allowing the more typical sources that contribute the
bulk of the submillimeter EBL to be found; 
(ii) make it possible to measure 
fluxes at rest wavelengths close to the peak of the blackbody distribution 
in the far-infrared (FIR); (iii) allow measurements of the FIR spectral energy
distributions (SEDs), when combined with the \afluxb data and with {\it Herschel}
and {\it Spitzer} data for isolated, brighter sources; and 
(iv) allow much more accurate positions to be obtained for the
sources than is possible at \afluxbr. 
With this positional accuracy, we can begin to identify 
optical/near-infrared (NIR) counterparts reliably, even without follow-up interferometric
observations.

The deepest blank-field \afluxa image is from the SCUBA-2 Ultra Deep Imaging 
EAO Survey (STUDIES; e.g., Wang et al.\ 2017; Lim et al.\ 2020), a multiyear 
JCMT Large Program that uses the CV DAISY scan pattern (Holland et al.\ 2013) 
and aims to reach the confusion limit at \afluxa in the COSMOS field.
Lim et al.\ (2020) report a central rms of 0.75~mJy at \afluxa for their
184~hr of on-sky observations.
They combine STUDIES with existing data from Casey et al.\ (2013;
Lim et al.\ quote 3.6~mJy for 38~hr; this uses the wider PONG-900 scan pattern) 
and Geach et al. (2013; Lim et al.\ quote 0.95~mJy for 150~hr in the deepest region; 
this is a mosaic of two CV DAISY maps with some overlap).
The rms of the combined datasets in the deepest region is 0.65~mJy (Lim et al.\ 2020).
Another field with deep \afluxa data is the
SCUBA-2 Cosmology Legacy Survey (S2CLS) Extended Groth Strip or EGS (this uses
the CV DAISY scan pattern), for which
Zavala et al.\ (2017) quote a central rms of 1.2~mJy at 450~$\mu$m.

In this paper, we focus on analyzing our deep SCUBA-2 \afluxa observations of both GOODS 
fields. We have been observing the GOODS fields for many years, most recently through our
SUbmillimeter PERspective on the GOODS fields (SUPER GOODS) program.
We chose these fields, because they are the most richly observed fields on the 
sky, with 4 optical bands ($B$ through $z$) from the
{\em HST}/ACS GOODS survey (Giavalisco et al.\ 2004) 
and two NIR bands from the {\em HST}/WFC3 CANDELS survey 
(Grogin et al.\ 2011; Koekemoer et al.\ 2011).  
The {\em Spitzer}/MIPS 70~$\mu$m survey (and associated 24 and 
160~$\mu$m data) and the deep {\em Herschel\/} FIR surveys 
are also among the deepest ever taken and provide sufficient 
sensitivity to constrain the dust SEDs of the SMGs
(e.g., Oliver et al.\ 2010; Lutz et al.\ 2011; Barger et al.\ 2012, 2014; Cowie et al.\ 2017, 2018). 
The GOODS-N has ultradeep 20~cm observations providing an rms noise in
the field center of 2.2~$\mu$Jy (Owen 2018), reflecting the power of the 
upgraded Karl G. Jansky Very Large Array (VLA). 
Meanwhile, the GOODS-S has VLA 20~cm observations that reach a best rms sensitivity 
of 6~$\mu$Jy (Miller et al.\ 2013). Finally, both fields have 
spectroscopic observations for many thousands of galaxies 
(e.g., Barger, Cowie, \& Wang 2008; Popesso et al.\ 2009; Balestra et al.\ 2010;
Barger et al.\ 2019 and references therein).  
{\em Chandra} obtained a 7~Ms X-ray image of the
{\em Chandra\/} Deep Field-South (CDF-S; Luo et al.\ 2017), which is by far the 
deepest X-ray image ever taken, followed only by the 2~Ms 
{\em Chandra\/} Deep Field-North (CDF-N; Alexander et al.\ 2003). 
Our SCUBA-2 observations cover the most sensitive regions of the 
{\em Chandra} images and almost the entire GOODS areas. 

This is the fifth paper in our SUPER GOODS series.
In the first two papers in the series (Cowie et al.\ 2017 and Barger et al.\ 2017), we analyzed the
SCUBA-2 850~$\mu$m observations of the GOODS-N/CANDELS/CDF-N,
complemented with targeted Submillimeter Array (SMA) interferometry, VLA interferometry,
{\em Herschel\/} imaging, and spectroscopy.
In the next two papers in the series (Cowie et al.\ 2018 and Barger et al.\ 2019),
we analyzed the SCUBA-2 850~$\mu$m observations of the 
GOODS-S/CANDELS/CDF-S,
complemented with targeted ALMA 870~$\mu$m interferometry, 
{\em Herschel\/} imaging, and spectroscopy.

In Section~\ref{secdata}, we describe our SCUBA-2 observations and data reduction, determine
our \afluxa selected samples, and
present the primary ancillary data used in this paper.
In Section~\ref{seccounts}, we construct the \afluxa number counts.
In Section~\ref{secfluxratios}, we use combinations of the \afluxar, \afluxbr, and 20~cm bands
to identify candidate high-redshift galaxies.
In Section~\ref{secSED}, we construct SEDs from our multiwavelength data and fit them 
with the publicly available Bayesian energy-balance SED-fitting code MAGPHYS
(da Cunha et al.\ 2015) to obtain SFRs and dust properties. We also construct gray body
fits for comparison.
In Section~\ref{secmagphys}, we examine the MAGPHYS SFRs and FIR luminosities
for galaxies with either a spectroscopic or photometric redshift $z>1$.
In Section~\ref{secdisc}, we discuss the redshift distribution of the \afluxa selected
sample and the dependence of dust temperature on
redshift and FIR luminosity. In Section~\ref{secsummary}, we summarize
our results.

%-----------------------------------------------------------------------------
\section{Data} 
\label{secdata}
%-----------------------------------------------------------------------------

%-----------------------------------------------------------------------------
\subsection{SCUBA-2 Observations} 
\label{scuba2}
%-----------------------------------------------------------------------------
In Cowie et al. (2017, 2018), we presented SCUBA-2 850~$\mu$m catalogs ($>4\sigma$)
of the GOODS-N (Cowie et al.\ 2017's Table~5) and
GOODS-S (Cowie et al.\ 2018's Table~2; this just covered the central 100~arcmin$^2$ region) 
through mid-2016 and through early 2017, respectively. 
We also presented a SCUBA-2 450~$\mu$m catalog ($>4\sigma$) of the 
GOODS-N (Cowie et al.\ 2017's Table~6), 

We have continued to deepen our observations.
To obtain maximum depth in the central region, we used the \textsc{CV Daisy} scan 
pattern, whose field size is $5\farcm5$ in radius (by this radius, the noise is twice the central noise).
To cover the outer regions to find brighter but rarer sources, we used the 
larger field \textsc{PONG-900} scan pattern, whose field size is $10\farcm5$ in radius 
(by this radius, the noise is twice the central noise).

%-----------------------------------------------------------------------------
% TABLE 1
%-----------------------------------------------------------------------------
\begin{deluxetable}{cccc}
\tablecaption{SCUBA-2 Observations \label{obstab}}
\tablehead{Field & Weather & Scan & Exposure\\ 
& Band & Pattern & (Hr)}
\startdata
GOODS-N   &   1   &   CV Daisy    &   140.84 \\
        &   1   &   PONG-900    &   20.01  \\
        &   2   &   CV Daisy    &   27.79  \\
        &   2   &   PONG-900    &   16.37  \\
\hline
GOODS-S   &   1   &   CV Daisy    &   50.12 \\
        &   1   &   PONG-900    &   16.20   \\
        &   2   &   CV Daisy    &   35.63  \\
        &   2   &   PONG-900    &   8.70  \\
\enddata
\label{tabobs}
\end{deluxetable}
%-----------------------------------------------------------------------------

We summarize the total weather band~1 ($\tau_{\rm 225~GHz}<0.05$) and
weather band~2 ($0.05<\tau_{\rm 225~GHz}<0.08$) SCUBA-2 observations 
in Table~\ref{obstab}. These are the only weather conditions 
under which \afluxa observations can be usefully obtained.
The GOODS-N observations have a central rms of
1.14~mJy at \afluxar, and the GOODS-S 1.86~mJy.
These values do not include confusion noise.

Our reduction procedures follow Chen et al.\ (2013b)
and are described in detail in Cowie et al.\ (2017). The 
galaxies are expected to appear as unresolved sources at the
$7\farcs5$ resolution of the JCMT at \afluxar.
We therefore applied a matched filter to our maps, which
provides a maximum likelihood estimate of the source strength for
unresolved sources (e.g., Serjeant et al.\ 2003).
Each matched-filter image has a PSF with
a Mexican hat shape and a FWHM corresponding to the 
telescope resolution.

For both fields, we chose sources in an area where the noise is less than 3.75~mJy
(roughly 3 times the central noise for the GOODS-N and 2 times for the GOODS-S).
For the GOODS-N, this corresponds to an area of 175~arcmin$^2$, 
and for the GOODS-S, 48~arcmin$^2$.
As described in Cowie et al.\ (2017), we extracted point sources by
identifying the peak signal-to-noise (S/N) pixel, subtracting this peak pixel
and its surrounding areas using the PSF scaled and centered
on the value and position of that pixel, and then searching
for the next S/N peak. We iterated this process until we
reached a S/N of 3.5. We then limited the sample
to the sources with a S/N above 4, giving 79
\afluxa sources in the GOODS-N and 16 in the GOODS-S.
Our S/N choice of $>4\sigma$ should minimize the number of spurious 
sources in our catalogs to $\le5$\% (e.g., Chen et al.\ 2013a; Casey et al.\ 2013).

The reason for the iterative process described above is to remove
contamination by brighter sources before we identify fainter 
sources and measure their fluxes. While critical
for the \afluxb data, it is somewhat less necessary for the 
\afluxa data due to its higher resolution and shallower
depth. However, we still followed the procedure.

We matched our \afluxa sources to $>4\sigma$ \afluxb sources within $4''$.
In cases where there was no \afluxb match, we measured the \afluxb
flux for the source at the \afluxa position. For these measurements,
we first removed all of the \afluxa sources with an \afluxb source match from 
the \afluxb matched-filter SCUBA-2 images using a
PSF based on the observed calibrators. This left residual images from which 
we measured the \afluxb fluxes (whether positive or negative) and statistical
uncertainties at the \afluxa positions.  This procedure minimizes contamination
by brighter \afluxb sources in the field.

In Tables~\ref{cdfntab} and \ref{cdfstab}, we 
give the \afluxa and \afluxb fluxes for, respectively, the GOODS-N 
and GOODS-S $>4\sigma$ \afluxa samples.
In the GOODS-N, there are two pairs of \afluxa galaxies that are blended at
\afluxbr, and in the GOODS-S, there is one. 
In these cases, we have assigned the \afluxb flux measured at the \afluxa
source position of the brighter member of the pair to that member,
and we have not assigned an \afluxb flux to the fainter member of the pair
(the fainter sources are labeled ``blend'' in the \afluxb flux column in the tables,
along with the source number of the brighter member of the pair). 

Lim et al.\ (2020) estimated the fraction of sources in their \afluxa sample 
that are also detected at \afluxb by considering how many \afluxb 
fluxes---all of which they measured at the \afluxa source positions---are five times 
higher than their estimate of the confusion noise (0.42~mJy), namely 2.1~mJy. 
They found that 99 of their 256 sources satisfied this criterion. This can be thought
of as them determining how many of their \afluxa sources would be contained in
a $>5\sigma$ \afluxb selected sample, which is a fairly low 38\%.
Using the same procedure, we find that 68 of our 92 sources
(after excluding the three fainter members of the blends)
have \afluxb fluxes above 2.1~mJy, or a much higher 74\%. 
This may reflect our slightly shallower \afluxa sample: When the \afluxa fluxes
are brighter, then they are easier to detect in a confusion-limited \afluxb image.

\begin{figure}
\centerline{\includegraphics[width=9.4cm]{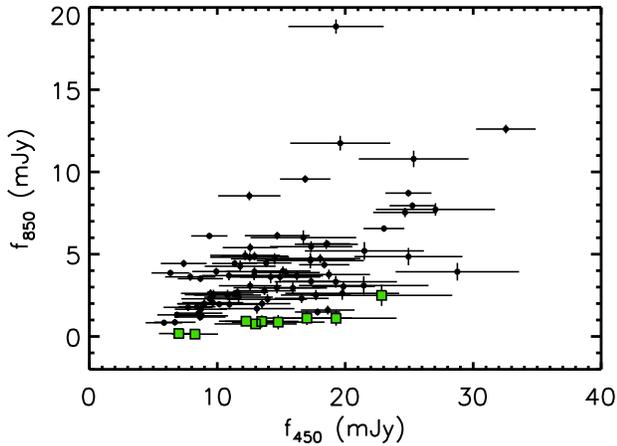}}
\caption{
Based on the combined GOODS-N (Table~\ref{cdfntab}) and GOODS-S (Table~\ref{cdfstab})
samples, SCUBA-2 \afluxb flux vs. SCUBA-2 \afluxa flux with uncertainties on both quantities.
The fainter members of the three blended \afluxa pairs have been excluded.
The green squares show the 9 \afluxa selected sources that are not also 
significantly ($>4\sigma$) detected at \afluxbr.
}
\label{figfvsf}
\end{figure}

However, another approach to determining the overlap between samples is to take 
advantage of our ability to probe below the \afluxb confusion limit with our predetermined
\afluxa sample. After excluding the three fainter members of the blends, all but 9 of our 
remaining 92 \afluxa sources are also detected at high significance ($>4\sigma$) at \afluxbr.
We show \afluxb flux versus \afluxa flux for the combined GOODS sample in 
Figure~\ref{figfvsf} with these nine sources denoted by green squares.
In order to quantify the level of contamination, we measured \afluxb fluxes
at randomized positions, finding that 7\% of these measurements result in 
a $>4\sigma$ \afluxb measurement. Allowing for this contamination, we conclude 
that 82\% of our \afluxa sample is detected at $>4\sigma$ at \afluxbr.

We also have two sources in the GOODS-N and one source in the GOODS-S
whose optical/NIR photometry is contaminated by a neighboring star or galaxy. 
We label these ``contam." in the redshift column in the tables because no 
redshifts could be determined for them.

We will hereafter refer to the sample of 89 sources,
after removal of the three fainter members of the blends
and the three optical/NIR contaminated sources, as our 
{\em \afluxa selected combined GOODS sample}.

%-----------------------------------------------------------------------------
% FIGURE 1 ; 450_f8_f4.ps
%-----------------------------------------------------------------------------
%-----------------------------------------------------------------------------

%---------------------------------------------------------------------
\subsection{VLA 20~cm Observations}
\label{vla}
%---------------------------------------------------------------------
The upgrade of the VLA greatly increased the sensitivity
of the decimetric continuum observations that can be obtained.
Here we use the extremely deep 20~cm image of the CDF-N field
with a best rms sensitivity of 2.2~$\mu$Jy (Owen 2018).
The image covers a $40'$ diameter region with an effective resolution of
$1\farcs8$. The absolute radio positions are known to $0\farcs1-0\farcs2$ rms.
The highest sensitivity region, about $9'$ in radius, is closely
matched to the full area of our SCUBA-2
survey and completely covers the deepest region of our map. 
There are 787 distinct $\ge5\sigma$ radio sources within the
$9'$ radius, excluding sources that appear to be parts of other sources.
We used these data in analyzing the SCUBA-2 \afluxb sources in
Cowie et al.\ (2017) and Barger et al.\ (2017). 
For the small number of sources outside the Owen (2018) field, 
we used radio fluxes from Morrison et al.\ (2010).

We find that 73 of the 79 \afluxa sources lie in the area covered
by the Owen (2018) 20~cm observations. Of these, 63, or 86\%, have
20~cm counterparts within $4''$. 
If we reduce the match radius to $3''$, then the number with 20~cm
counterparts drops to 57.
Following Cowie et al.\ (2017), we have included HDF850.1
(originally detected in the SCUBA map by Hughes et al.\ 1998; the galaxy's 
redshift of $z=5.183$ was eventually measured from CO observations by 
Walter et al.\ 2012) in these numbers.
It lies just below the Owen (2018) flux threshold
and is offset from the radio position by $1\farcs8$.

We checked the absolute astrometric pointing of the \afluxa map
by comparing the positions of the \afluxa sources with 
radio counterparts with the radio positions.
We found offsets of $-0\farcs76$ in R.A. and 
$-0\farcs33$ in Decl., which are small compared to the positional uncertainties 
of the \afluxa data. We applied these corrections to the \afluxa positions.

\begin{figure}
\centerline{\includegraphics[width=9.4cm]{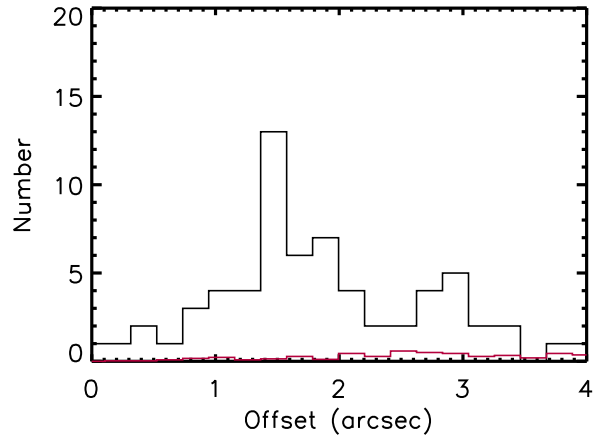}}
\caption{
Offsets between the 65 SCUBA-2 \afluxa centroid positions and the radio positions 
for counterparts within $4''$.
The mean offset is $2\farcs0$, and the median offset is $1\farcs7$.
The red curve shows the histogram of offsets that would be expected between the positions of
a random distribution of \afluxa sources and the radio positions for counterparts
within $4''$, based on 64 simulations.
}
\label{figoffsets}
\end{figure}

In Figure~\ref{figoffsets}, we show the offsets of the radio counterparts.
To compare with expectations for offsets between the radio sources and a random 
distribution of \afluxa sources, we generated 64 simulated \afluxa maps by removing 
the 79 true sources from the image and each time randomly populating this cleaned 
map with 79 sources with the same fluxes as the real sample. 
The red curve shows the offset results from the simulations. 
The number of random identifications only becomes comparable to the number
of actual identifications above about $3\farcs5$, showing that most of the closer
identifications are real.

%-----------------------------------------------------------------------------
% FIGURE 2 ; offset.ps
%-----------------------------------------------------------------------------
%-----------------------------------------------------------------------------

For the GOODS-S, we use the VLA 20~cm catalog of Miller et al.\ (2013) of the
Extended CDF-S, which covers an area of about a third of a square degree
and reaches a best rms sensitivity of 6~$\mu$Jy. Their catalog
contains 883 distinct radio sources. For this field, we do our absolute
astrometric pointing check and offset calculation using the ALMA data 
(see Section~\ref{secalma}).

%---------------------------------------------------------------------
\subsection{ALMA 870~$\mu$m Observations of the GOODS-S}
\label{secalma}
%---------------------------------------------------------------------
In the most sensitive 100~arcmin$^2$ area, all of the $>2.25$~mJy 
SCUBA-2 \afluxb sources ($>4\sigma$)
have been observed with ALMA in band~7 (870~$\mu$m), 
together with a number of fainter SCUBA-2 sources. 
Details of these observations and their reductions may be found in Cowie et al.\ (2018). 
We restrict the area of the individual ALMA images to their
FWHM radius of $8\farcs75$. With this restriction, the ALMA
images cover a (non-contiguous) total area of 7.2~arcmin$^2$.
We used these data in analyzing the SCUBA-2 \afluxb sources in
Cowie et al.\ (2018) and Barger et al.\ (2019).

Note that some of the ALMA sources are off the \afluxa area that we 
use in this paper (see Section~\ref{scuba2}).
Based on the mean offset of 11 isolated ALMA sources
with 870~$\mu$m fluxes $>2.25$~mJy from the nearest SCUBA-2 \afluxa peak, 
we found an absolute astrometric offset of $1\farcs44$ in R.A. and 
$1\farcs28$ in Decl., which we applied to the SCUBA-2 \afluxa 
positions in Table~\ref{cdfstab}. This is well within the expected uncertainty 
in the absolute SCUBA-2 \afluxa astrometry. 

\begin{figure}
\centerline{\includegraphics[width=11.5cm,angle=0]{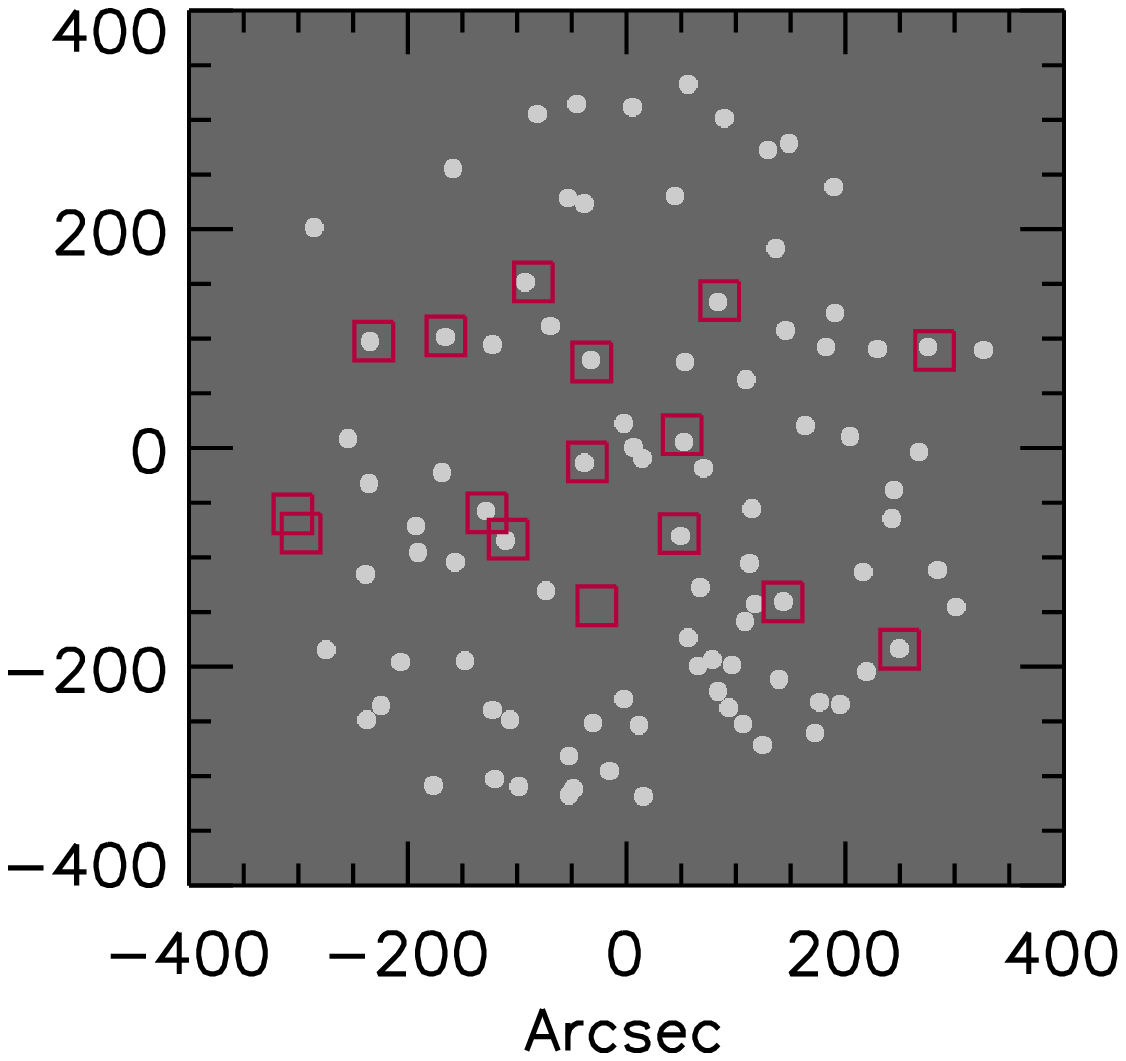}}
\vskip -1.3cm
\centerline{\includegraphics[width=11.5cm,angle=0]{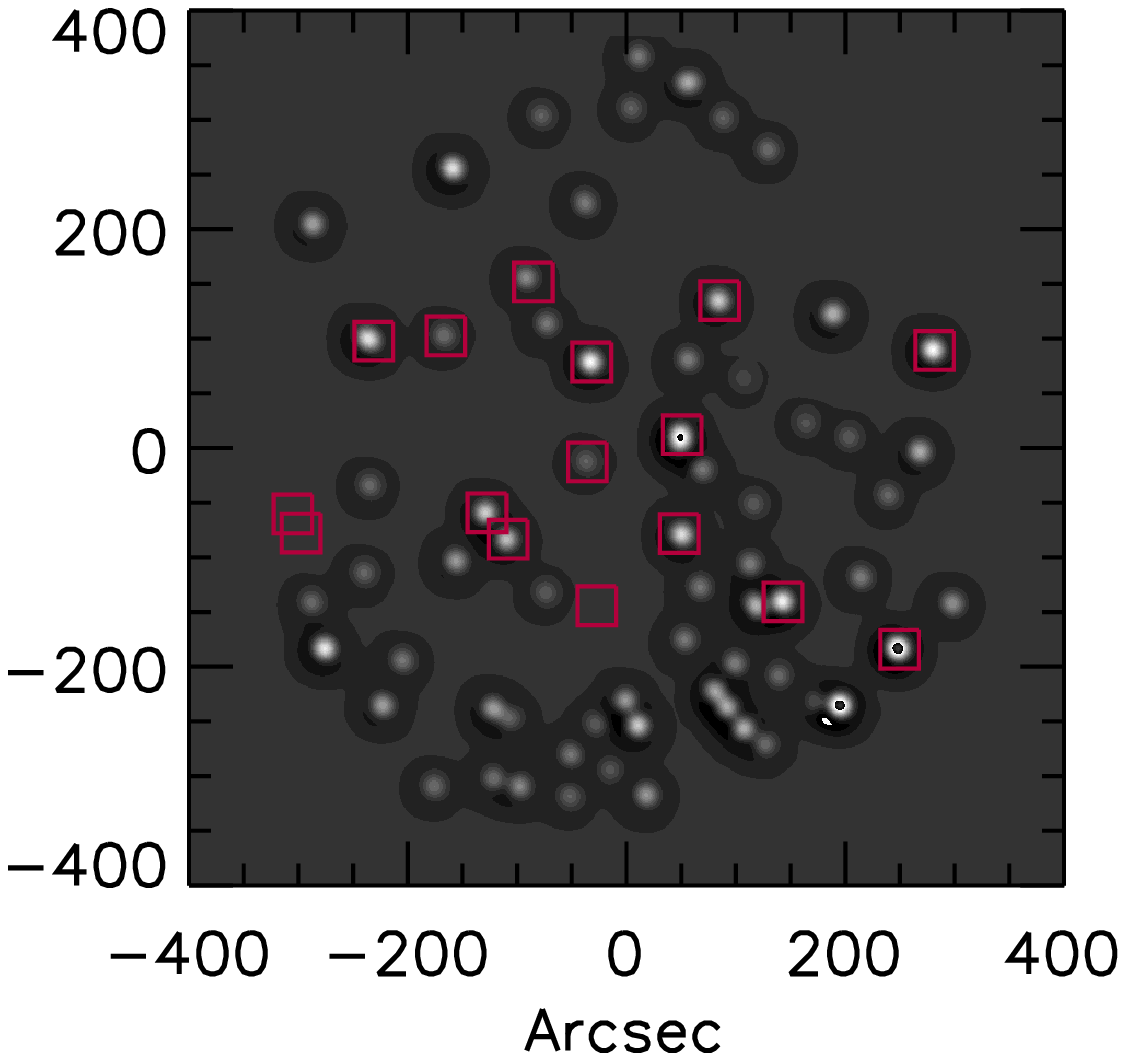}}
\vskip -1.3cm
\centerline{\includegraphics[width=11.5cm,angle=0]{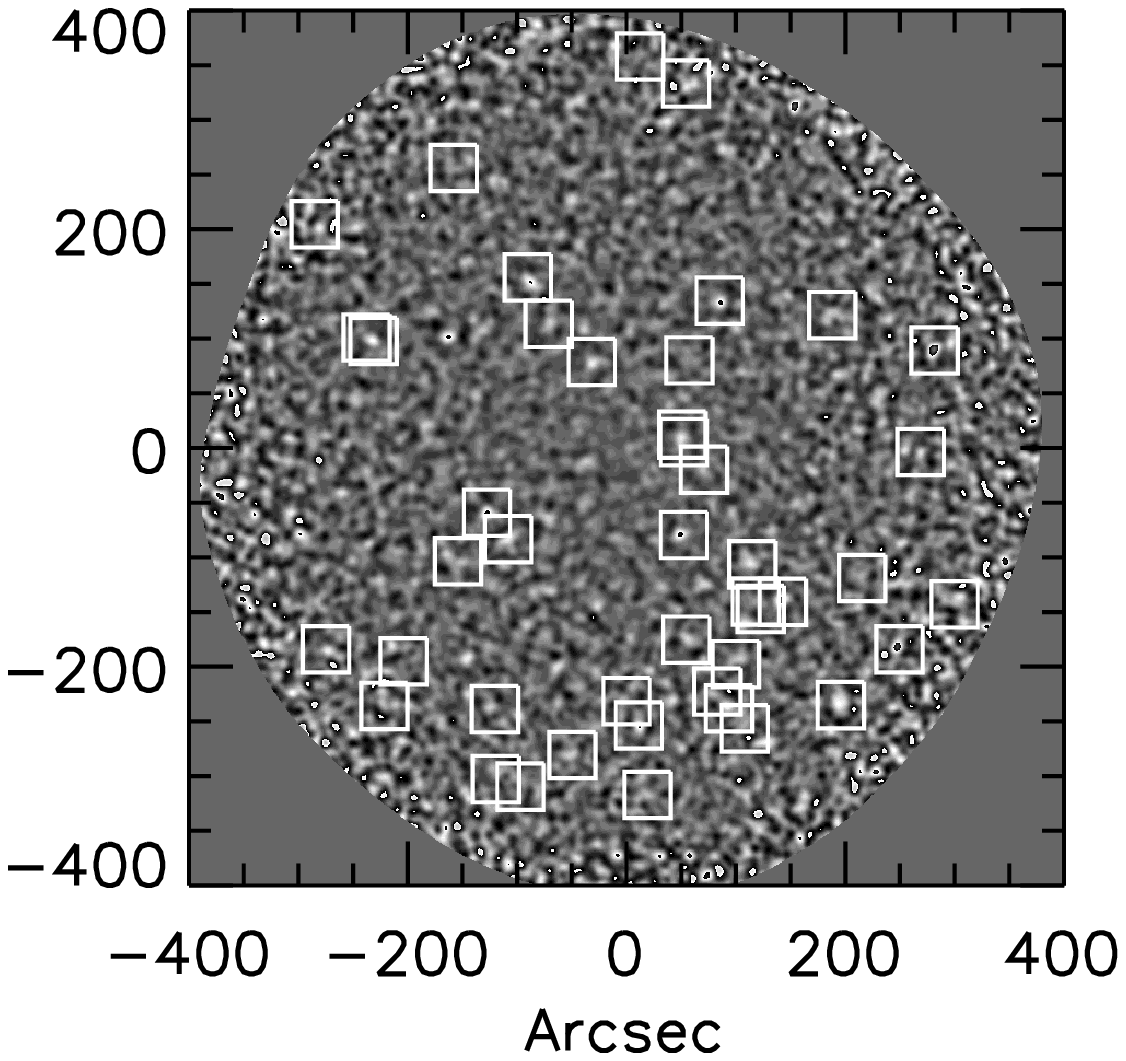}}
\vskip -1.1cm
\caption{
(a) ALMA observed areas 
and (b) ALMA sources convolved with SCUBA-2 \afluxb PSF.
The \afluxa source positions are overlaid in both panels (red squares).
(c) \afluxa image with the ALMA $>2.25$~mJy source positions 
overlaid (white squares).
The images are centered at R.A. $=3^{\rm h}  32^{\rm m}  26^{\rm s}$,    
decl. $= -27^\circ   48'   17''$,
with the x-axis (y-axis) corresponding to offsets in R.A. (decl.)
from this position.
}
\label{ALMAon450}
\end{figure}

In Figure~\ref{ALMAon450}(a), we show the ALMA observed areas (gray circles)
with the SCUBA-2 \afluxa source positions overlaid (red squares). 13 of the 16
\afluxa sources lie on an ALMA image. In Figure~\ref{ALMAon450}(b), we show 
the ALMA sources convolved with the SCUBA-2 \afluxb PSF with the
\afluxa source positions again overlaid (red squares).
12 of the 13 \afluxa sources that lie on an ALMA image have an ALMA source 
counterpart within $4''$; the remaining \afluxa source lies on the edge of an ALMA image,
as noted in Table~\ref{cdfstab}.
In Figure~\ref{ALMAon450}(c), we show the \afluxa image with the
ALMA $>2.25$~mJy source positions overlaid (white squares).
Some of the \afluxa sources do not correspond to the ALMA $>2.25$~mJy 
sample, and some of the ALMA $>2.25$~mJy sample do not have
\afluxa counterparts at our current sensitivity level.

%-----------------------------------------------------------------------------
% FIGURE 3 ; show_alma_area.ps, show_alma_detect.ps, show_450_alma.ps
%-----------------------------------------------------------------------------
%-----------------------------------------------------------------------------

%-----------------------------------------------------------------------------
% FIGURE 4 ; f450_f500.ps, f500_f450.ps
%-----------------------------------------------------------------------------
\begin{figure}
\centerline{\includegraphics[width=12cm,angle=0]{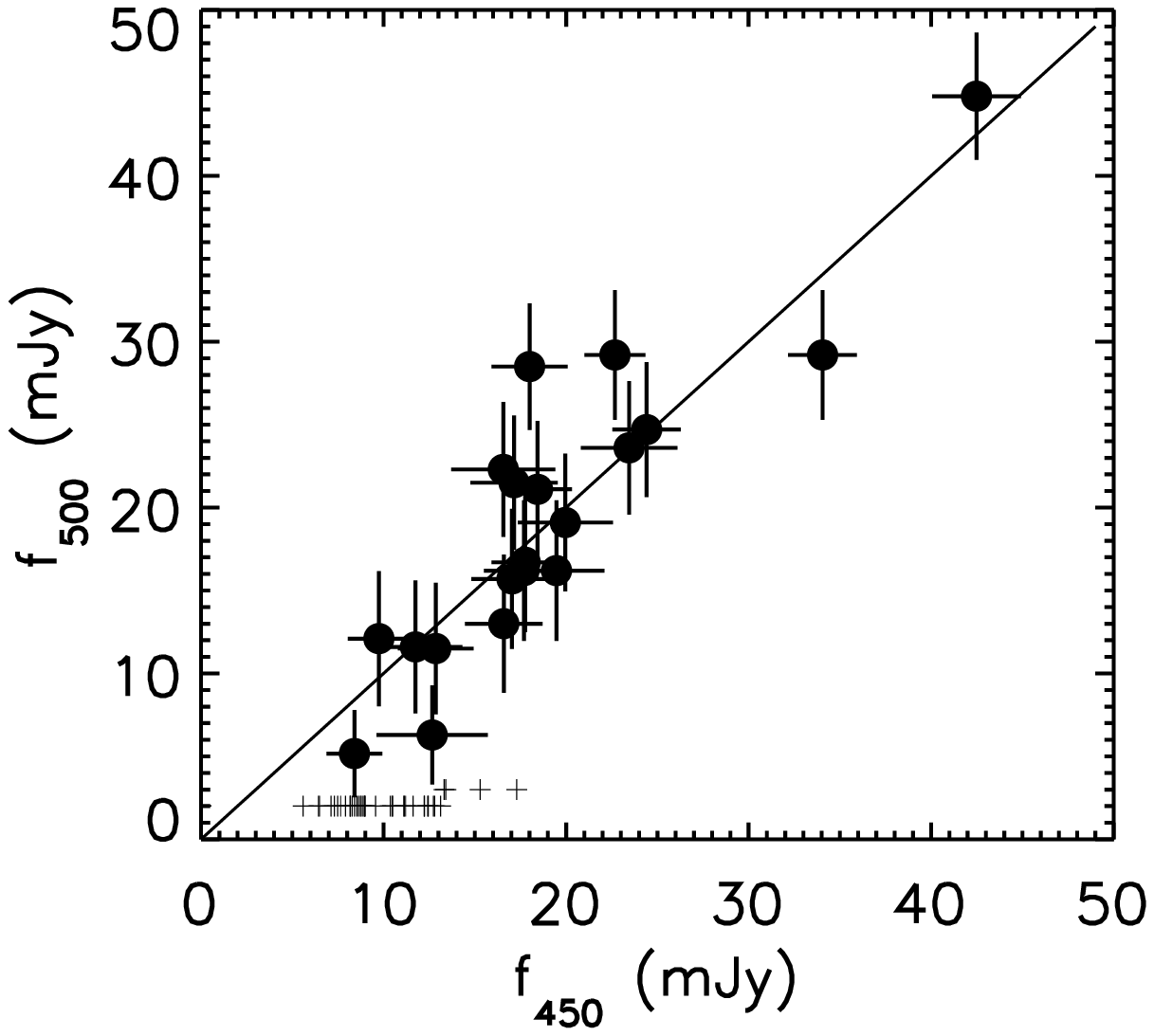}}
\vskip -0.95cm
\centerline{\includegraphics[width=12cm,angle=0]{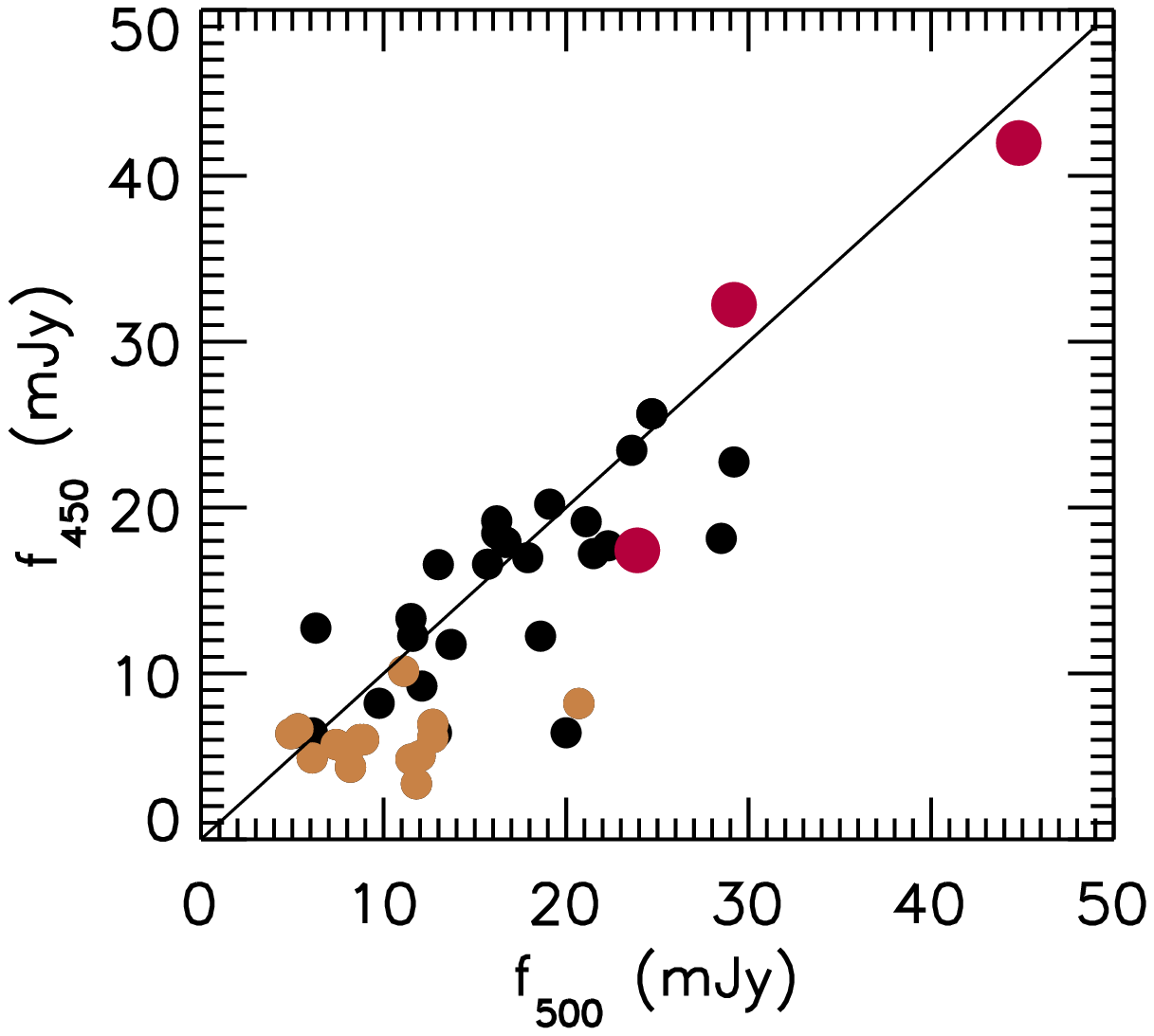}}
\vskip -0.95cm
\caption{
(a) {\em Herschel\/} 500~$\mu$m flux from the E11
GOODS-N catalog vs. SCUBA-2 \afluxa 
flux from Table~\ref{cdfntab} for the 63 \afluxa sources 
with noise $<2.8$~mJy (twice the minimum noise).
Black circles denote \afluxa sources with a
500~$\mu$m counterpart within $3\farcs5$ of the \afluxa position.
Plus signs show the remaining sources
at the bottom (a slight vertical offset separates
sources $>12$~mJy from those $<12$~mJy).
For sources 1 and 6 in Table~\ref{cdfntab}, each of which has a very 
nearby second source, we show the combined \afluxa flux.
The plotted flux uncertainties are $1\sigma$ at both \afluxa and 500~$\mu$m.
(b) SCUBA-2 \afluxa flux vs. {\em Herschel\/} 500~$\mu$m flux for all 
sources in the E11 catalog 
that lie in the SCUBA-2 area with \afluxa noise $< 2.8$~mJy.
Black circles show 500~$\mu$m sources with a \afluxa counterpart
within $12''$ of the 500~$\mu$m source position. Red circles show
500~$\mu$m sources with multiple 
\afluxa counterparts within $12''$ of the 500~$\mu$m source; in these
cases we use the combined \afluxa flux. 
Gold circles show 500~$\mu$m sources with no \afluxa counterparts 
in Table~\ref{cdfntab};
in these cases, we measured the \afluxa fluxes directly from the image
at the {\em Herschel\/} catalog positions.
}
\label{500on450}
\end{figure}
%-----------------------------------------------------------------------------

%---------------------------------------------------------------------
\subsection{Optical/NIR Fluxes}
\label{secoptobs}
%---------------------------------------------------------------------
The GOODS fields are defined by the deep {\em HST}/ACS optical imaging of 
Giavalisco et al.\ (2004). These fields are also covered by deep CANDELS 
{\em HST}/WFC3 ultraviolet/optical and infrared channel imaging 
(Grogin et al.\ 2011; Koekemoer et al.\ 2011), and there are 
{\em Spitzer}/IRAC observations from the GOODS {\em Spitzer\/} Legacy
Program (PI: M. Dickinson) and from the SEDS and S-CANDELS surveys 
(Ashby et al.\ 2013, 2015).

In the GOODS-N, we used the CANDELS/SHARDS multiwavelength catalog of 
Barro et al.\ (2019) for our photometry, after identifying counterparts to
our \afluxa selected sample using a $1''$ search radius if there is an
accurate position from the radio or the Submillimeter Array, and a
$3''$ search radius otherwise.
55 of the 79 sources lie within the Barro et al.\ area and have counterparts
within $1''$.

In the GOODS-S, we used the multiwavelength catalog of Guo et al.\ (2013)
for our photometry, after identifying counterparts to
our \afluxa selected sample using a $1''$ search radius if there is an
accurate position from the radio or ALMA, and a $3''$ search radius otherwise.
All of the sources with accurate positions lie within the Guo et al.\ area
and have counterparts within $1''$.

%---------------------------------------------------------------------
\subsection{MIR/FIR Fluxes}
\label{secherschel}
%---------------------------------------------------------------------
We used the catalogs of Magnelli et al.\ (2011) 
and Elbaz et al.\ (2011; hereafter, E11),
respectively, to obtain the {\em Spitzer}/MIPS 24~$\mu$m fluxes for the 
GOODS-N and GOODS-S fields.
These catalogs were constructed using the deep 24~$\mu$m MIPS images
from the {\em Spitzer\/} GOODS Legacy program (PI: M. Dickinson).
We used the catalogs of Magnelli et al.\ (2013)
and E11, respectively,
to obtain the {\em Herschel}/PACS 100~$\mu$m and 160~$\mu$m fluxes
for the GOODS-N and GOODS-S fields.
Magnelli et al.\ (2013) constructed 
the GOODS-N catalog using the combined data 
sets of the PACS Evolutionary Probe (PEP; Lutz et al.\ 2011) 
guaranteed time key program and the GOODS-{\em Herschel\/} 
(E11) open time key program, while E11 constructed the GOODS-S
catalog using the GOODS-{\em Herschel\/} program data.
We also used the wider area {\em Herschel}/SPIRE 
250~$\mu$m, 350~$\mu$m, and 500~$\mu$m
catalogs from E11 for both fields.

E11 and Magnelli et al.\ (2013) both used 24~$\mu$m priors 
to deblend the {\em Herschel\/} data when constructing their catalogs.
They provide a detailed discussion of the robustness
of the 24~$\mu$m priors in deblending the longer wavelength data.
They also provide flags to assess whether sources are contaminated
by nearby brighter sources.

We cross-identified the \afluxa samples with the $>20~\mu$Jy 24~$\mu$m
sources, which are what E11
used as priors in measuring the {\em Herschel}/PACS 100~$\mu$m fluxes. 
For the unmatched sources, we measured the fluxes ourselves in the 
24~$\mu$m images at the \afluxa positions using a $3''$ diameter aperture.
We aperture corrected these 24~$\mu$m fluxes using an average correction 
inferred from the matched sources. We then measured fluxes
in the longer wavelength data using matched-filter images, after first
removing the sources in the Magnelli et al.\ (2011) and E11
catalogs from the images. 

We examined the relative calibration of the SCUBA-2 \afluxa data 
versus the {\em Herschel\/} 500~$\mu$m data using the GOODS-N data.
In Figure~\ref{500on450}(a), we plot 500~$\mu$m flux from the E11 catalog 
versus \afluxa flux from Table~\ref{cdfntab} for the 63
sources with noise $<2.8$~mJy.
Above a \afluxa flux of 12~mJy, there are 19 sources in the figure,
all but 4 of which are detected at 500~$\mu$m. Three of the non-detections 
at 500~$\mu$m (non-detections are shown with plus signs in the figure)
have measured 350~$\mu$m fluxes, so the 
500~$\mu$m fluxes are simply too faint for the catalog.
The fourth (source 26 in Table~\ref{cdfntab}) appears to have had its
500~$\mu$m flux allocated by E11
to a 24~$\mu$m source that is a neighbor to the
correct 24~$\mu$m counterpart. 

For these bright sources, the median \afluxa to 500~$\mu$m 
flux ratio is 1.00, and the median \afluxa to 350~$\mu$m
flux ratio is 0.68. Interpolation of the {\em Herschel\/} data gives a 
median SCUBA-2 \afluxa to interpolated {\em Herschel} \afluxa flux ratio of 0.90,
so the SCUBA-2 data are slightly fainter on average than the {\em Herschel\/}
data. However, the difference is well within the calibration uncertainty
in both data sets. There is also uncertainty due to the {\em Herschel\/} deblending,
which is hard to quantify.

Below a \afluxa flux of 12~mJy, there are 38 non-detections at 500~$\mu$m 
in the figure.
This shows SCUBA-2's ability to probe deeper than the {\em Herschel\/} data.

Conversely, we can measure the \afluxa counterparts to the 500~$\mu$m 
sources in order to understand the {\em Herschel\/} selection.
In Figure~\ref{500on450}(b), we plot \afluxa flux 
versus 500~$\mu$m flux for all  sources in the E11 GOODS-N catalog 
that lie in the SCUBA-2 area with \afluxa noise $< 2.8$~mJy.
Blends of multiple \afluxa sources at the {\em Herschel\/} resolution 
are seen for some of the brightest sources (red circles).
Sources with offsets from the diagonal line may have flux 
contributions from \afluxa 
counterparts that are below our flux threshold.
The gold circles show sources detected in the {\em Herschel\/} 500~$\mu$m catalog
that are not present in our SCUBA-2 \afluxa catalog. For these sources, we
measured the \afluxa fluxes directly from the image at the {\em Herschel\/} 
catalog positions.
These \afluxa sources may be multiples, where the {\em Herschel\/} flux
is combining sources that are below our \afluxa flux threshold, and where
single-position measurements continue to underestimate the total \afluxa flux.
While {\em Herschel\/} may be able to pick up such sources, which are not
included in our catalog, it is important to stress that one is not learning about 
individual SMGs from such data.

%-----------------------------------------------------------------------------
\subsection{Redshifts}
\label{subsecz}
%-----------------------------------------------------------------------------
We have been compiling spectroscopic redshifts (speczs) from our own Keck observations and 
from the literature for sources in the GOODS fields for many years, so we have a large
database from which to draw from to find redshifts.

In the GOODS-N, many of the speczs were already presented in the SCUBA-2
\afluxb and \afluxa tables of Cowie et al.\ (2017) (i.e., their Tables~5 and 6, respectively). The
literature references from which these were drawn include Cohen et al.\ (2000), 
Cowie et al.\ (2004, 2016), Swinbank et al.\ (2004), 
Wirth et al.\ (2004, 2015), Chapman et al.\ (2005), Treu et al.\ (2005), 
Reddy et al.\ (2006), Barger et al.\ (2008), Pope et al.\ (2008), Trouille et al.\ (2008), 
Daddi et al.\ (2009a,b), Cooper et al.\ (2011), Walter et al.\ (2012), Bothwell et al.\ (2013), 
Momcheva et al.\ (2016), and Riechers et al.\ (2020).
There are also three new NOEMA redshifts from Jones et al.\ (2021).

In the GOODS-S, all of the speczs were already presented in the ALMA redshift table 
of Cowie et al.\ (2018) (i.e., their Table~5), apart from one, which we obtained from
later Keck/DEIMOS observations.
The references for the other redshifts are Szokoly et al.\ (2004),
Casey et al.\ (2012), Kurk et al.\ (2013), Inami et al.\ (2017), and 
Franco et al.\ (2018; including B.~Mobasher 2018, private communication and
G.~Brammer 2018, private communication).

In cases where we do not have a specz, we use photometric
redshift estimates (photzs) from the literature, though we caution that there is considerable
scatter in the estimates from different literature catalogs for optical/NIR
faint SMGs (see, e.g., Cowie et al.\ 2018). 
For the GOODS-N, we adopt the photzs from the 3D-HST survey of Momcheva et al.\ (2016), 
who used a combination of multiband photometry and G141 2D spectral data.
We supplement these with photzs from Yang et al.\ (2014), 
who used the EAZY code (Brammer et al.\ 2008) to fit 15 broadbands
from the $U$-band to the infrared (IRAC 4.5~$\mu$m) over the 
Hawaii-Hubble Deep Field North (Capak et al.\ 2004).
For these, we only use the photzs that satisfy their quality flag $Q_z<1$.

For the GOODS-S, we adopt the photzs from Straatman et al.\ (2016),
who used EAZY to fit the extensive ZFOURGE catalog from 0.3 to 8~$\mu$m. 
None of the photzs in our sample exceed their quality flag $Q<3$.

\begin{figure}
\centerline{\includegraphics[width=9cm,angle=0]{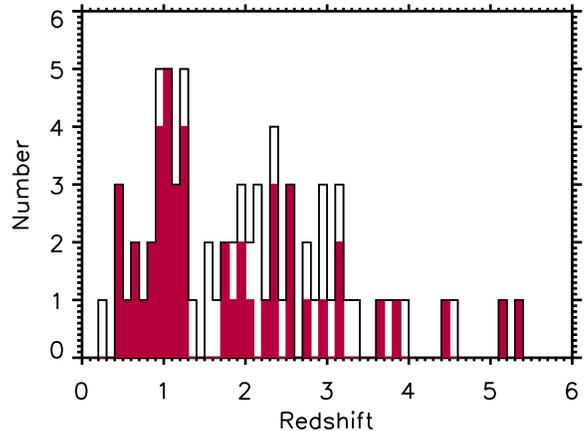}}
\caption{
Histogram of the redshift distribution for the \afluxa selected combined GOODS sample. 
There are 16 sources without either
a specz (red shaded histogram) or a photz (open histogram).
}
\label{zdist}
\end{figure}

We show the redshift distribution for the \afluxa selected combined GOODS sample
in Figure~\ref{zdist}. After excluding the 16 sources without either a
specz or a photz, the median redshift for the remaining 73 sources is $z=1.99$ 
with a 68 percent confidence range from 1.55 to 2.10. This is consistent with 
the median redshifts of $z=1.95\pm0.19$ from 
Casey et al.\ (2013) and $z=1.66\pm0.18$ from Zavala et al.\ (2018).
However, as we will discuss in Section~\ref{secfluxratios}, median redshifts are not too 
meaningful when there is such a wide spread in the redshift distribution of the sample.

%-----------------------------------------------------------------------------
% FIGURE 5 ; 450_zhist.ps
%-----------------------------------------------------------------------------
%-----------------------------------------------------------------------------

%-----------------------------------------------------------------------------
\section{\afluxa Number Counts} 
\label{seccounts}
\begin{figure}
\centerline{\includegraphics[width=9cm,angle=0]{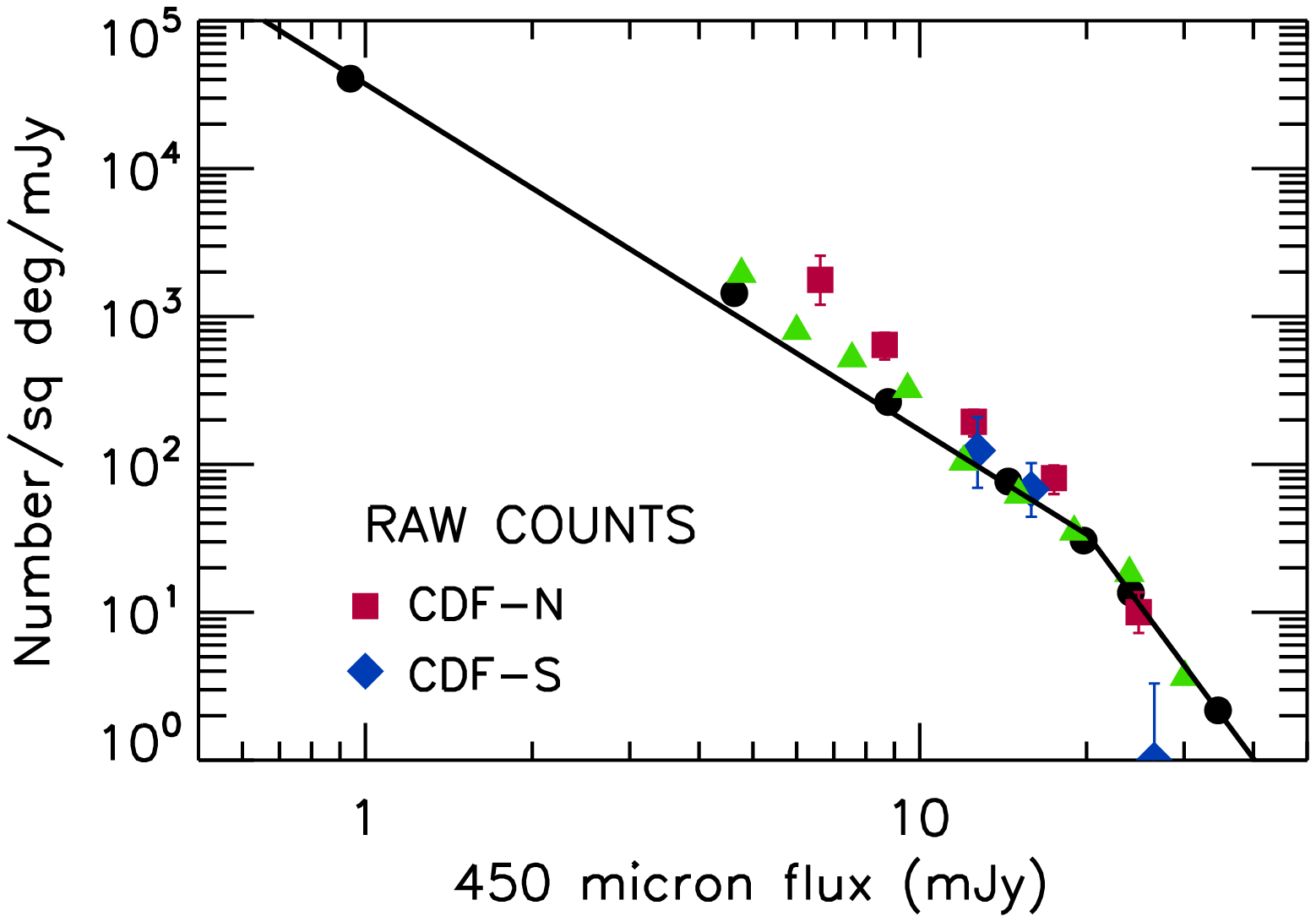}}
\centerline{\includegraphics[width=9cm,angle=0]{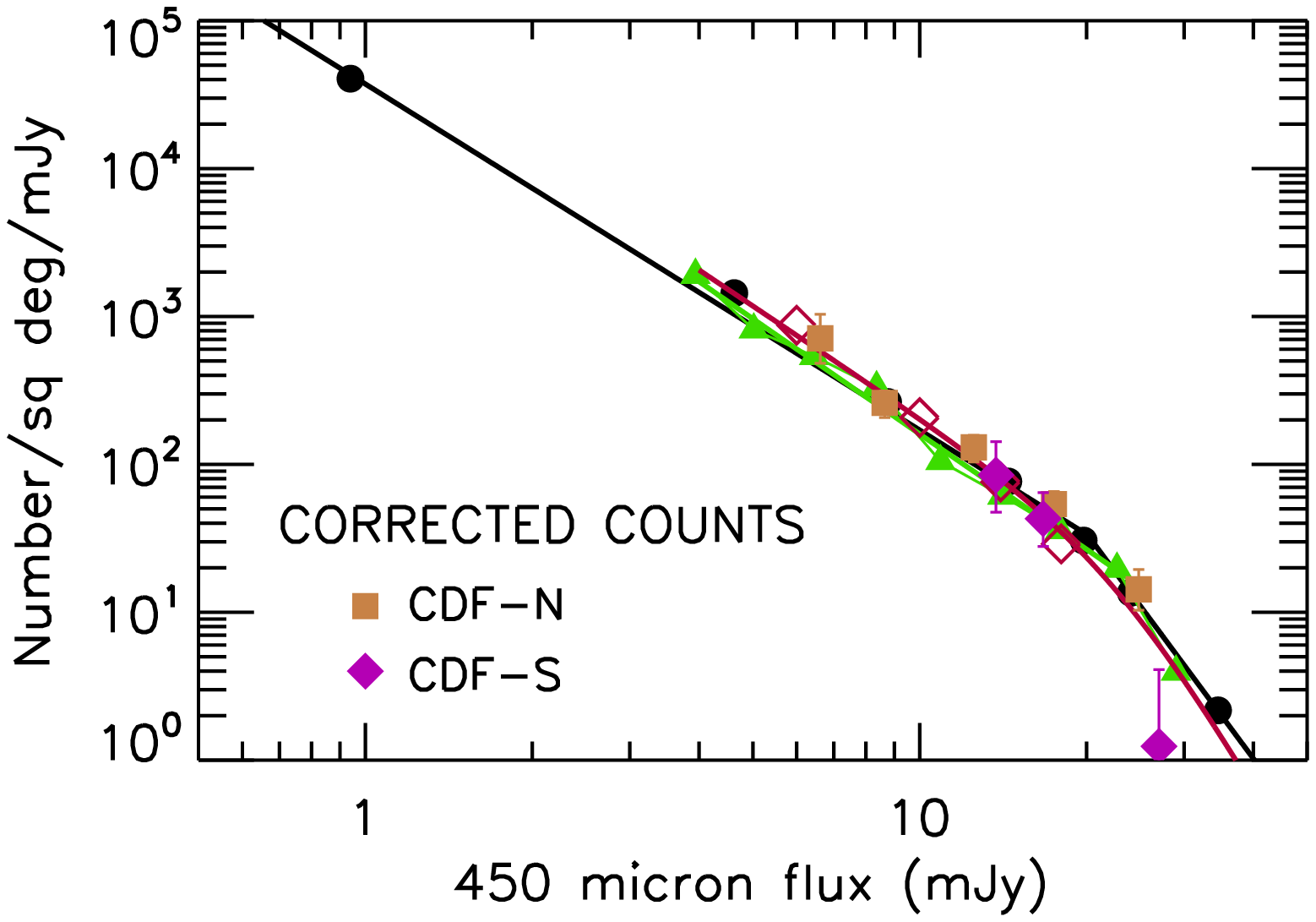}}
\caption{
Differential \afluxa (a) raw number counts (red solid squares---CDF-N; 
blue solid diamonds---CDF-S; 
green solid triangles---STUDIES from Wang et al.\ 2017) 
and (b) corrected number counts (orange solid squares---CDF-N; purple solid diamonds---CDF-S;
green solid triangles---STUDIES from Wang et al.; open diamonds---S2CLS EGS from Zavala et al.\ 2017).
In both panels, the Hsu et al.\ (2016) corrected number counts (black solid circles) and best-fit
broken power law (black lines) are shown.
}
\label{numcounts}
\end{figure}

%-----------------------------------------------------------------------------
In order to test the flux calibration and consistency with previous number counts,
we construct the differential number counts at \afluxar.
Each source contributes $1/(A_e(S)dS)$ to the counts, where $A_e$ is the effective area where
the source can be detected at $>4\sigma$ given its measured flux density $S$, and $dS$
is the flux density interval. The error is taken to be Poissonian.
We show the raw differential \afluxa number counts for both fields in Figure~\ref{numcounts}(a), 
where we compare them with the raw STUDIES counts (Wang et al.\ 2017; green triangles).

There are observational biases that affect the raw counts. These include flux boosting from
Eddington bias (Eddington 1913), spurious sources, source blending,
and detection incompleteness. In order to correct the raw counts for these effects to obtain
the intrinsic counts, we need to perform Monte Carlo simulations.
We do this by first generating source-free 
maps with only pure noise for each of our fields. 
These pure noise maps are sometimes referred to as jackknife maps 
in the literature (see, e.g., Cowie et al.\ 2002). 
Following Chen et al.\ (2013a, 2013b), for each field, we subtract two 
maps that were each produced by coadding roughly half of the flux-calibrated 
data. This process subtracts off the real sources,
leaving residual maps that are free of any sources. 
We then rescale the value of each pixel by a factor of 
$\sqrt{t_1\times t_2}/(t_1+t_2)$, with $t_1$ and $t_2$ representing
the integration time of each pixel from the two maps. 

Next, we generate 40 simulated maps for each field by randomly populating our pure noise maps
with the same number of sources as in our $>4\sigma$ samples.
We draw these sources from the best-fit broken power law count model of Hsu et al.\ (2016),
\begin{equation}
{dN\over dS}=
\begin{dcases} 
N_0\Bigl({S\over S_0}\Bigr)^{-\alpha} \quad \text{if} \quad S\le S_0 \\
N_0\Bigl({S\over S_0}\Bigr)^{-\beta} \quad \text{if} \quad S> S_0  \\
\end{dcases}
\end{equation}
where $N_0=33.3$, $S_0=20.1$, $\alpha=2.34$, and $\beta=5.06$.

For each simulated map, we rerun our source extraction down to the $4\sigma$ threshold
and compute the recovered counts using the same method and flux bins used for the 
science map. We compare the ratio of the total number of sources from the simulated maps 
with the input counts. We use all 40 simulations to determine the average ratios as a function 
of flux, which we then apply to the raw counts to get the corrected counts.
We show the corrected number counts for both fields in Figure~\ref{numcounts}(b), 
where we compare
them with the corrected STUDIES counts (Wang et al.\ 2017; green triangles).

By using SCUBA-2 observations of cluster fields
(A1689, A2390, A370, MACSJ0717.5+3745, MACSJ1149.5+2223, and MACSJ1423.8+2404), 
Hsu et al.\ (2016) were able to construct the \afluxa number counts for
flux detections ($>3\sigma$) down to fainter than 1~mJy,
thanks to gravitational lensing effects. 
They used three blank fields (CDF-N, CDF-S, and COSMOS) to cover the
brighter fluxes. We show their corrected number counts (black solid circles) and best-fit 
broken power law (black lines) in both panels of Figure~\ref{numcounts}.
The consistency with Hsu et al.\ (2016) means it is not necessary to iterate on the input 
counts. A multiplicative error of 1.15 in the flux normalization in either direction would 
result in a significant discrepancy relative to previous counts.
We note that the \afluxa counts are very uniform across fields, which is also seen by
W.-H.~Wang (private communication).

%-----------------------------------------------------------------------------
% FIGURE 6 ; raw_diff_counts.ps, corr_diff_counts.ps
%-----------------------------------------------------------------------------
%-----------------------------------------------------------------------------

%-----------------------------------------------------------------------------
\section{Flux Ratios}
\label{secfluxratios}
%-----------------------------------------------------------------------------
The empirical observation of
a tight correlation between thermal dust emission and radio continuum emission
(e.g., Helou et al.\ 1985; Condon et al.\ 1991), known as the FIR-radio correlation,
is thought to result from both quantities being linearly related to the massive SFR (Condon 1992),
although we have no good theoretical understanding of the correlation. If this correlation 
continues to hold at high redshifts, as evidence suggests it does at least out to $z\sim 5$
(e.g., Barger et al.\ 2012, 2015; Thomson et al.\ 2014; Lim et al.\ 2020), 
then very sensitive radio observations 
can be used to localize distant galaxies detected in the submillimeter. In the early days of trying 
to understand the SCUBA sources, 20~cm data from the VLA were used for just this purpose
(e.g., Barger et al.\ 2000; Smail et al.\ 2000; Chapman et al.\ 2003).
In combination with the submillimeter data, the 20~cm data have the advantage of allowing crude 
redshift estimates (sometimes called millimetric redshifts)
to be made, based on the opposing spectral slopes of the blackbody 
spectrum in the submillimeter and the synchrotron spectrum in the radio 
(e.g., Carilli \& Yun 1999; Barger et al.\ 2000).

%-----------------------------------------------------------------------------
% FIGURE 7 ; 450_ratiox2.ps, 450_radratz.ps, 450_smmratz.ps
%-----------------------------------------------------------------------------
\begin{figure*}
\centerline{\includegraphics[width=9cm,angle=0]{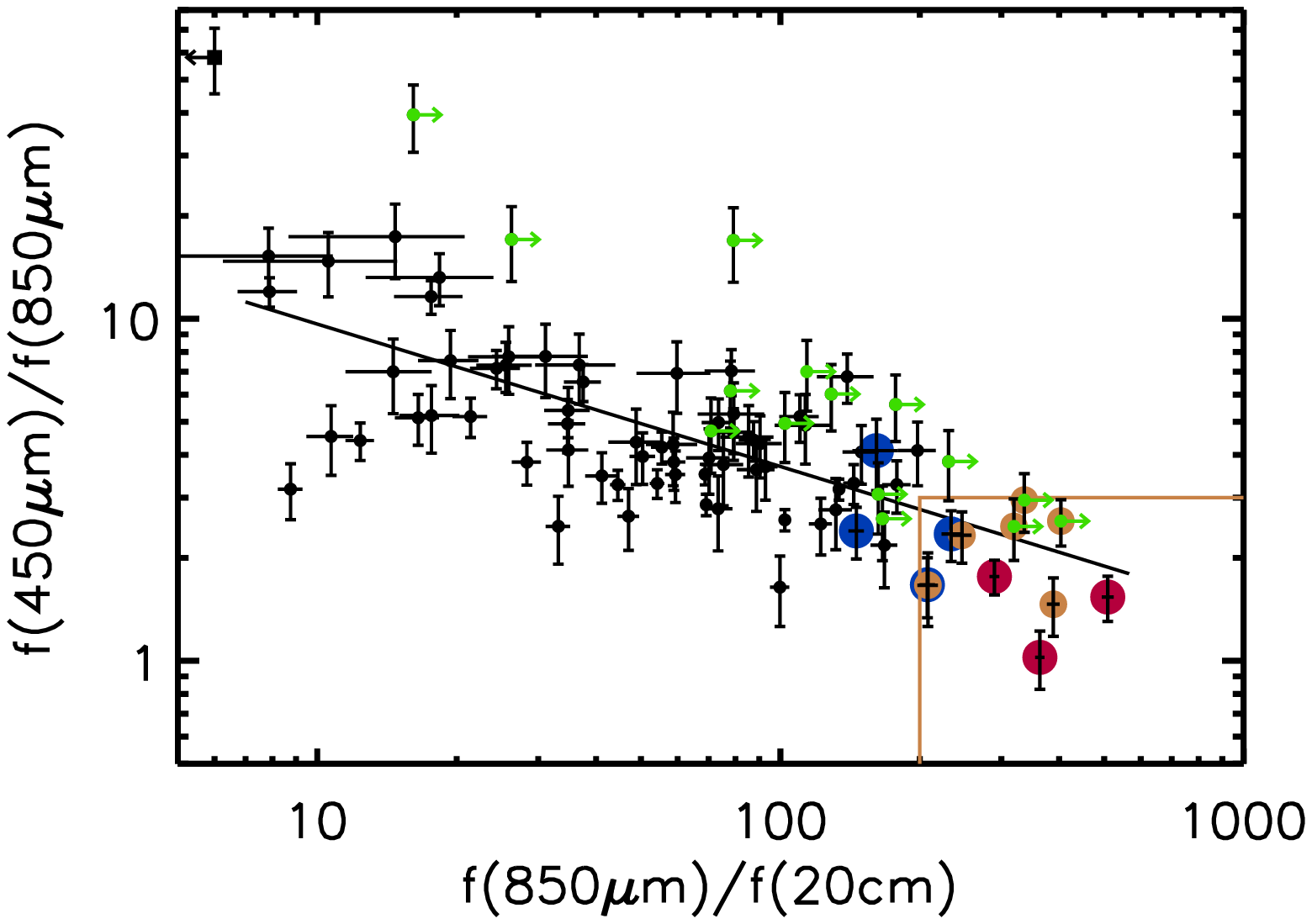}}
\centerline{\includegraphics[width=9cm,angle=0]{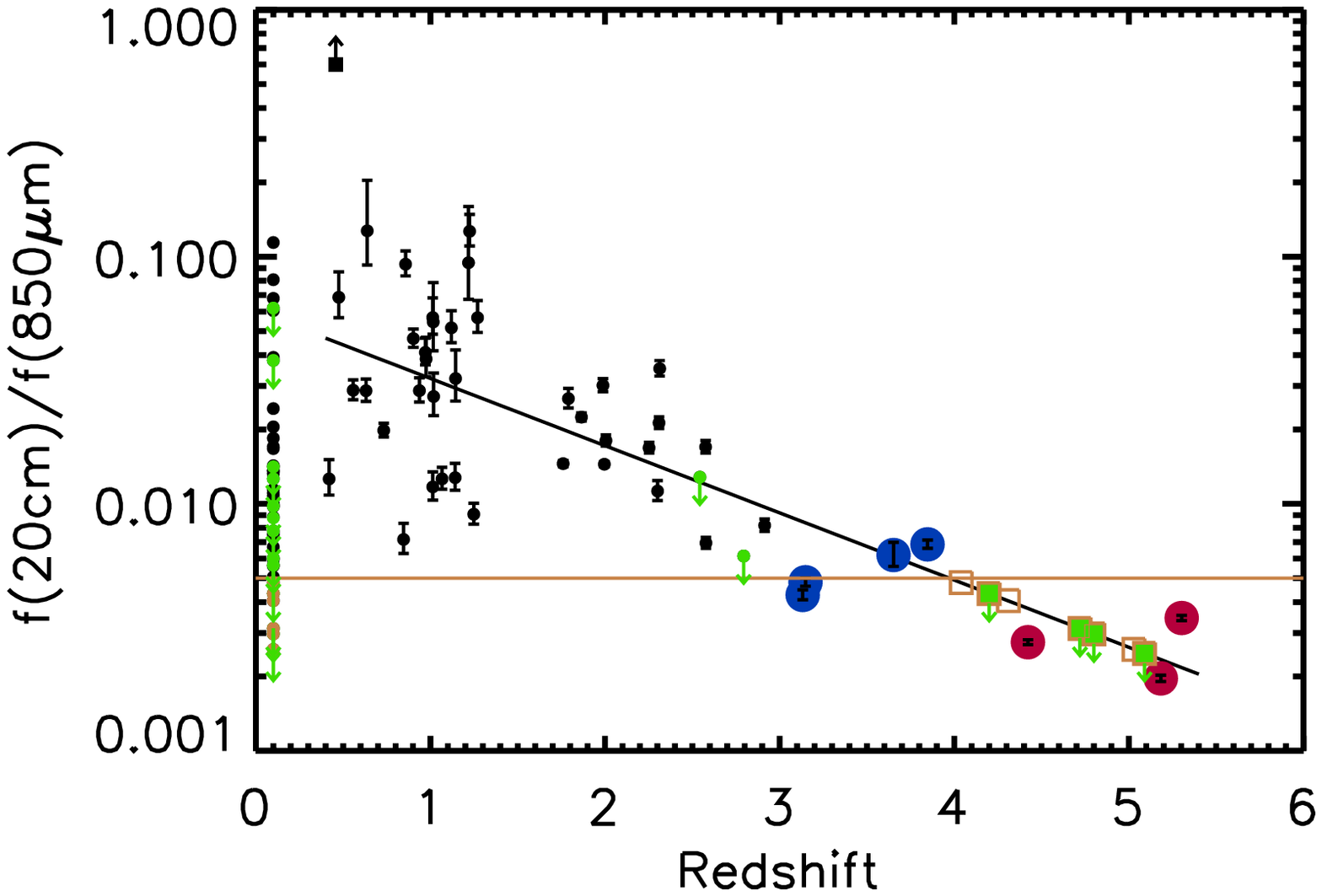}
\includegraphics[width=9cm,angle=0]{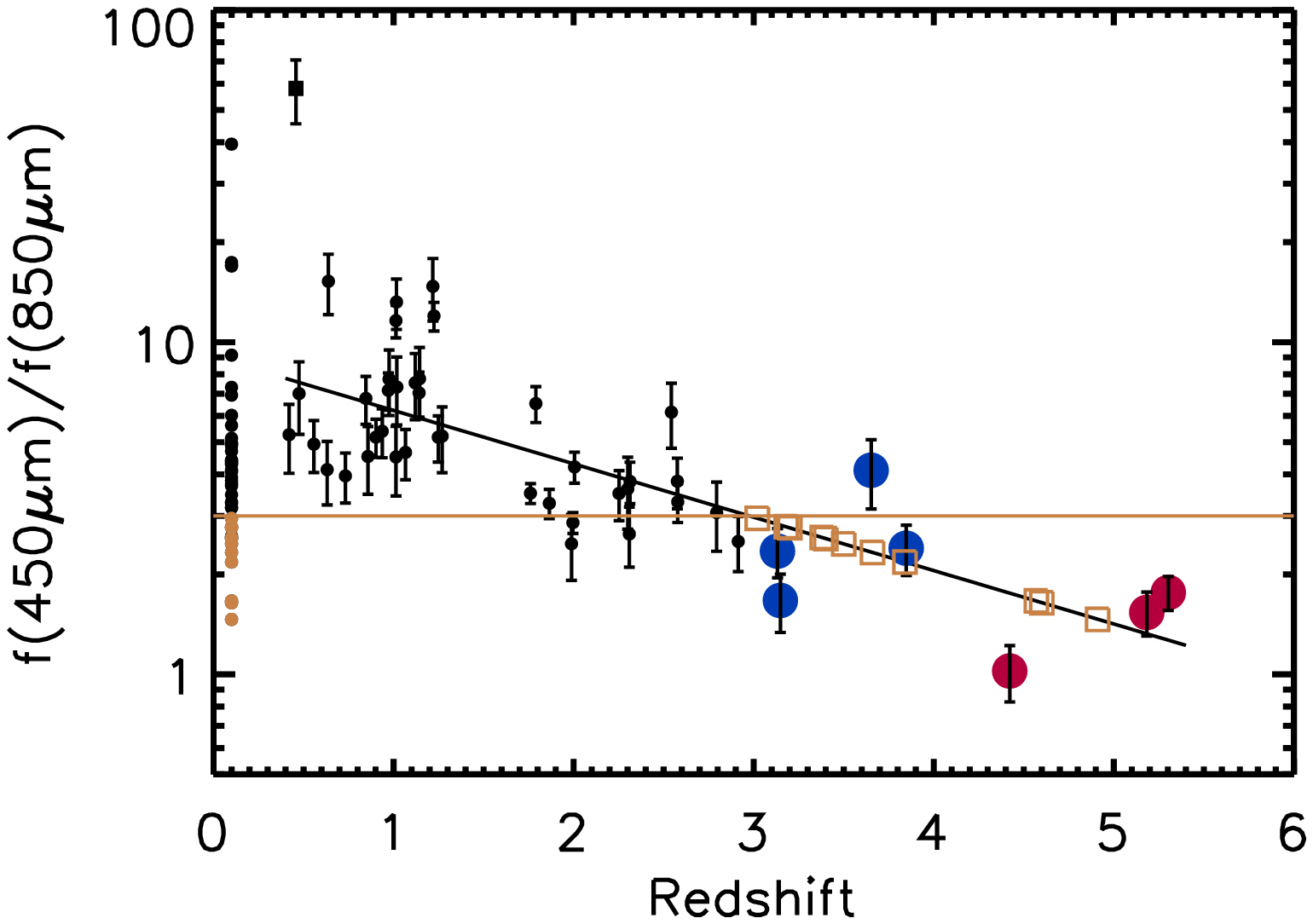}}
\caption{
Based on our \afluxa selected combined GOODS sample:
(a) \afluxa to \afluxb flux ratio (uncertainties from the \afluxa measurements)
vs. \afluxb to 20~cm flux ratio (uncertainties from the \afluxb measurements) for sources
with radio data,
(b) 20~cm to \afluxb flux ratio vs. redshift
(uncertainties from the \afluxb measurements) for sources with radio data, and (c)
\afluxa to \afluxb flux ratio vs. redshift (uncertainties from the \afluxa measurements).
Sources with only radio limits are shown in green with (a) rightward-pointing arrows
and (b) downward-pointing arrows.
In (b) and (c), sources without speczs are placed at a 
nominal redshift of $z=0.1$ and shown without uncertainties, for clarity. In all panels,
solid lines show simple linear fits to the data, without considering uncertainties and after 
excluding GOODS-N source~70, 
which is at the top-left of each plot (black square; arrows denote when its 
value is off the plot). Sources with $3\le z_{spec}<4$
are shown in blue, while those with $z_{spec}\ge4$ are shown in red. 
Our adopted high-redshift selection criteria are denoted by gold lines, 
and the sources that satisfy the (a) combined or (b) and (c) individual
criteria but do not have speczs are colored gold. In (b) and (c), we estimate
where these sources would lie in redshift based on the given linear fit (gold open squares).
\label{fluxratios}
}
\end{figure*}
%-----------------------------------------------------------------------------

FIR/submillimeter flux ratios have also been used to estimate redshifts.
For example, Figure~19 of Cowie et al.\ (2017) showed that
for their \afluxb sample in the GOODS-N, \flipradfluxratio\ and \firfluxratio\
were well correlated, 
suggesting that both provide a measure of the redshifts. They also found
that those with speczs were well segregated by
redshift interval. Interestingly, they determined that the scatter in specz
versus \flipradfluxratio\ was considerably larger than the scatter in 
specz versus \firfluxratio\ (their Figure~20). This is a clear
indication that the radio power is not always an accurate measure of the star formation
(see Barger et al.\ 2017 for a detailed analysis of the star-forming galaxies and AGNs 
in the GOODS-N that lie on the FIR-radio correlation).
In contrast, the scatter in specz versus \firfluxratio\ is primarily caused by 
variations in the SEDs of the sources.

Other authors (e.g., Casey et al.\ 2013; Wang et al.\ 2019; Lim et al.\ 2020)
have constructed the flux ratios for SMGs in COSMOS
and compared them with optical/NIR photzs. 
Of particular interest is the identification of
high-redshift sources. For example, Wang et al.\ plotted \almafluxratio\
versus $z_{phot}\gtrsim3$ for their ALMA-detected $H$-dropouts in  their 
Extended Data Figure~4.

To look for high-redshift candidates in our \afluxa selected combined GOODS  sample,
in Figure~\ref{fluxratios}(a), we plot  \flipsmmfluxratio\ versus \radfluxratio.
We color-code the $3\le z_{spec}<4$ (blue large circles) and $z_{spec}\ge4$ sources (red large circles).
We adopt the uncertainties on the \afluxa fluxes in the y-axis, since we expect them to be 
larger than the uncertainties on the \afluxb fluxes, and we adopt the uncertainties on
the \afluxb fluxes in the x-axis, since we expect them to be larger than the
uncertainties on the 20~cm fluxes. We assign the 20~cm ($>5\sigma$) flux limits
of the two samples 
to the sources in the radio areas without radio detections (green rightward-pointing arrows).

We can see the good relation between the two flux ratios.  
A linear fit without considering uncertainties and after excluding source 70, which is located at the
top-left of the plot, gives
$\log($\flipsmmfluxratio)$=-0.42\log($\radfluxratio$)+1.40$ (solid line).

We can also see the generally good 
segregation of the high-redshift sources with speczs from the rest of the population.
Thus, we define high-redshift selection criteria based on this plot, namely, 
\flipsmmfluxratio$<3$
and \radfluxratio$>200$ (gold lines). Based on these criteria, we have
six candidate high-redshift sources (gold large circles).

In Figures~\ref{fluxratios}(b) and (c), respectively, we show the two axes from (a)
separately versus redshift. We again color-code in blue and red the high-redshift sources with speczs.
We plot the sources without speczs at a nominal redshift of $z=0.1$ and without uncertainties, for clarity.
In (b), we again show the sources in the radio areas without radio detections as limits
(green downward-pointing arrows).
In both panels, we show the linear fits as solid lines without considering 
uncertainties and after excluding source 70, but we only include in the fits those sources having speczs. 
The resulting fits are (b) $\log($\flipradfluxratio)$=-0.27 z -1.22$ and
(c) $\log($\flipsmmfluxratio)$=-0.16 z+0.96$.

%-----------------------------------------------------------------------------
% FIGURE 8 ; 450_z_f4.ps, 450_z45.ps
%-----------------------------------------------------------------------------
\begin{figure*}
\centerline{\includegraphics[width=9cm,angle=0]{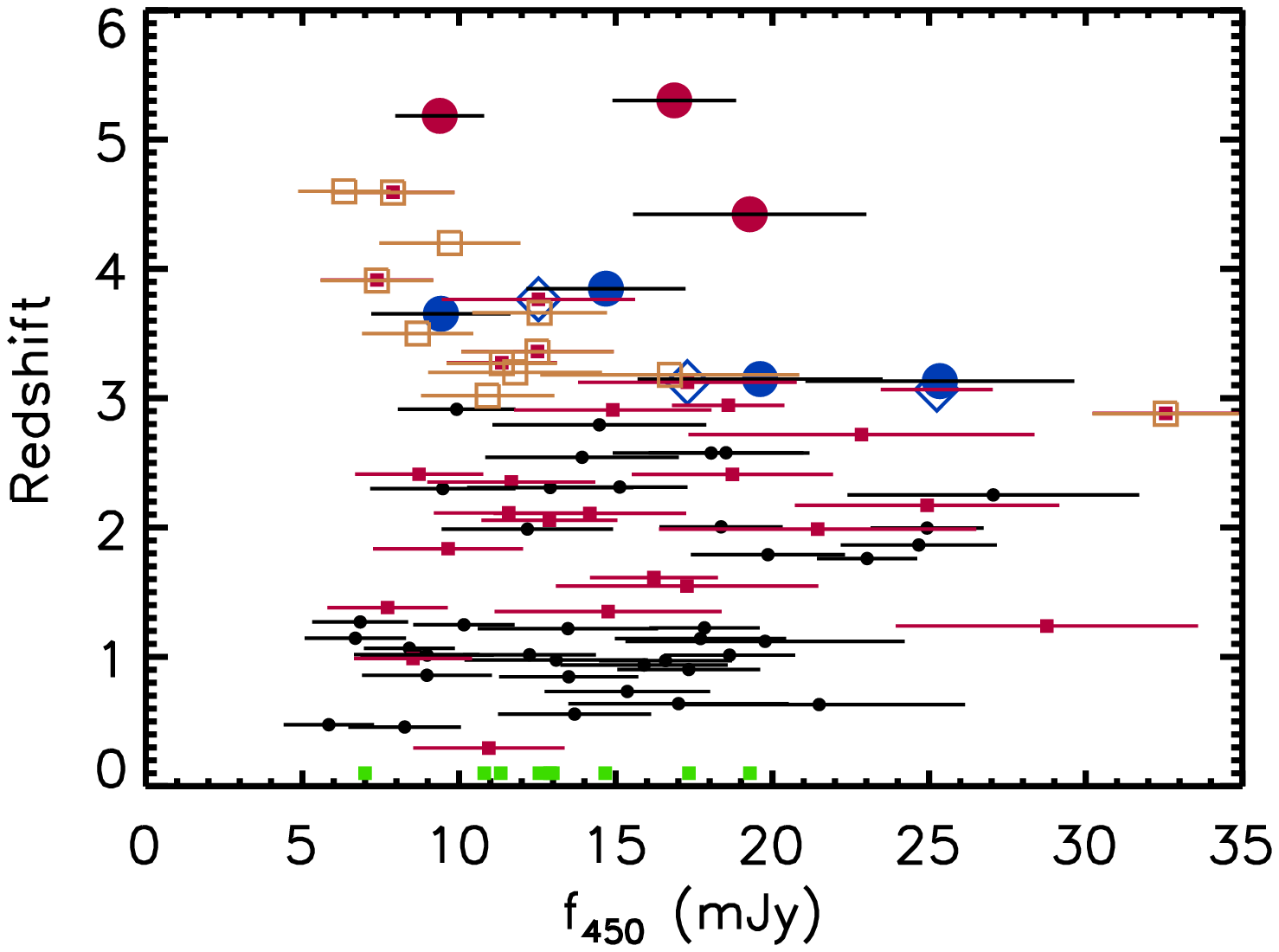}
\includegraphics[width=9cm,angle=0]{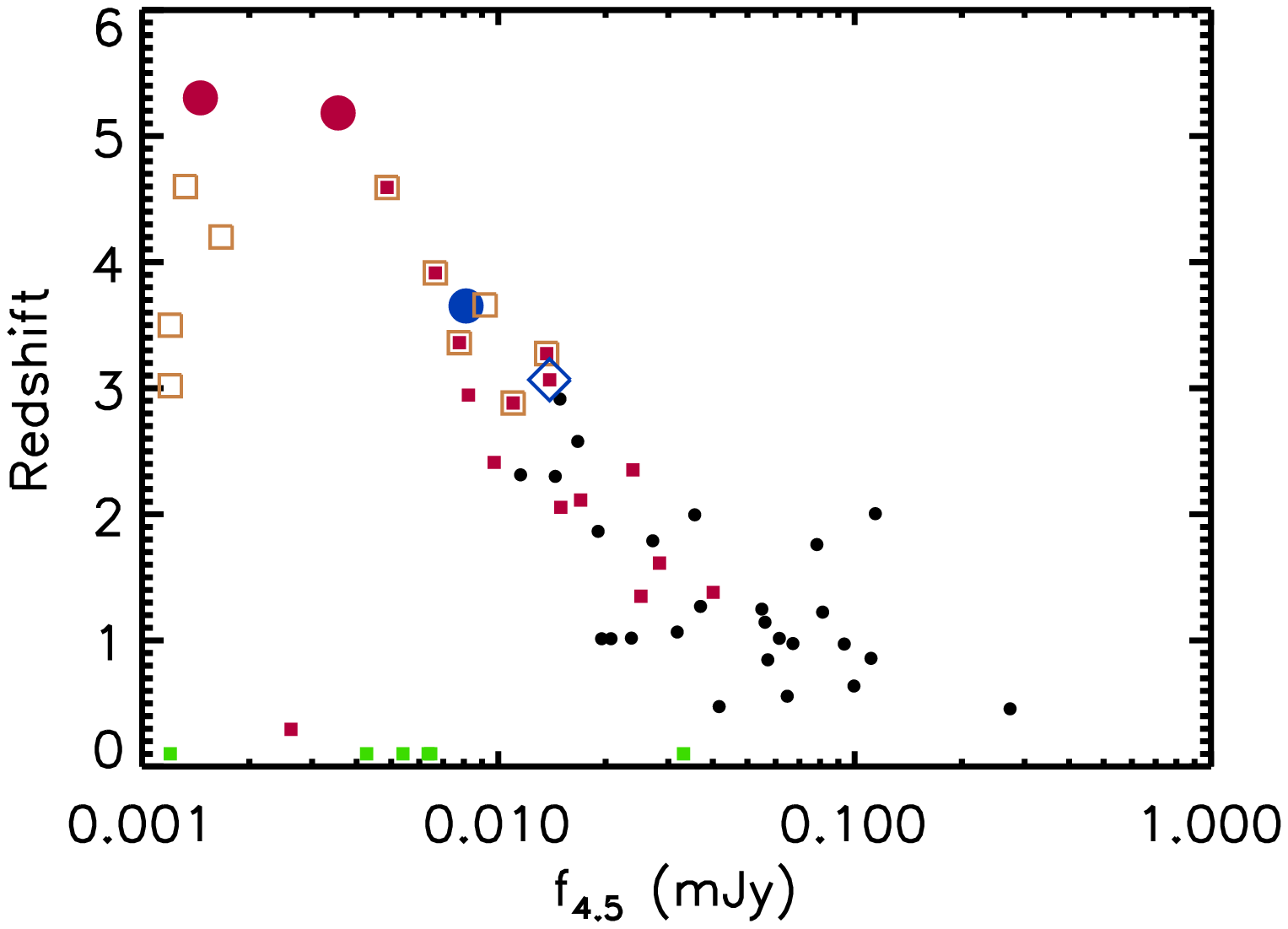}}
\caption{
(a) Redshift vs. \afluxa flux (with uncertainties)
for the \afluxa selected combined GOODS sample.
(b) Redshift vs. 4.5~$\mu$m flux for the sources in the 
\afluxa selected GOODS-N sample 
that lie within the {\em HST}/ACS footprint. We use the
Barro et al.\ (2019) catalog for most of the 4.5~$\mu$m fluxes,
but for the small number of sources not in that catalog, we measured
the fluxes ourselves. Note that
the uncertainties on the 4.5~$\mu$m fluxes are smaller than the data points.
In both panels, sources with speczs $<3$ are shown as black circles, while sources 
with photzs are shown as red squares. Sources with $3\le z_{spec}<4$
are denoted by blue large circles, while those with $z_{spec}\ge4$ are denoted by red
large circles. 
The high-redshift candidates from Table~\ref{highztab} are denoted by gold open squares
(we use the photzs for the five that have a photz).
Sources with $3\le z_{phot}<4$ that were not also identified as high-redshift candidates
from the FIR are enclosed in blue large open diamonds.
Finally, sources without either a specz or a photz and not identified as high-redshift candidates
from the FIR are placed at a nominal redshift of $z=0.1$
and shown as green squares without uncertainties, for clarity.
\label{zsmm}
}
\end{figure*}
%-----------------------------------------------------------------------------

In each figure, we show the relevant high-redshift criterion from Figure~\ref{fluxratios}(a) 
as a gold line, and we color in gold the sources that satisfy it.
In (b), we have the same high-redshift candidates
as in (a), plus one additional candidate. Note that we conservatively
consider only a radio limit that is already below the gold line to be a candidate high-redshift source.
In (c), we have the same high-redshift candidates as in (a), plus
five additional candidates, none of which is the additional candidate found in (b).
In both (b) and (c), we plot gold open squares at the redshifts we estimate from the linear fits.

%-----------------------------------------------------------------------------
% FIGURE 9
%-----------------------------------------------------------------------------
% Number scheme changed from GN: figs have table numbers not IDL numbers
% Generated using s450_magphys_sh.pro in cdfs_scuba2
% Only have images for those sources on the CANDELS region
% Source 4 is contaminated, so no SED is shown
%
\begin{figure*}
\centering
\gridline{\boxedfig{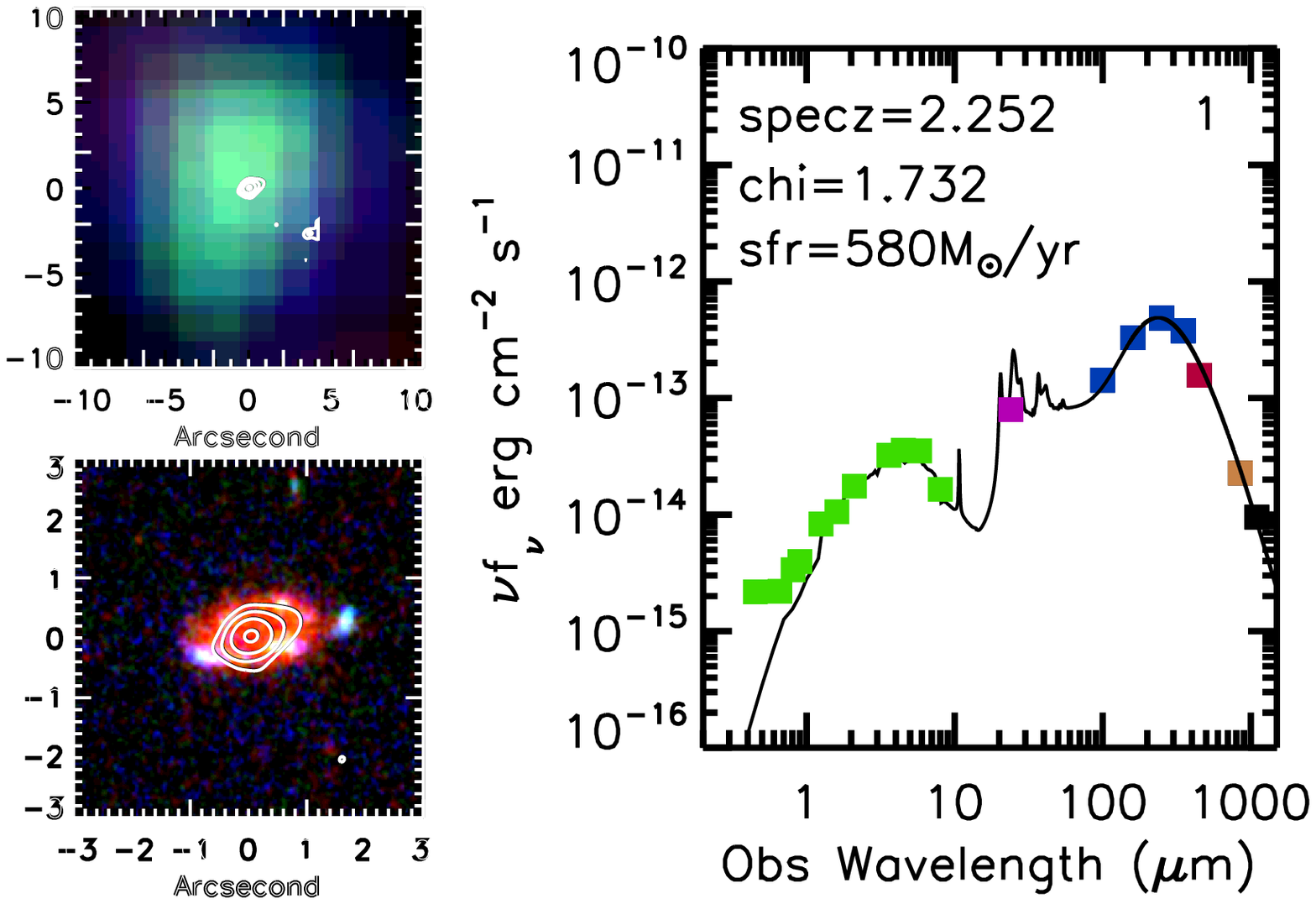}{0.325\textwidth}{(1)}
\boxedfig{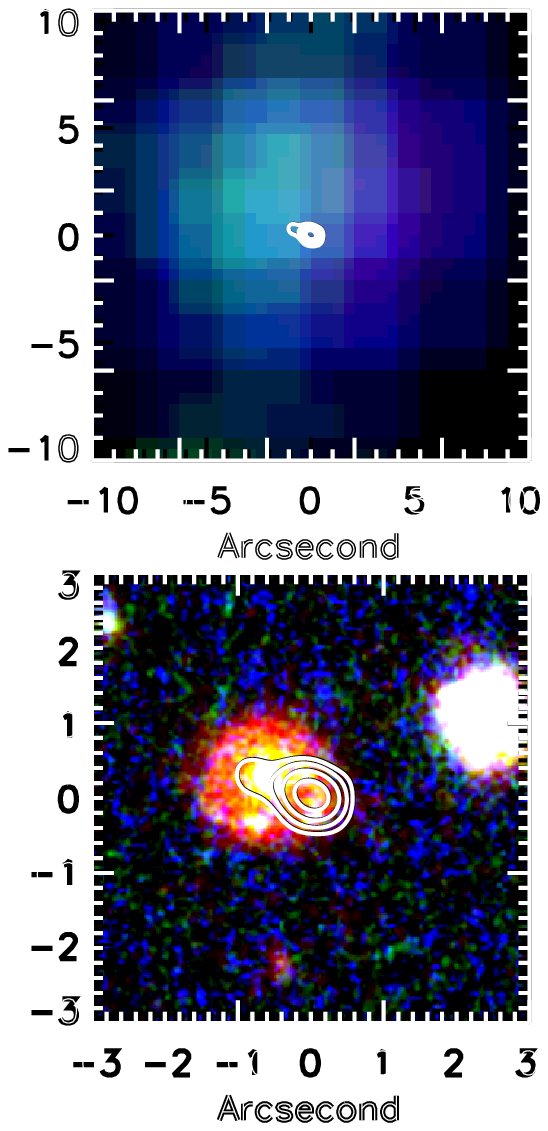}{0.325\textwidth}{(4)}
\boxedfig{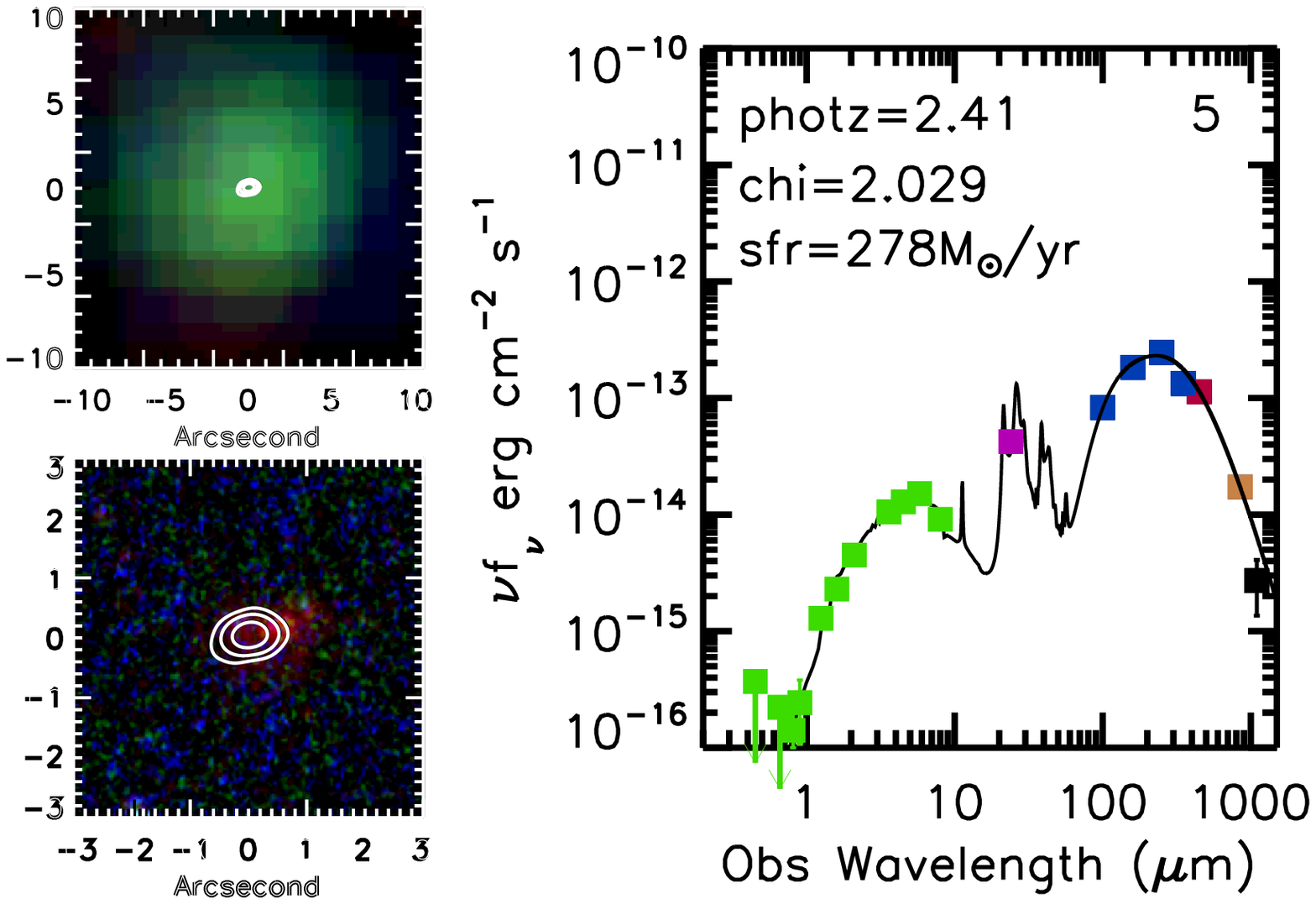}{0.325\textwidth}{(5)}
}c
\gridline{\boxedfig{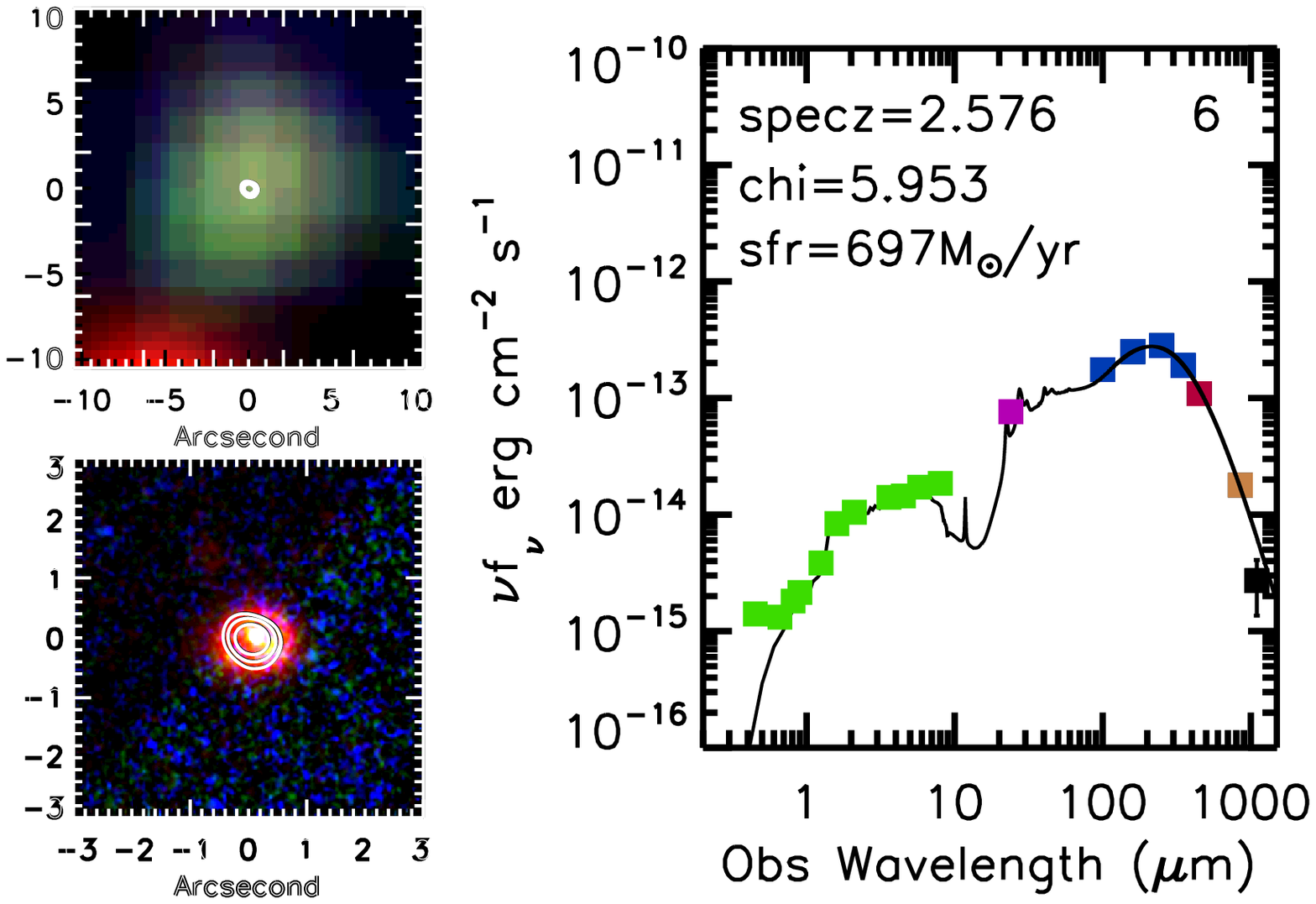}{0.325\textwidth}{(6)}
\boxedfig{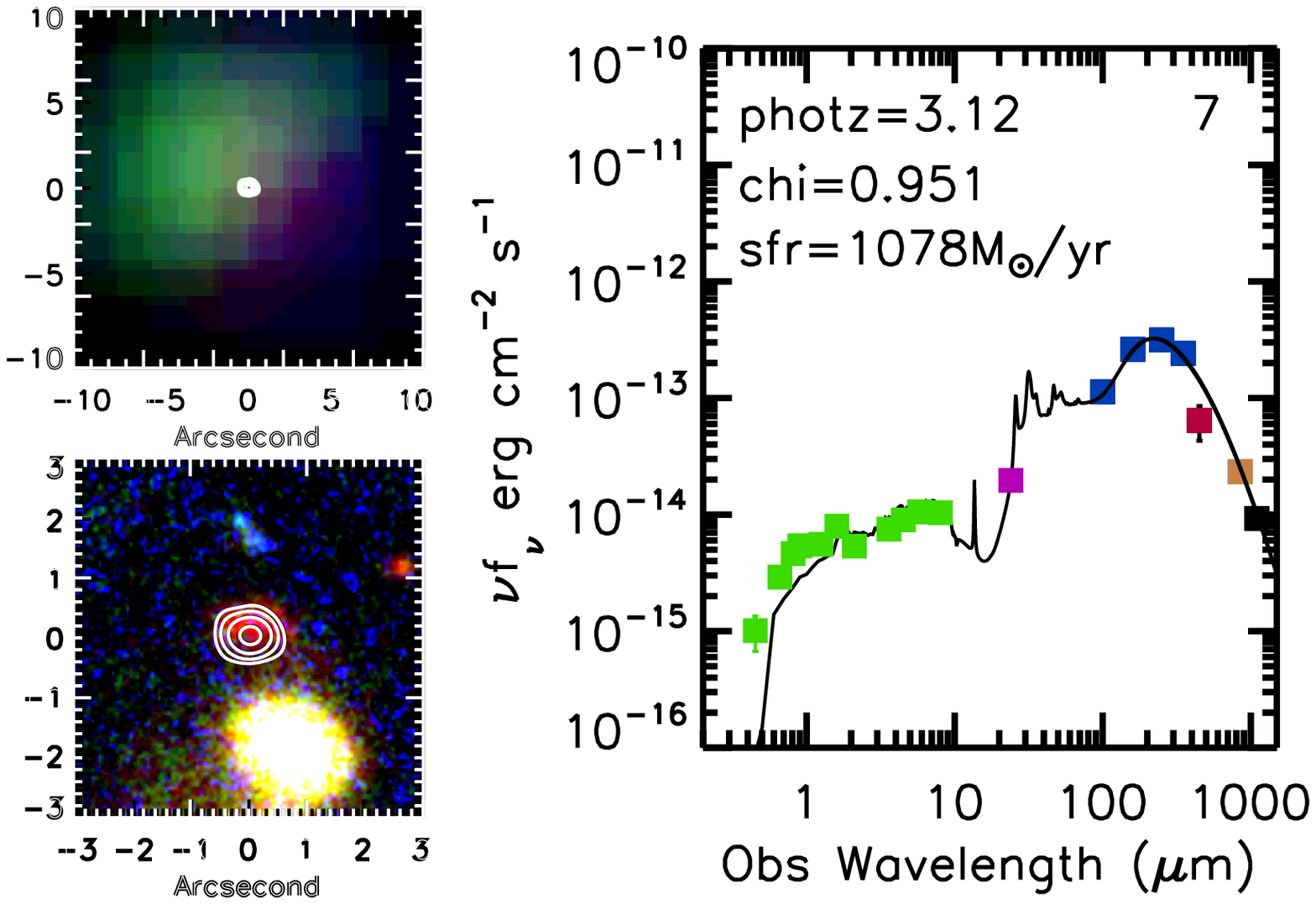}{0.325\textwidth}{(7)}
\boxedfig{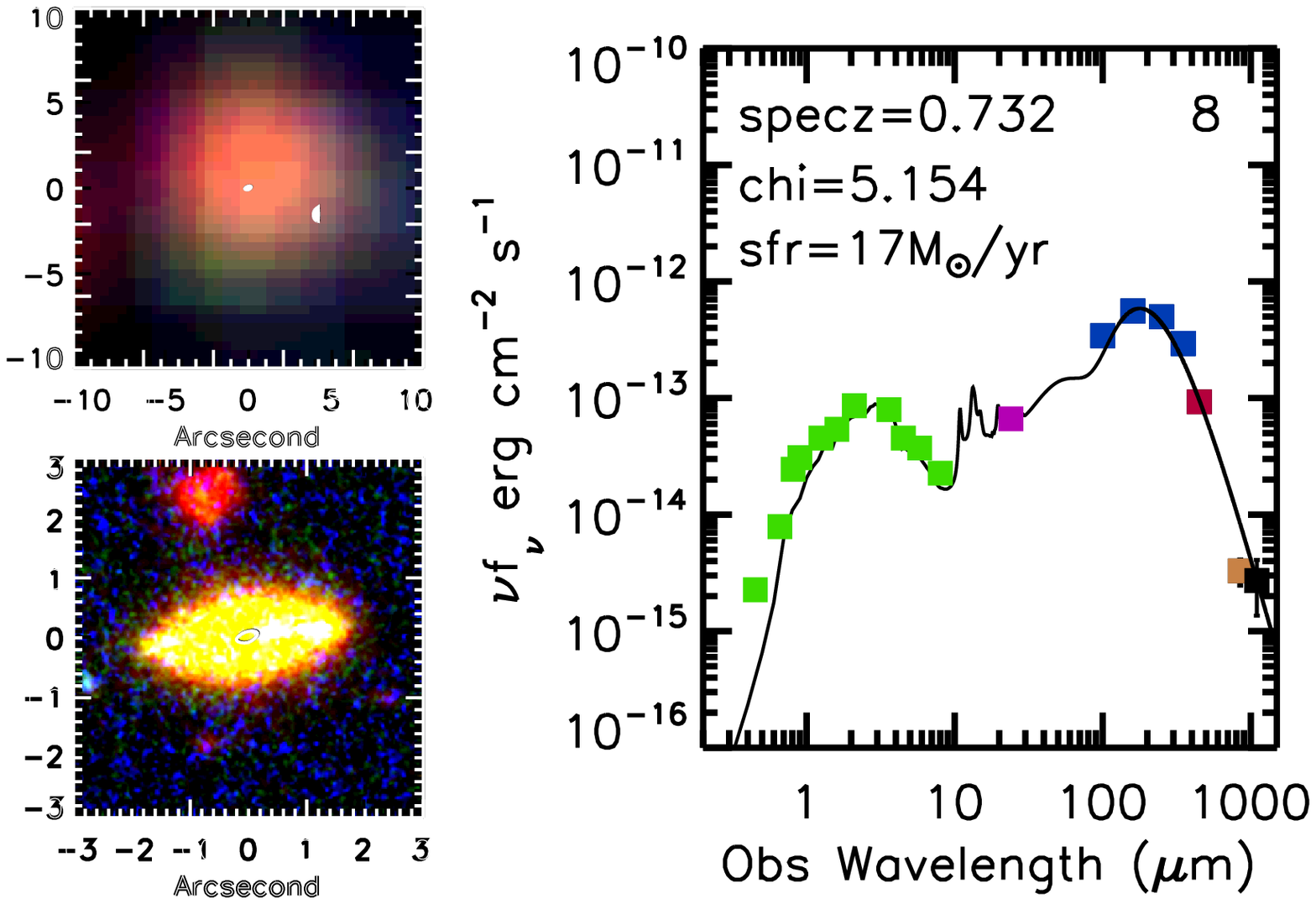}{0.325\textwidth}{(8)}
}
\gridline{\boxedfig{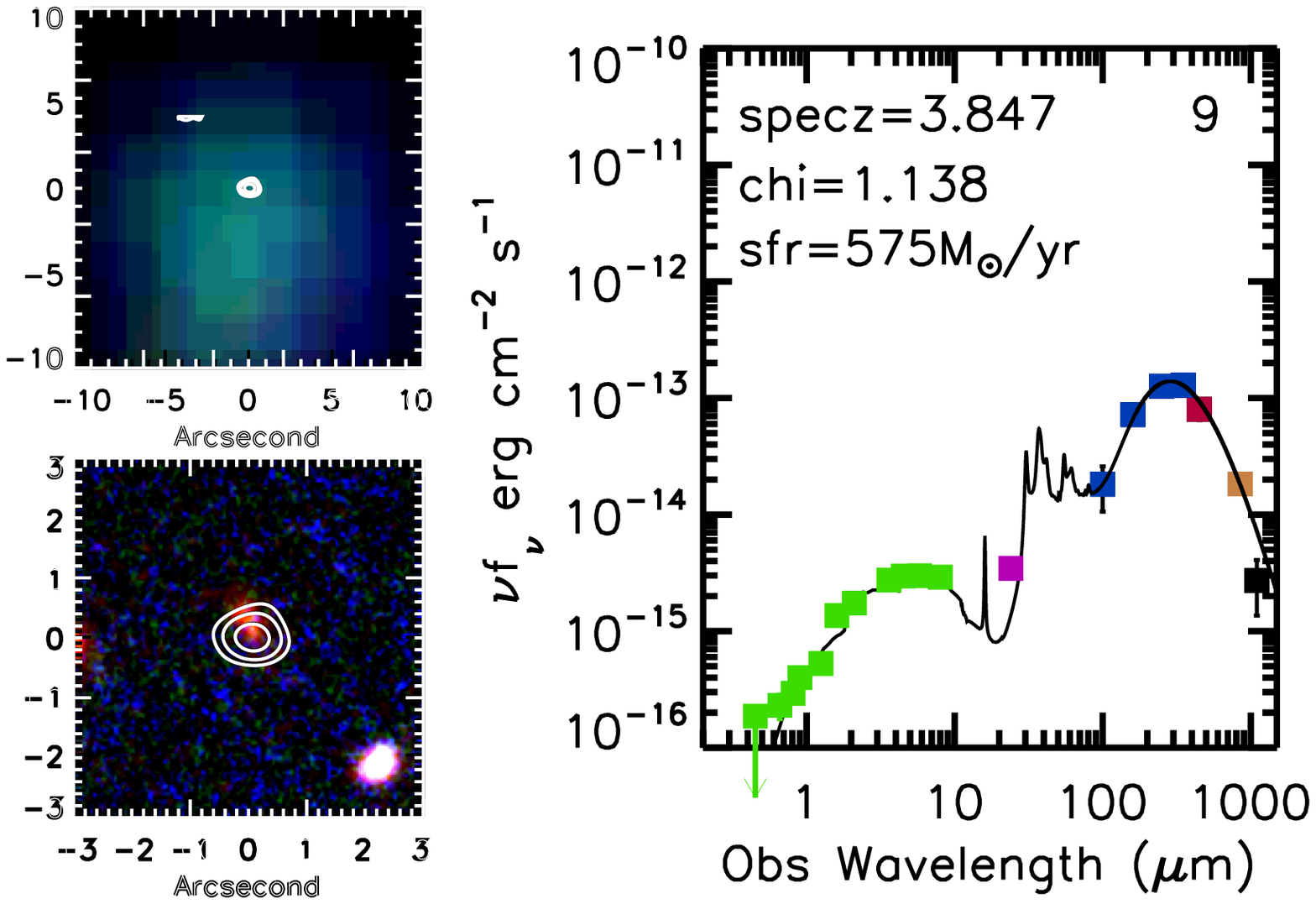}{0.325\textwidth}{(9)}
\boxedfig{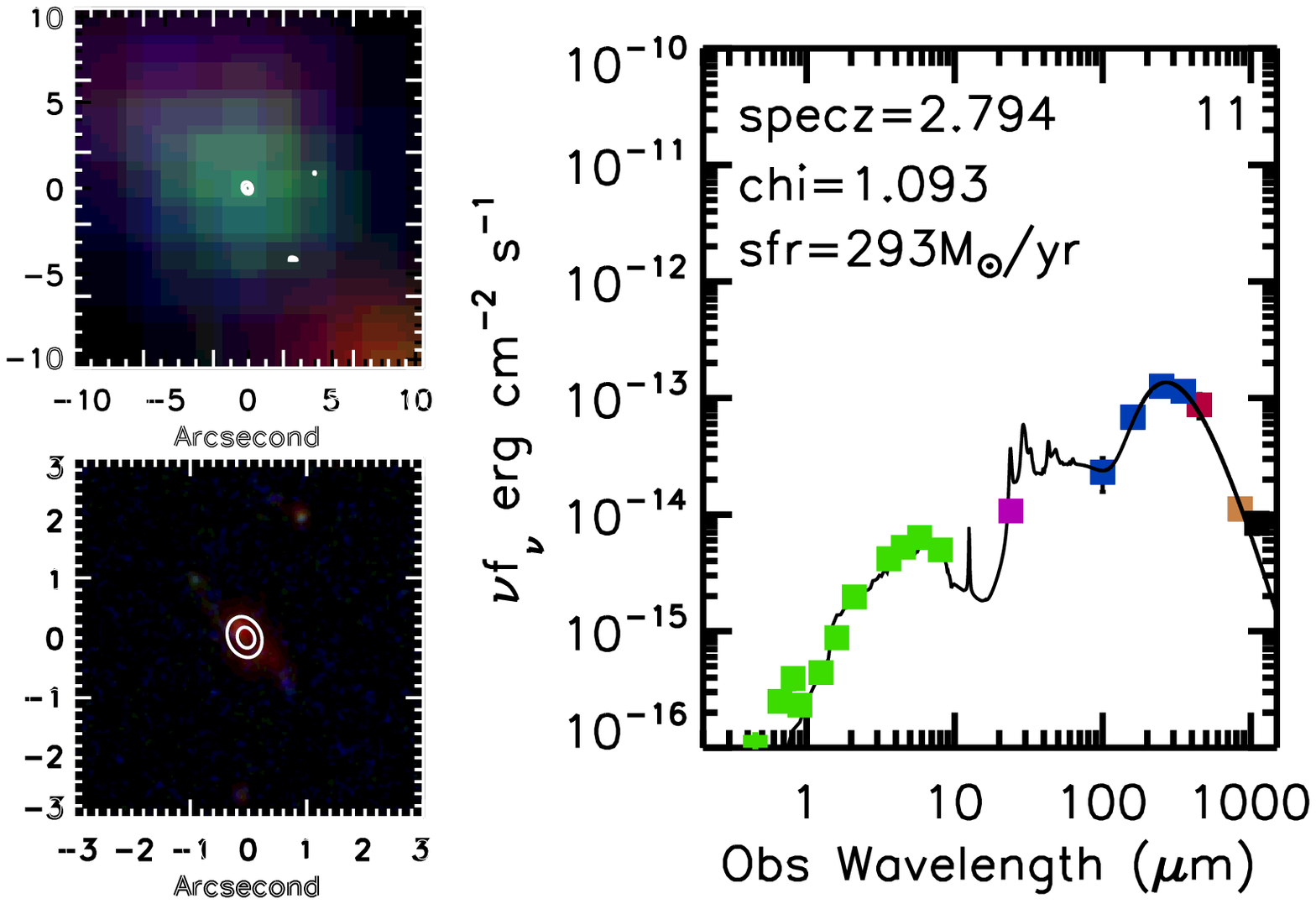}{0.325\textwidth}{(11)}
\boxedfig{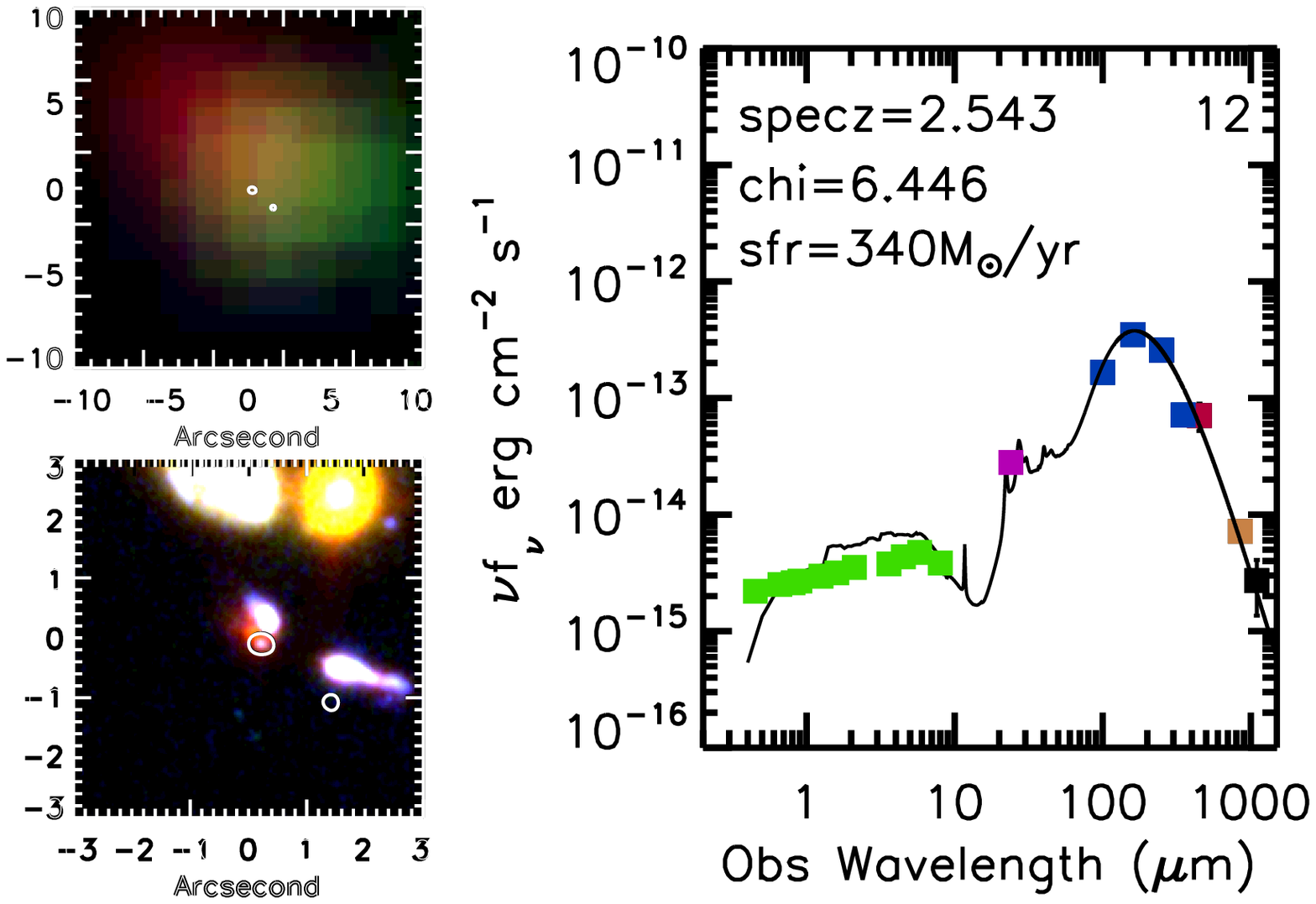}{0.325\textwidth}{(12)}
}
\gridline{\boxedfig{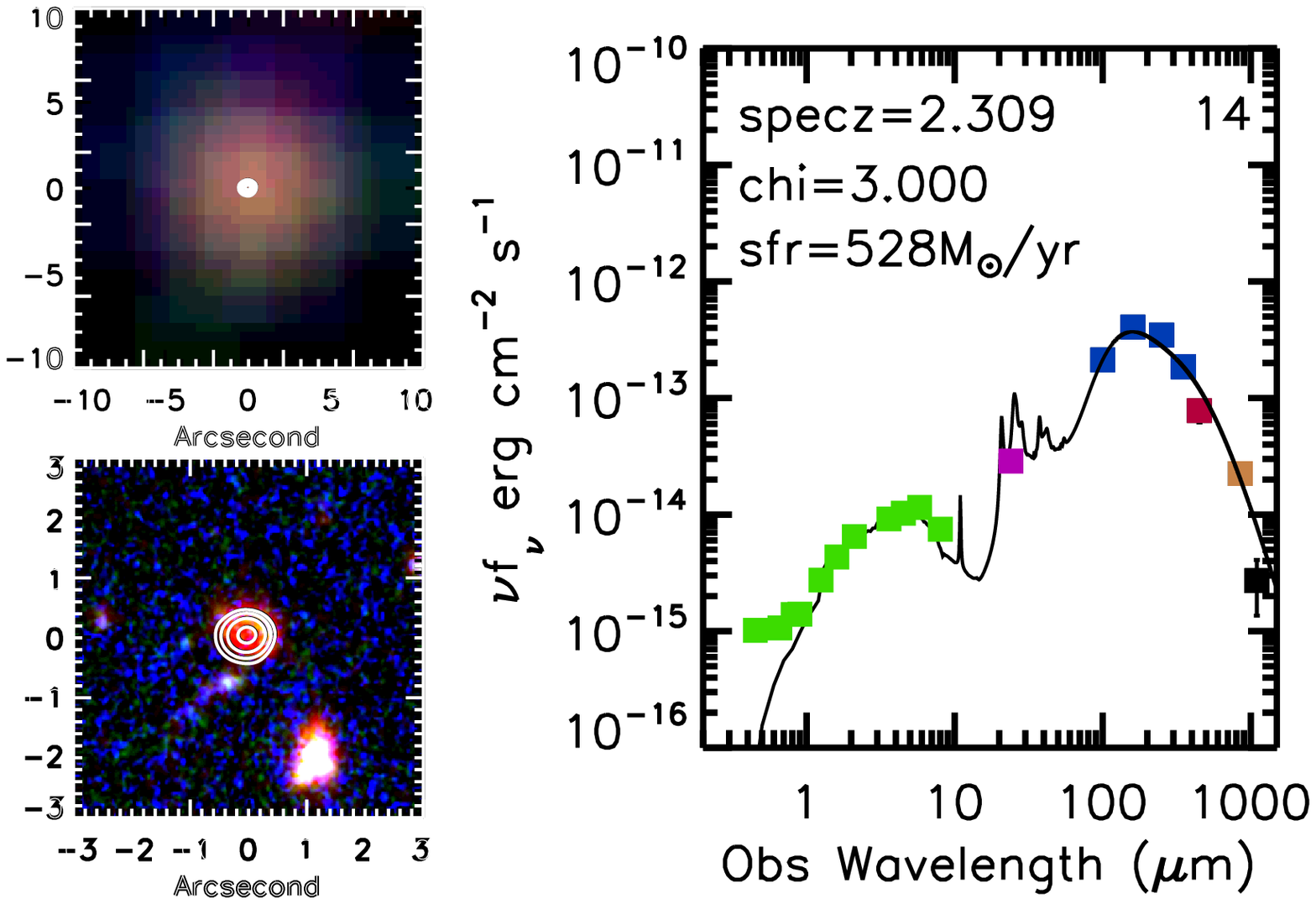}{0.325\textwidth}{(14)}
\boxedfig{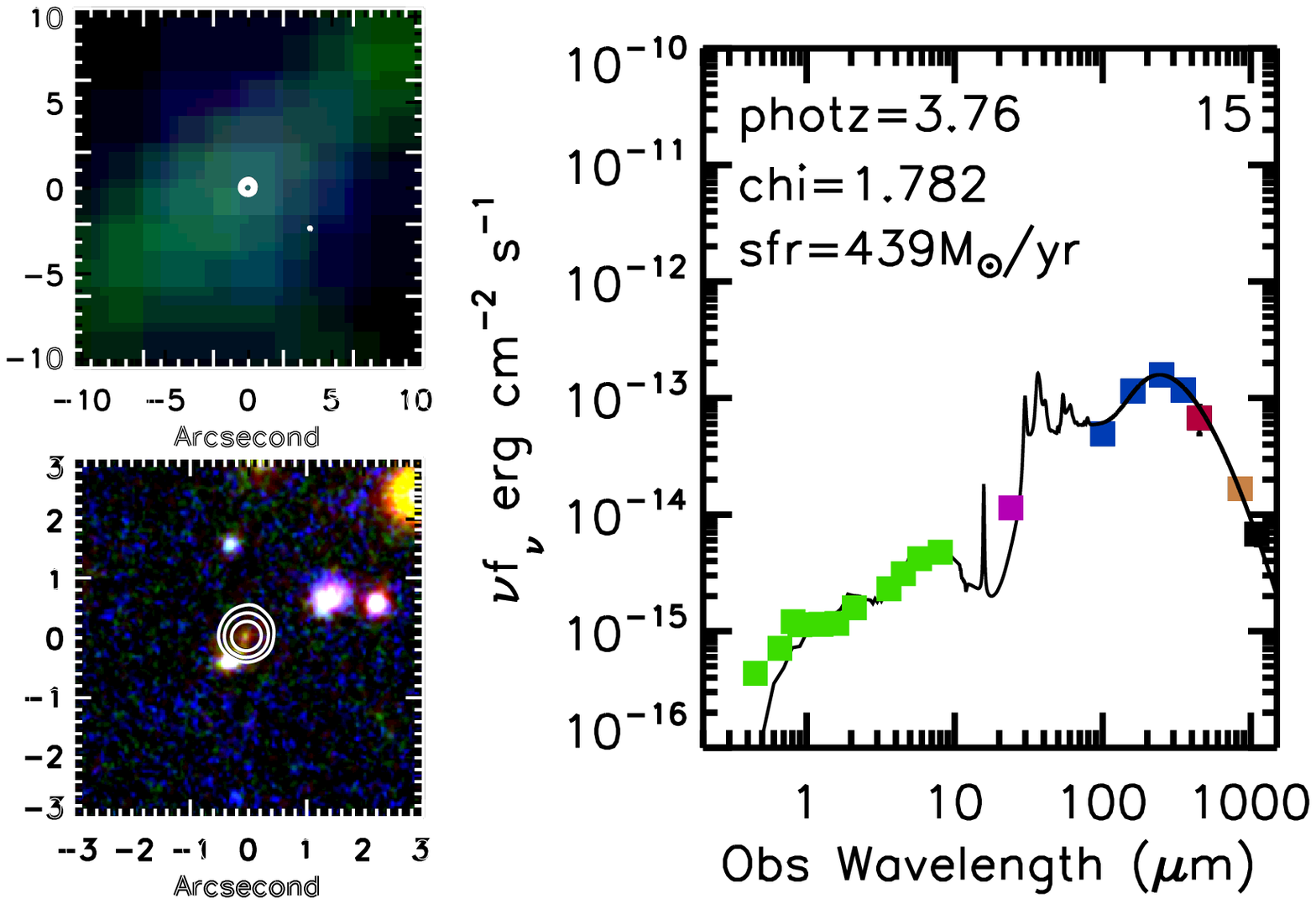}{0.325\textwidth}{(15)}
\boxedfig{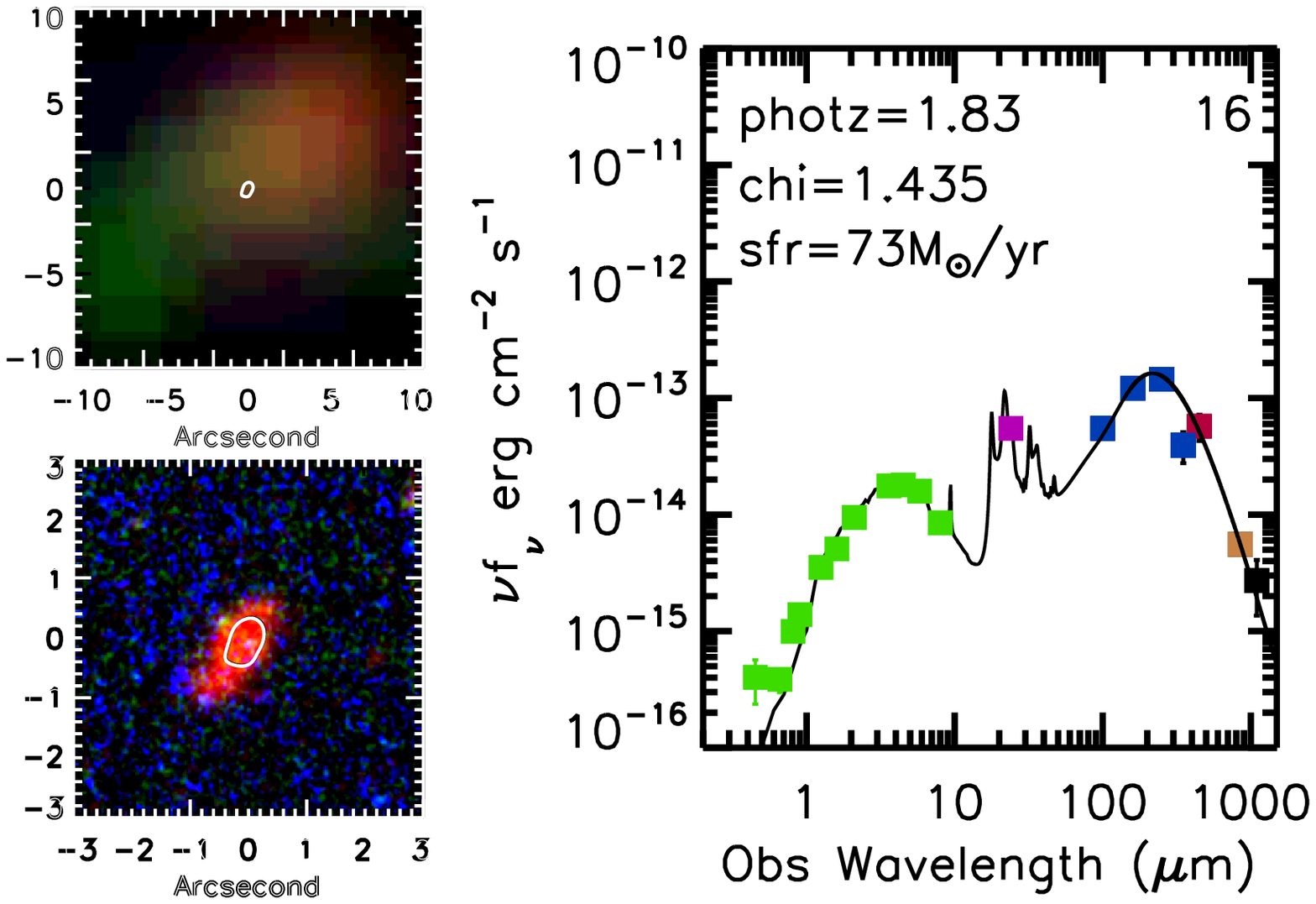}{0.325\textwidth}{(16)}
}
\caption{
All \afluxa selected sources in the GOODS-S 
(i.e., Table~\ref{cdfstab}; source numbers on bottom) 
that lie in the CANDELS region.
{\em (Right in each panel)\/}: Multiwavelength SEDs 
(green---{\em HST\/} and {\em Spitzer}/IRAC optical/NIR;
pink---{\em Spitzer}/MIPS 24~$\mu$m;
blue---{\em Herschel\/} FIR; red---SCUBA-2 450~$\mu$m;
gold---ALMA 870~$\mu$m; black---ALMA 1.1~mm)
and MAGPHYS fit (black curve). The SED plots are labeled at the top-right
with the source number and at the top-left with
the specz or the photz, the $\chi^2$ value for the fit, and the SFR.
{\em (Left in each panel):\/} two thumbnail images showing
{\em (top)\/} a combination of {\em Herschel\/}/PACS 100~$\mu$m (blue),
SCUBA-2 \afluxa (green), and SCUBA-2 \afluxb (red) data, 
and {\em (bottom)\/} a combination of
{\em HST\/} F435W (blue), F814W (green), and 
F160W (red) data.
The contours are ALMA 870~$\mu$m. 
Note that the ALMA position for source~4 is on the edge of another 
source (see bottom-left thumbnail), so no
SED was constructed due to contamination of the photometry.
\label{magphysSEDCDFS}
}
\end{figure*}
%-----------------------------------------------------------------------------

In Table~\ref{highztab}, we list the 12 high-redshift candidates that satisfy either
the 20~cm/\afluxb selection or the \afluxar/\afluxb selection that we defined
in Figure~\ref{fluxratios}.
All of these candidates come from the GOODS-N. We also give
their \afluxar, \afluxbr, and 20~cm fluxes; their 20~cm/\afluxb and \afluxar/\afluxb flux ratios;
their photzs from Table~\ref{cdfntab}, where available; and the redshifts we estimate from the linear
fits in Figures~\ref{fluxratios}(b) and (c), respectively, {\em if} the source is a 
high-redshift candidate based on the figure in question.

Most of the candidates are very faint in the optical/NIR and do not have reliable photzs, but
for the five that have photzs, the photzs are consistent with the sources being 
at high redshift. In combination with three sources with $z_{phot}\ge3$ that were
not identified as high-redshift candidates in the FIR and
seven sources with known $z_{spec}\ge3$, we have a total of 22 possible high-redshift sources
(we list all of these sources in Table~\ref{highztab}).

In Figure~\ref{zsmm}(a), we show redshift versus \afluxa flux for our \afluxa selected combined 
GOODS sample, with the high-redshift spectroscopic sources denoted by blue and red circles,
the high-redshift candidates selected from Figure~\ref{fluxratios} denoted by gold squares 
(5 of these also have photzs, so in those cases we use the photzs; we use $z_{450/850}$ for 6 sources; 
and for the last source, which was not identified as a high-redshift candidate from the 
\afluxar/\afluxb flux ratio, we use $z_{20/850}$), and the three additional high-redshift photz 
sources denoted by blue diamonds.  The redshift distribution at every 
flux is very large, demonstrating why quoting a median redshift for the sample as a whole
is not particularly meaningful.

High-redshift sources can also be picked out by their faintness in the MIR. 
In Figure~\ref{zsmm}(b), we plot redshift versus 4.5~$\mu$m flux for the \afluxa selected 
sample lying within the {\em HST\/}/ACS footprint of the GOODS-N. We used the 
Barro et al.\ (2019) catalog for the 4.5~$\mu$m fluxes.
For the small number of sources not in the catalog,
we measured the 4.5~$\mu$m fluxes on the {\em Spitzer}/IRAC image using a $3''$
diameter aperture and obtained the normalization by matching to
sources in the catalog.

For the sources with a measured 4.5~$\mu$m flux and either a specz, a photz, or 
a FIR redshift from Table~\ref{highztab}, there is a strong 
correlation of redshift with 4.5~$\mu$m flux, with all of the 
$z\ge 3$ sources having fluxes below 0.014~mJy. This alternate high-redshift diagnostic provides 
a strong confirmation of the high-redshift candidate selection from the FIR.

It is clear that using FIR methods to identify high-redshift candidates that cannot be found with photzs
is critical, or else the redshift distribution for \afluxa selected samples will not be complete and 
will be biased to lower redshifts.

%-----------------------------------------------------------------------------
\section{SED Fitting}
\label{secSED}
%-----------------------------------------------------------------------------
We next want to determine the SFRs and dust temperatures of the \afluxa 
selected samples.
We obtain best fits to the SMGs' full SEDs using the publicly available 
Bayesian energy-balance SED-fitting code MAGPHYS 
(da Cunha et al.\ 2015). 
We fit at the adopted redshifts (see Section~\ref{subsecz}), and we do not
allow the redshift to vary to optimize the SED fit.
The SED fits give SFRs (from the total bolometric luminosities; 
these are 100~Myr averages for a Chabrier 2003 initial mass function, or IMF) 
and dust properties, together with their error ranges. We note that
SFRs are relatively insensitive to redshift for dust-dominated galaxies.

For illustrative purposes,
in Figure~\ref{magphysSEDCDFS}, we show the SEDs (colored symbols) and the
MAGPHYS fits (black curve) for each \afluxa source in the GOODS-S
sample (i.e., Table~\ref{cdfstab}) that lies in the CANDELS region, alongside two
multiwavelength thumbnails (FIR/submillimeter on top and optical/NIR on bottom). 
The SED plots are labeled at the top with the specz or photz,
the $\chi^2$ value for the fit, and the SFR. Note that there is no SED plot for source~4,
since the ALMA position for that source is on the edge of another source, and thus the 
photometry is contaminated.

The MAGPHYS infrared SEDs contain
multiple temperature components, 
which provide an average,
luminosity-weighted dust temperature
(see da Cunha et al.\ 2015's Equation~8).
We shall generally adopt these temperatures.
However, we note that at low redshifts,
where the short wavelength FIR data are
relatively unconstraining, MAGPHYS may
insert hot components that raise
the temperatures and FIR luminosities.
In order to check the reality of these components, we also constructed
gray body fits for our \afluxa selected GOODS-N sample using an
optically thin, single temperature modified blackbody, 
$S_\nu \propto \nu^\beta B_\nu (T)$ (e.g., see Klaas et al.\ 1997).
Both the MAGPHYS and the gray body fits assume $\beta = 1.5$.

%-----------------------------------------------------------------------------
% FIGURE 10 ; gray_19.ps, gray_30.ps, gray_24.ps, gray_38.ps
%-----------------------------------------------------------------------------
\begin{figure*}
\centerline{\includegraphics[width=9cm,angle=0]{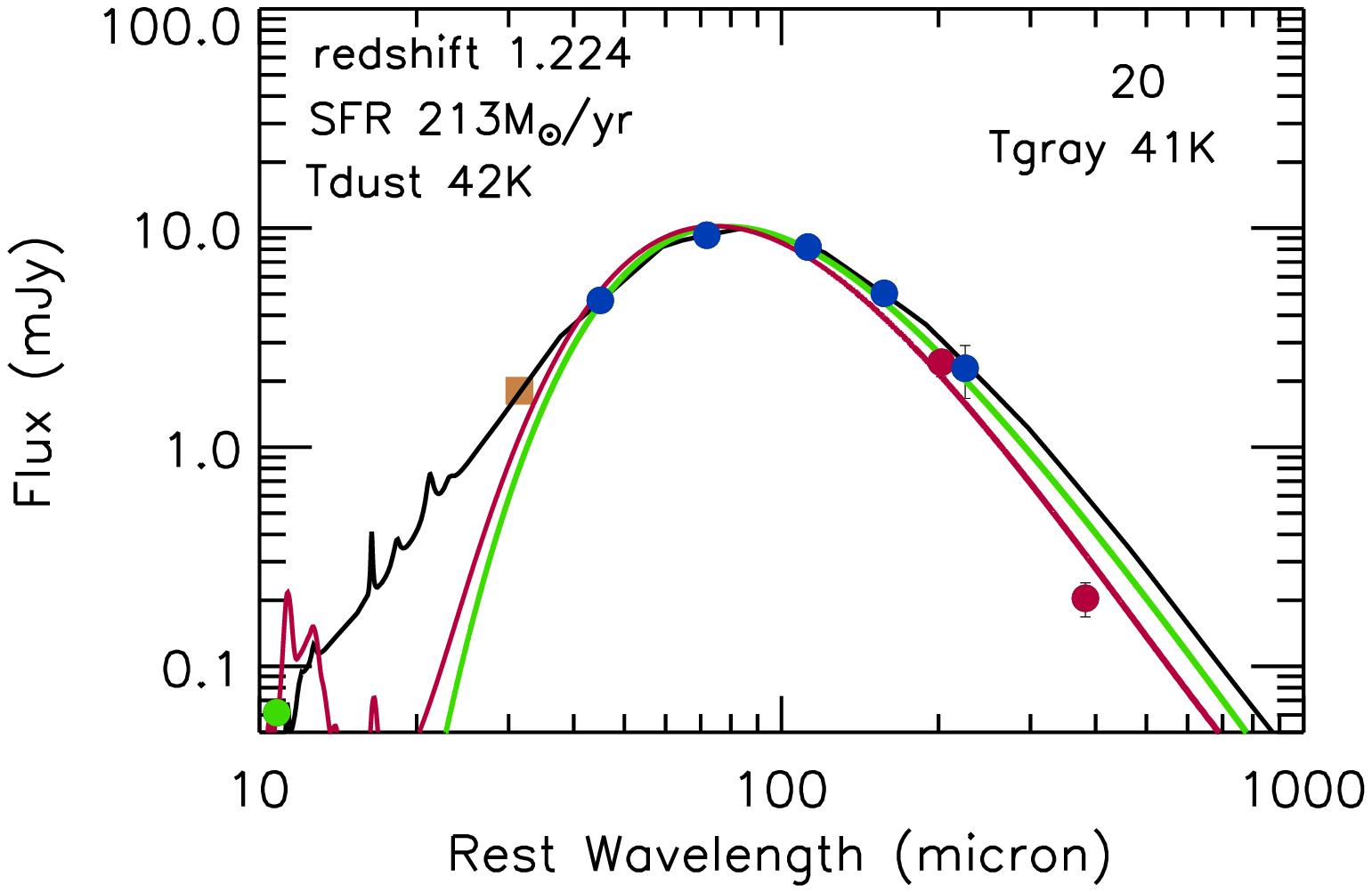}
\includegraphics[width=9cm,angle=0]{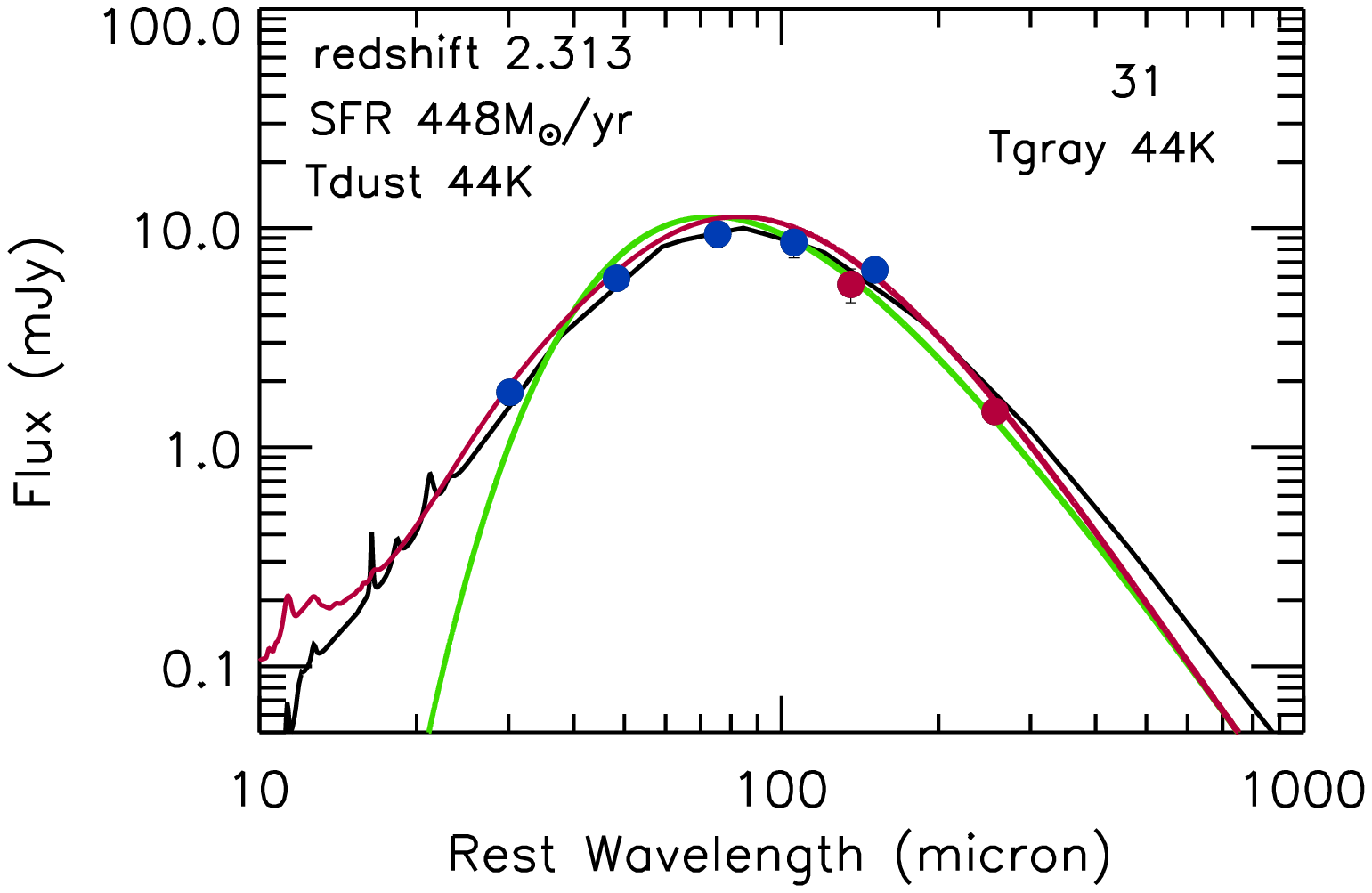}}
\centerline{\includegraphics[width=9cm,angle=0]{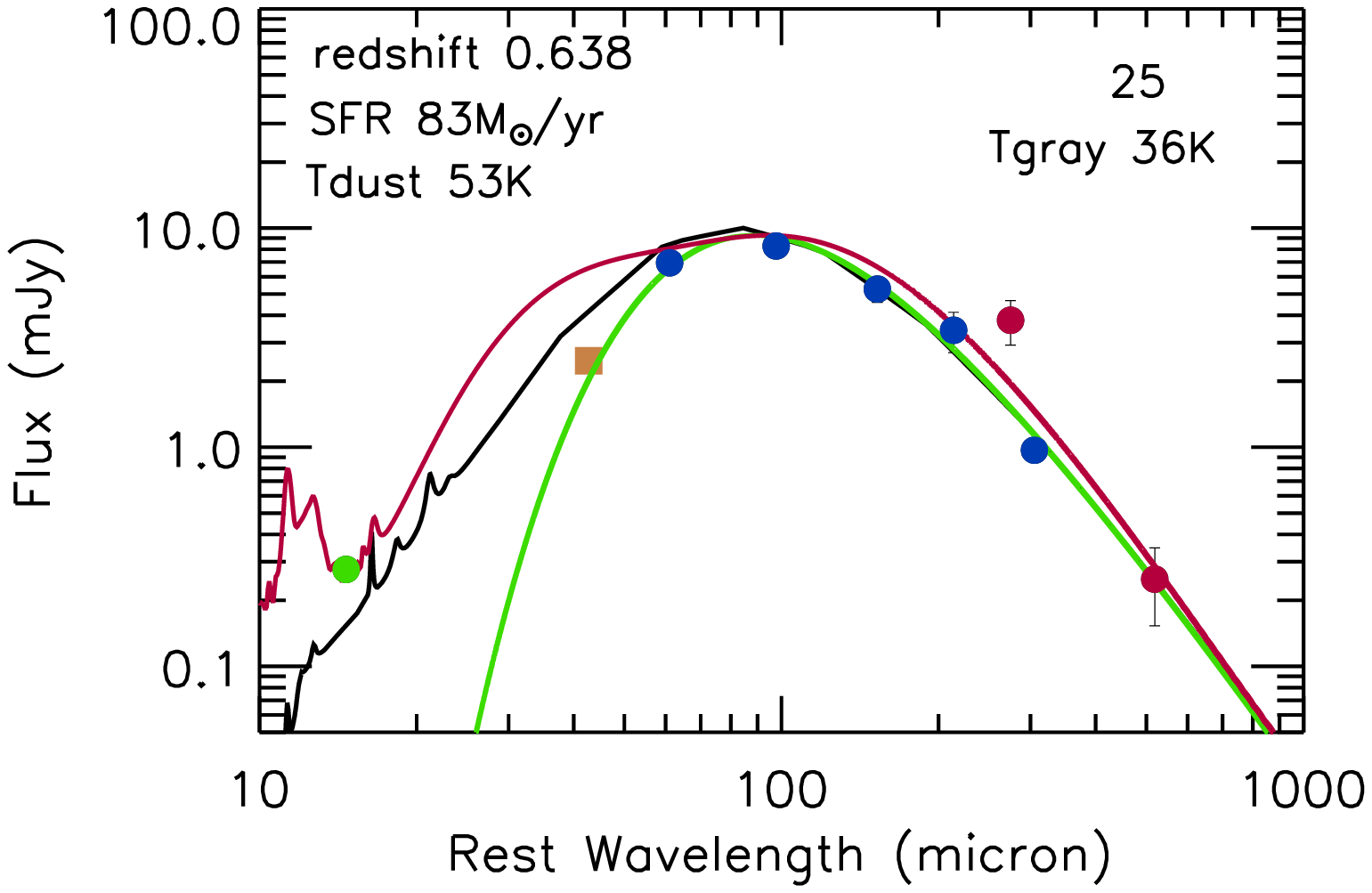}
\includegraphics[width=9cm,angle=0]{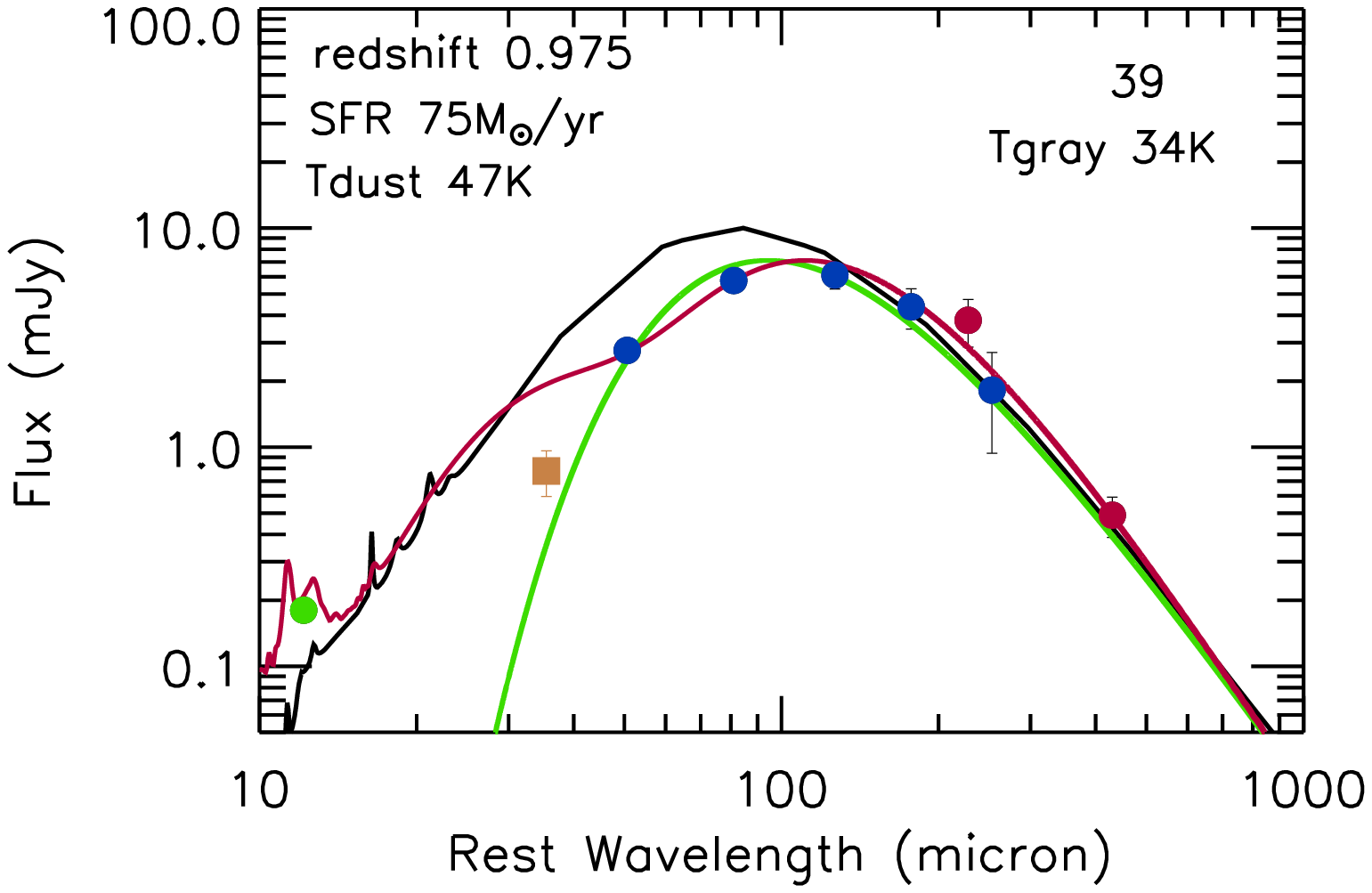}}
\caption{
SEDs for four example sources from the \afluxa selected
GOODS-N sample. Where available, we show the 70~$\mu$m flux
with gold squares.
The curves are Arp~220 (black), MAGPHYS fits (red), and 
gray body fits (green).
{\em (Top row)\/} Two higher-redshift examples where the MAGPHYS
and gray body fits are consistent and also similar to Arp~220.
{\em (Bottom row)\/} Two lower-redshift examples where the MAGPHYS 
fit includes a hot component that is inconsistent with the 70~$\mu$m flux. 
\label{graybody}
}
\end{figure*}
%-----------------------------------------------------------------------------

Fourteen of the sources have observed-frame 70~$\mu$m fluxes, 
and these fluxes were not used in either the MAGPHYS or the gray body fits.
In Figure~\ref{graybody}, we show example comparisons of the two fits
(MAGPHYS in red and gray body in green), along with an Arp~220 SED (black).
For the two higher-redshift sources shown (top row;
sources~20 and 31), the MAGPHYS and gray body fits
are similar and return nearly identical dust temperatures.
Source~20 has a 70~$\mu$m measurement (gold square) 
consistent with the fitted SED; source~31 does not have
a 70~$\mu$m measurement. However, for the two
lower-redshift sources (bottom row; sources~25
and 39), MAGPHYS includes a hot component
that substantially raises the temperature above the
gray body fit.
Both of these sources have
70~$\mu$m measurements that are strongly inconsistent
with the MAGPHYS SED but agree roughly with
the gray body fits. 

We conclude that the
gray body fits are preferred at lower redshifts,
noting that this also reduces the FIR luminosities
and SFRs for these sources. In our subsequent
analyses, we will adopt the SFRs and dust temperatures derived from the gray
body fits when we extend to sources below $z=1$.

%-----------------------------------------------------------------------------
\section{MAGPHYS Star Formation Rates and FIR Luminosities}
\label{secmagphys}
%-----------------------------------------------------------------------------
In this section, we examine the SFRs and FIR luminosities that MAGPHYS outputs
for the \afluxa selected combined 
GOODS sample sources with $z>1$. 

\begin{figure*}
\centerline{\includegraphics[width=9cm,angle=0]{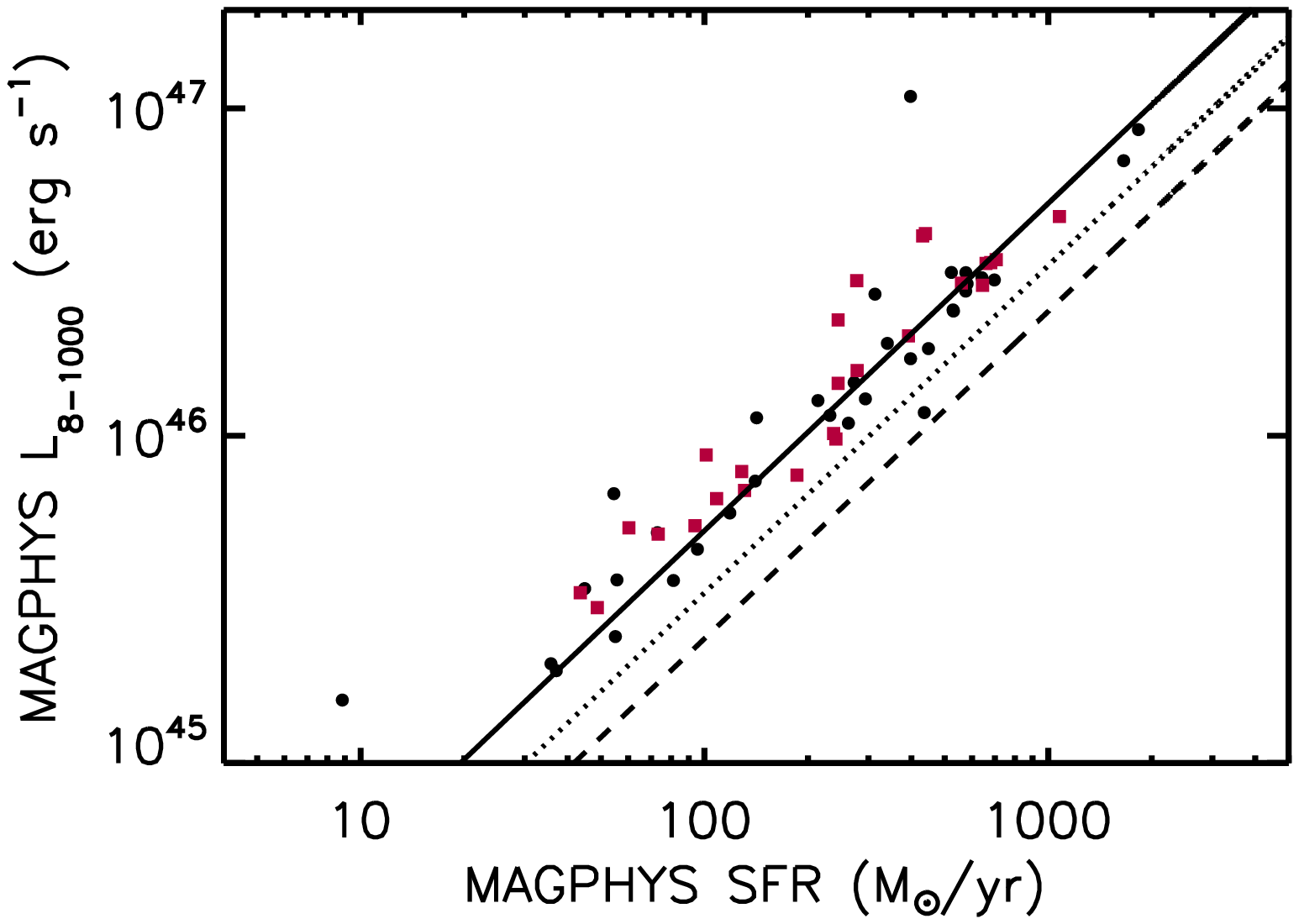}
\includegraphics[width=9cm,angle=0]{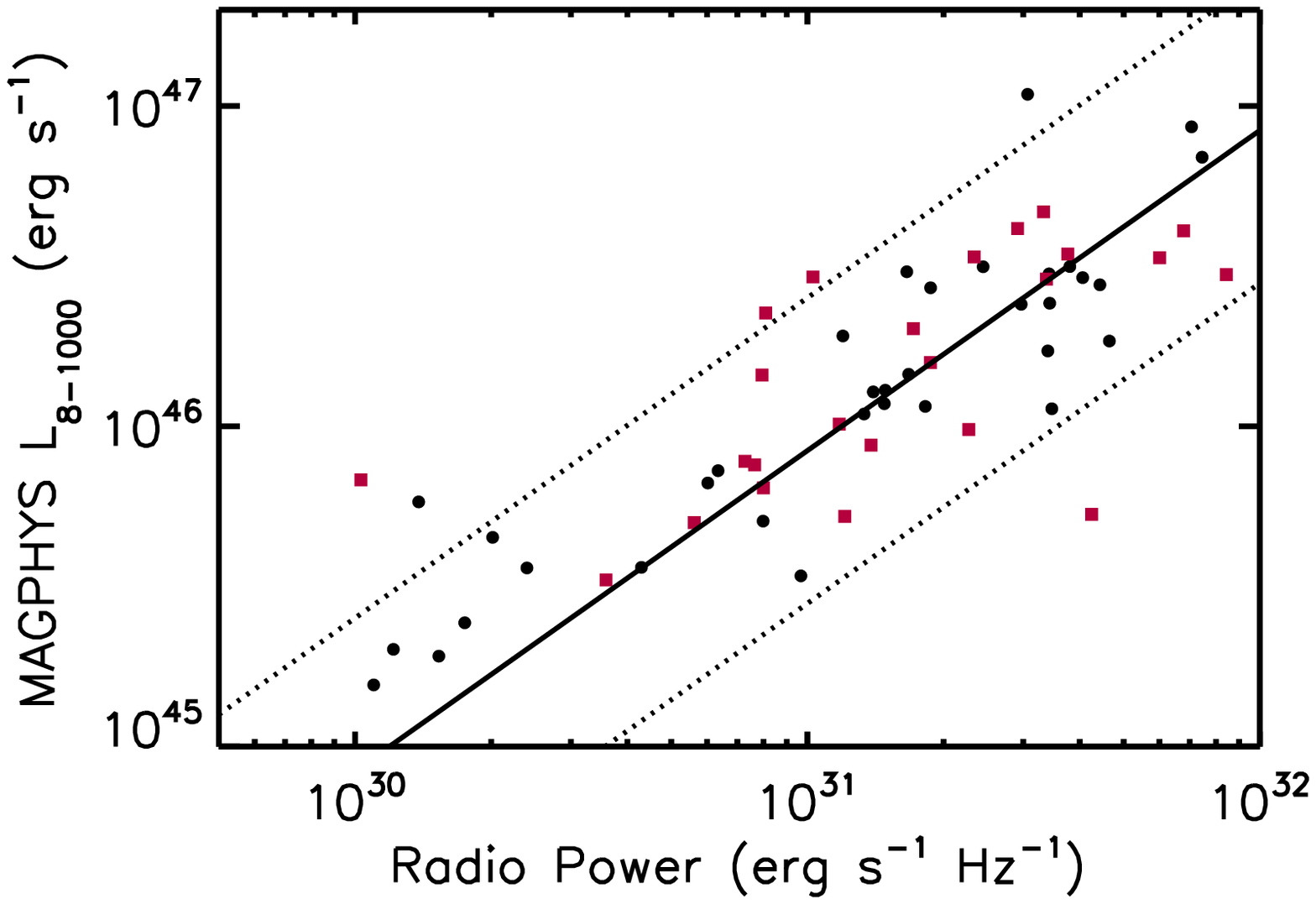}}
\centerline{\includegraphics[width=9cm,angle=0]{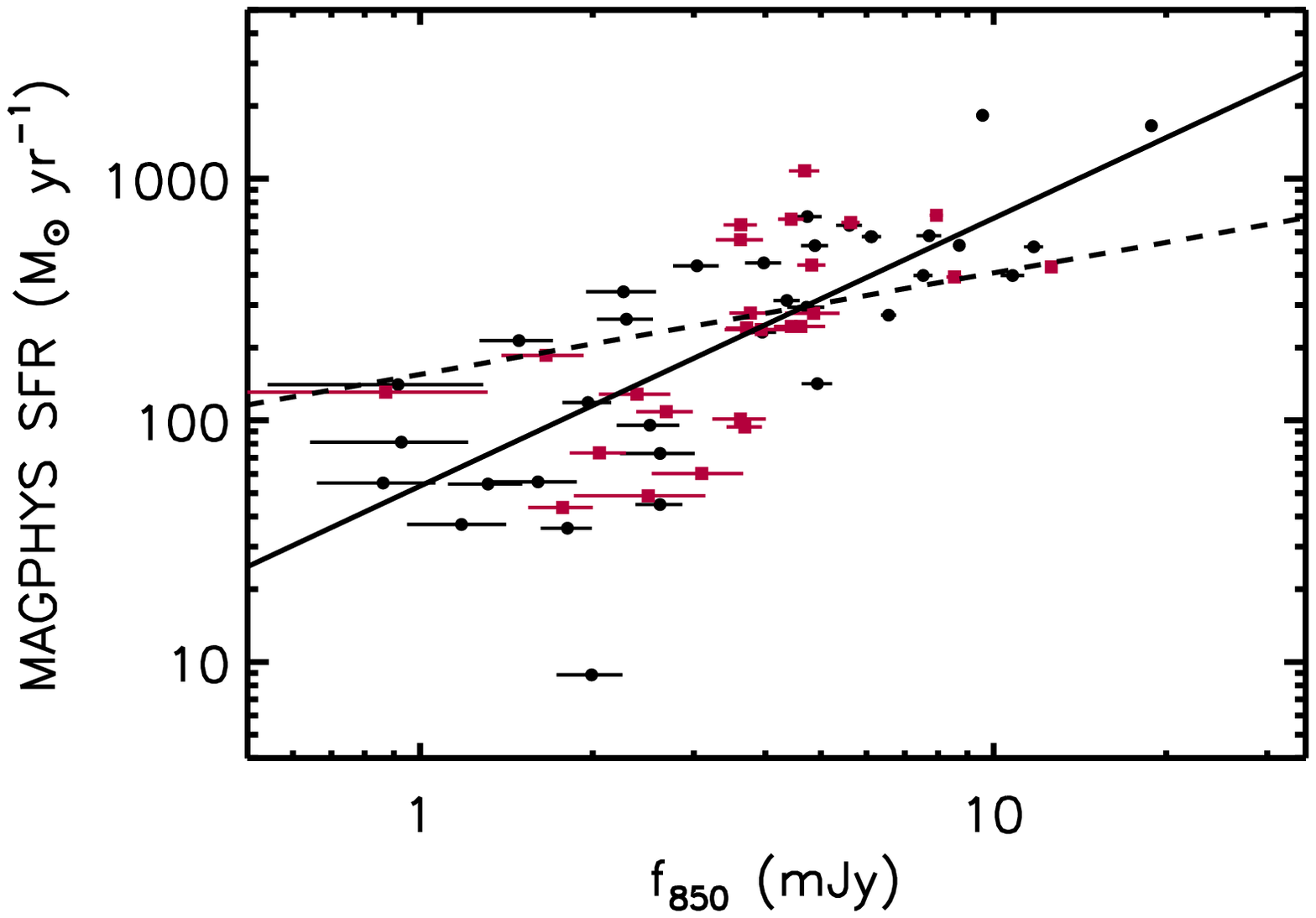}
\includegraphics[width=9cm,angle=0]{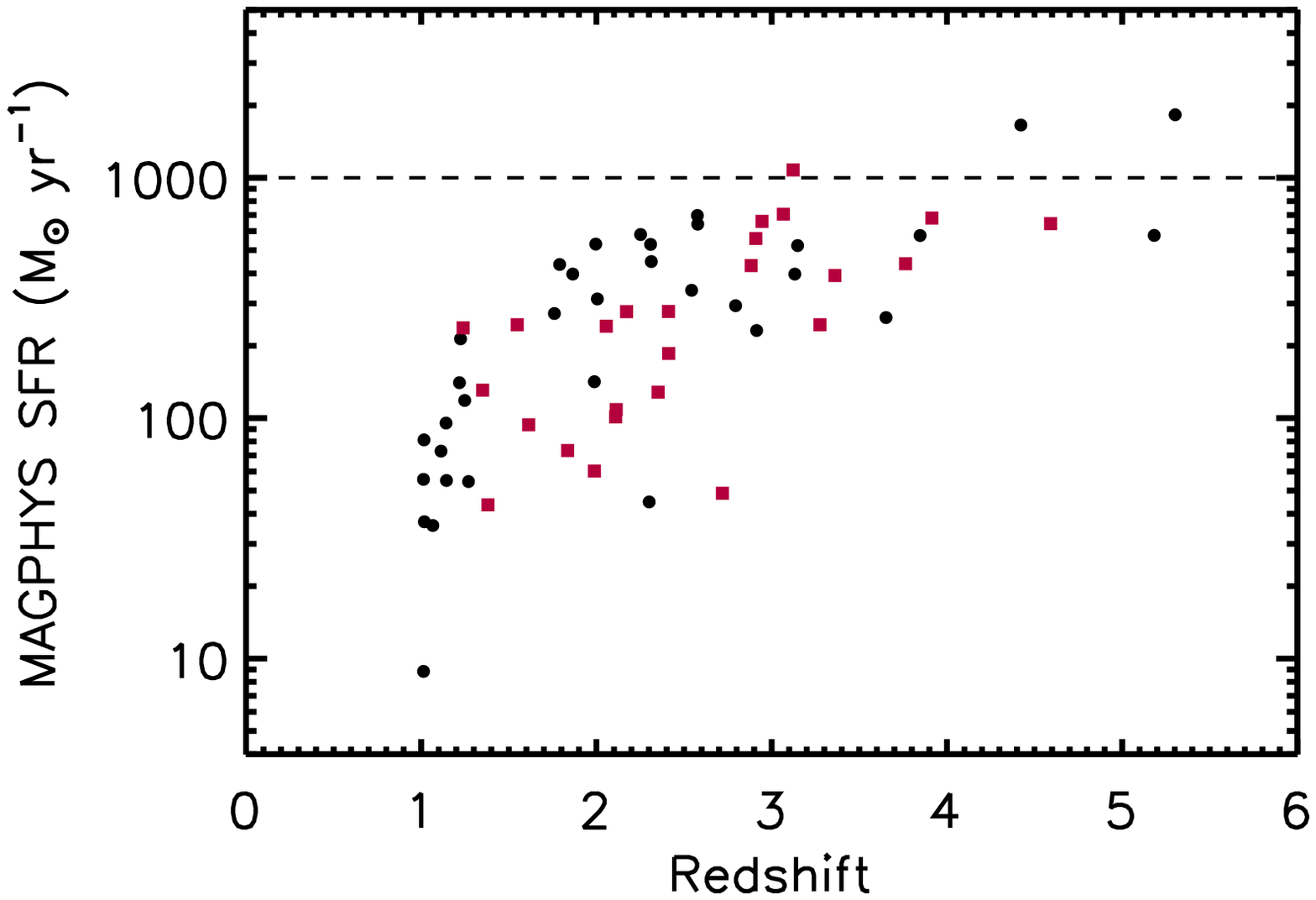}}
\caption{
Based on the \afluxa selected combined GOODS sample
with either $z_{spec}>1$ or a $z_{phot}>1$:
(a) MAGPHYS FIR luminosity vs. MAGPHYS SFR,
(b) MAGPHYS FIR luminosity vs. radio power,
(c) MAGPHYS SFR vs. SCUBA-2 \afluxb flux (with uncertainties on the flux),
and (d) MAGPHYS SFR vs. redshift.
Sources with speczs are shown as black circles, while sources with photzs are shown 
as red squares. In (a), the solid line shows a simple linear fit to the two MAGPHYS outputs, 
while the dashed line shows the relation (after conversion to a Chabrier IMF) 
from Murphy et al.\ (2011; their Equation~4),
and the dotted line shows the relation (after conversion to a Chabrier IMF) adopted 
by Madau \& Dickinson (2014), which is based on Kennicutt (1998). In (b), 
the solid line shows the median FIR-radio correlation for star-forming galaxies from Barger et al.\ (2017)
(calculated from their Equation~3, using a parameterization of 2.35, and their Equation~4), 
and the dotted lines show a multiplicative factor of 3 of this value, which they considered 
to be the region where sources lie on the FIR-radio correlation.
In (c), the solid line shows shows a simple linear fit to the data
(this paper's Equation~\ref{ourSFRreln}), 
and the dashed line shows, for comparison, the linear relation
from Dudzevi\u{c}i\={u}t\.{e} et al.\ (2020) 
(given in this paper's Equation~\ref{DSFRreln}).
In (d), the dashed line shows the maximum SFR 
for galaxies from Barger et al.\ (2017)
based on the Murphy et al.\ (2011) relation and a Kroupa IMF, which is very similar to
a Chabrier IMF (see text in Section~\ref{secmagphys}).
\label{magphysSFR}
}
\end{figure*}

In Figure~\ref{magphysSFR}(a), we plot \lfir\ versus SFR. A simple linear fit gives
\begin{equation}
\log {\rm SFR}(M_\odot~{\rm yr^{-1}})=\log L_{8-1000~\mu{\rm m}} {\rm (erg~s^{-1})} - 43.71 \,,
\label{magphysSFReq}
\end{equation}
which we denote on the figure by a solid line.
For comparison, Murphy et al.\ (2011) determined
a theoretical conversion of $-43.41$ between the two terms, which they computed from
a Starburst99 (Leitherer et al.\ 1999) constant SFR model and a Kroupa (2001) IMF.
For a Chabrier IMF, this would correspond to $-43.38$. We denote this relation by the
dashed line. Meanwhile, Madau \& Dickinson (2014)
adopted $-43.52$ using Kennicutt (1998),
after conversion to a Chabrier IMF from their Salpeter (1955) IMF.
We denote this relation by the dotted line.
Both of these relations give slightly lower SFRs than the MAGPHYS relation.
This emphasizes the uncertainty in the \lfir\ to SFR calibration.

In Figure~\ref{magphysSFR}(b), we verify empirical expectations
for a correlation between \lfir\ and radio power. We determine the latter 
independently of MAGPHYS using
\begin{equation}
P_{\rm 20~cm}=4\pi {d_L}^2 f_{\rm 20\,cm} 10^{-29} (1+z)^{\alpha - 1}~{\rm erg~s}^{-1}~Hz^{-1} \,.
\end{equation}
Here $d_L$ is the luminosity distance (cm) and $f_{20\,{\rm cm}}$ is the 20~cm flux in units
of $\mu$Jy. This equation assumes that the radio flux density goes as $\nu^{-\alpha}$.
Following Barger et al.\ (2017), we adopt a radio spectral index of $\alpha=0.8$ 
(Condon 1992; Ibar et al.\ 2010; An et al.\ 2021). 
The solid line shows the median FIR-radio correlation for star-forming galaxies from Barger et al.,
and the dotted lines show a multiplicative factor of 3 of this value, which Barger et al.\ considered 
to be the region where sources lie on the FIR-radio correlation. The current data are well-represented
by the Barger et al.\ correlation.

In Figure~\ref{magphysSFR}(c), we plot SFR versus SCUBA-2 \afluxb flux.
A simple linear fit to the data, not considering uncertainties on the data, gives
\begin{equation}
\log{\rm SFR} = 1.12\log f_{850~\mu{\rm m}} + 1.70 \,,
\label{ourSFRreln}
\end{equation}
which we denote on the figure by a solid line.
There may be a shift to higher SFRs for brighter sources in our sample.
The slope in our relation is steeper than that in Dudzevi\u{c}i\={u}t\.{e} et al.\ (2020),
who found
\begin{equation}
\log{\rm SFR} = 0.42\log f_{870~\mu{\rm m}} + 2.19 
\label{DSFRreln}
\end{equation}
for 517 AS2UDS\footnote{AS2UDS  (Stach et al.\ 2019) is an ALMA 870~$\mu$m follow-up survey of 
716 $>4\sigma$ SCUBA-2 sources (corresponding to observed \afluxb fluxes $\ge3.4$~mJy)
detected in the SCUBA-2 Cosmology Legacy Survey (S2CLS; Geach et al.\ 2017) 
\afluxb map of the UKIDSS Ultra Deep Survey field.}
sources with redshifts between $z=1.8$ and $z=3.4$.
We denote their relation on the figure as a dashed line. 
The reason for the disagreement is not clear.

In Figure~\ref{magphysSFR}(d), we plot SFR versus redshift.
The present results are broadly consistent with the maximum SFR
from Barger et al.\ (2017), namely
1000~M$_\odot$~yr$^{-1}$ (dashed line), above which there are relatively few galaxies
(see also Karim et al.\ 2013 and Barger et al.\ 2014).
The Barger et al.\ (2017) value is based on the Murphy et al.\ (2011) relation and 
a Kroupa IMF, and the primary uncertainty is the calibration of the SFR.

%-----------------------------------------------------------------------------
% FIGURE 11 ; 450_lfir_sfr.ps, 450_lfir_radpow.ps, 450_sfr_850.ps, 450_sfr_z.ps
%-----------------------------------------------------------------------------
%-----------------------------------------------------------------------------

%-----------------------------------------------------------------------------
\section{Discussion}
\label{secdisc}
%-----------------------------------------------------------------------------
Most comparisons of \afluxa and \afluxb
selected samples have argued that the \afluxa
samples peak at lower redshifts
(e.g., Casey et al.\ 2013; Zavala et al.\ 2018; Lim et al. 2020)
and may be biased to higher dust temperatures 
(Casey et al.\ 2013; Lim et al.\ 2020).
Such sample differences are clearly introduced
by the SED shape, where \flipsmmfluxratio\ 
becomes smaller as one moves to higher redshifts (see Figure~\ref{fluxratios}(c))
and larger as one moves to higher dust temperatures.
However, as Lim et al.\ (2020) emphasize, these selection
biases are very dependent on the relative
depths of the \afluxa and \afluxb samples,
since the dust temperature is strongly dependent on the FIR
luminosity, and submillimeter samples are biased to higher FIR luminosities 
at higher redshifts.

%-----------------------------------------------------------------------------
\subsection{Are the Redshift Distributions Different for a \afluxa versus an \afluxb Selection?}
\label{secdisczdist}
\begin{figure*}
\centerline{\includegraphics[width=9cm]{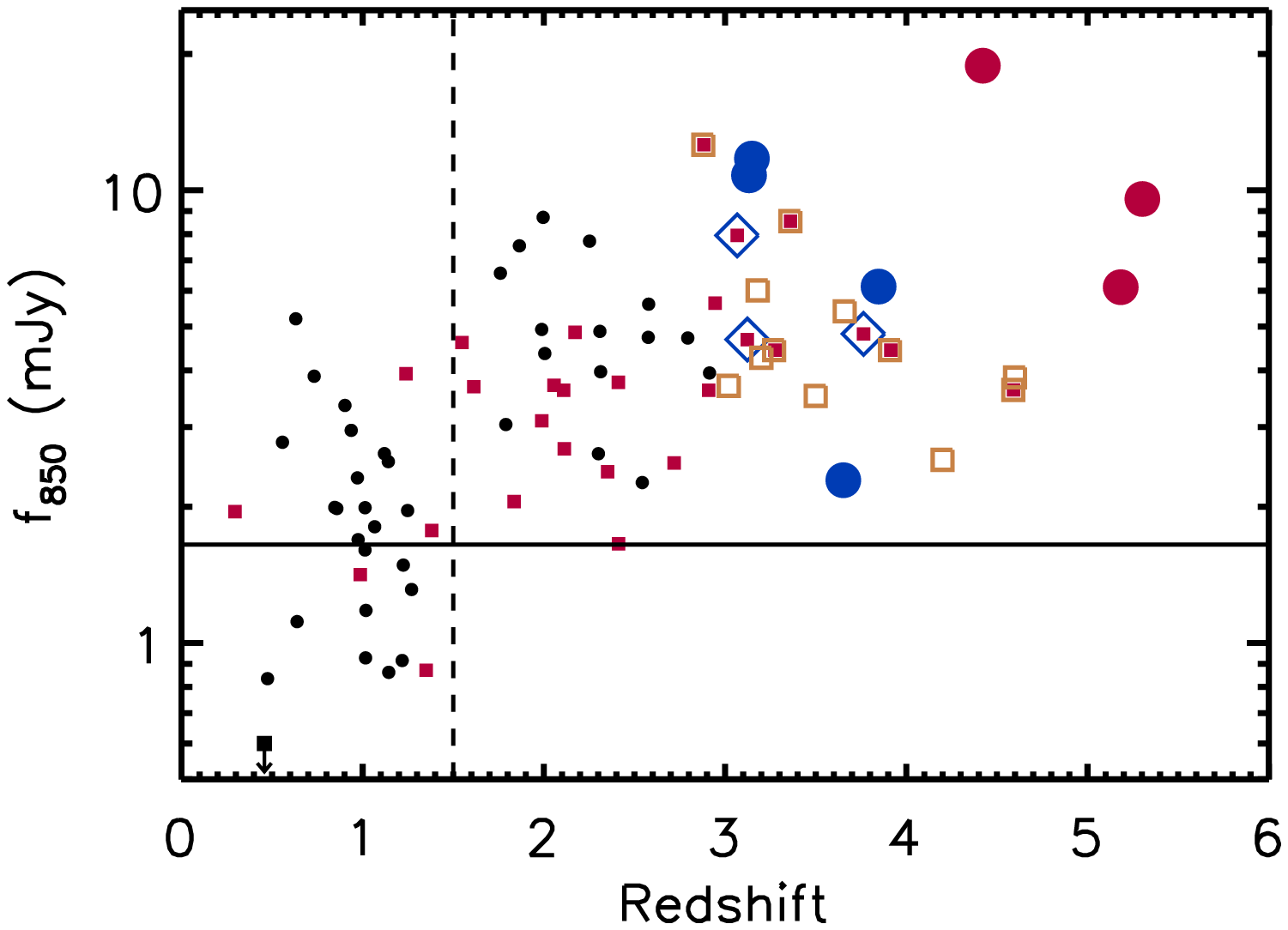}
\includegraphics[width=9cm]{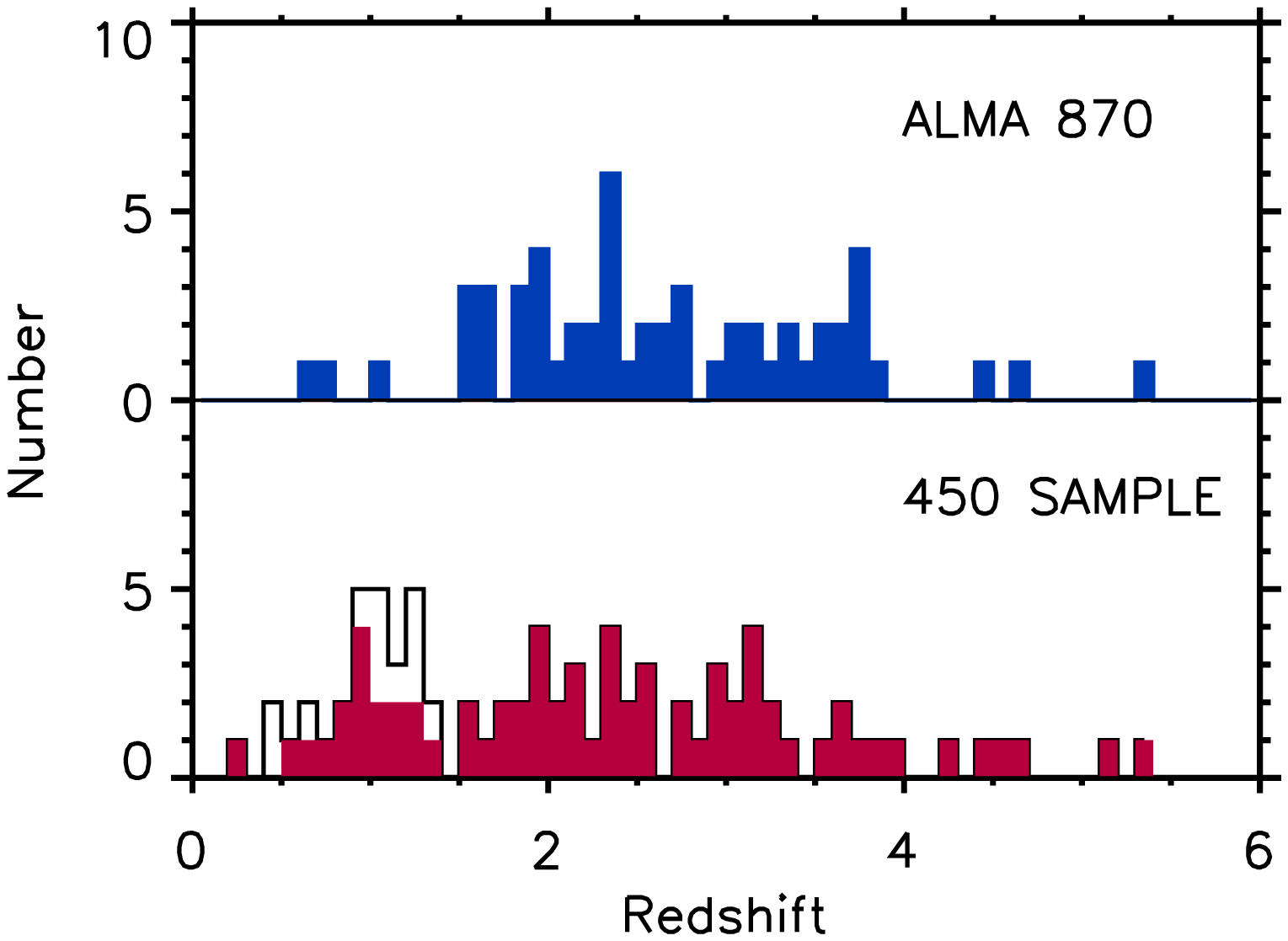}
}
\caption{
(a) SCUBA-2 \afluxb flux vs. redshift for the \afluxa selected combined GOODS sample.
Sources with $z_{spec}<3$ are shown as black circles, while sources 
with $z_{phot}$ are shown as red squares. Sources with $3\le z_{spec}<4$
are denoted by blue large circles, while those with $z_{spec}\ge4$ are denoted by red
large circles. 
The high-redshift candidates from Table~\ref{highztab} are denoted by gold open squares
(we use the photzs for the five that have a photz).
Sources with $3\le z_{phot}<4$ that were not also identified as high-redshift candidates
from the FIR are enclosed in blue large open diamonds.
GOODS-N source~70 (bottom-left of the plot)
is denoted by a black square with a downward pointing arrow, indicating it is off the plot.
The horizontal line shows the SCUBA-2 1.65~mJy ($4\sigma$) confusion limit at 850~$\mu$m,
and the dashed vertical line shows the redshift $z=1.5$ above which all of the \afluxa 
sources would be found in an \afluxb sample.
(b) Redshift distributions for the \afluxa selected combined GOODS sample 
(bottom histogram; shaded region shows the sample that has SCUBA-2
\afluxb fluxes $>1.65$~mJy)
and for the $>1.65$~mJy ALMA 870~$\mu$m GOODS-S sample of
Cowie et al.\ (2018) (top histogram). 
We do not show the eight \afluxa sources without any redshift information in either
(a) or (b).
\label{850_lims}
}
\end{figure*}

%-----------------------------------------------------------------------------
In Figure\ref{850_lims}(a), we show the \afluxb fluxes from Tables~\ref{cdfntab} and \ref{cdfstab}
measured for the \afluxa selected combined GOODS sample versus redshift.
The horizontal line shows the SCUBA-2 1.65~mJy ($4\sigma$) confusion limit at \afluxbr.
As was also noted by Zavala et al.\ (2018), all of the sources above
$z=1.5$ (vertical line) would be included in an \afluxb sample, and the \afluxa selection 
only adds sources that would not be found in an \afluxb sample below $z=1.5$.

In Figure~\ref{850_lims}(b) (bottom histogram), we show the redshift distribution of
the \afluxa selected combined GOODS sample (black line histogram). 
The red shading shows the sources in this sample that are detected
above 1.65~mJy at \afluxbr. In Figure~\ref{850_lims}(b) (top histogram), we show
the redshift distribution of the 58 sources with fluxes $>1.65$~mJy in the GOODS-S
ALMA 870~$\mu$m sample of
Cowie et al.\ (2018) (blue histogram). A Mann-Whitney test shows no statistically 
significant difference between the blue histogram and the red histogram.

%-----------------------------------------------------------------------------
% FIGURE 12 ; 450_f8_z.ps, 850_hist.ps
%-----------------------------------------------------------------------------
%-----------------------------------------------------------------------------

Restricting the \afluxa selected sample to sources that 
would be present in the SCUBA-2 confusion-limited ($>1.65$~mJy) 
\afluxb selected sample raises the
median redshift to $z=2.30$ from the $z=1.99$ value noted previously
for the full \afluxa sample. However, as we have emphasized, 
there is a wide distribution of redshifts, and the median is a poor 
characterization of this distribution. 

This situation is best summarized
by noting that the two distributions are quite similar for the higher redshift
sources, but the \afluxa sample adds in lower redshift sources, which
results in the reduction of the median.

%-----------------------------------------------------------------------------
\subsection{Do Dust Temperatures Evolve with Redshift?}
\label{secdiscevol}
%-----------------------------------------------------------------------------
Many previous papers have noted strong evolution in dust
temperature with both redshift and FIR luminosity
(e.g., Magnelli et al.\ 2013; Swinbank et al.\ 2014; B\'ethermin et al.\ 2015; 
Schreiber et al.\ 2018; Zavala et al.\ 2018). 
If the effective area over which the galaxy radiated were constant as a function of 
both redshift and FIR luminosity, then there would be a simple monotonic relation between 
FIR luminosity and dust temperature.
However, when we measure the evolution of FIR luminosity, we are also measuring the
evolution in the galaxy properties. There is
a dependence of FIR luminosity on dust temperature, but it is not the sole dependence.

\begin{figure*}
\hskip -0.5cm
\centerline{\includegraphics[width=9.5cm,angle=0]{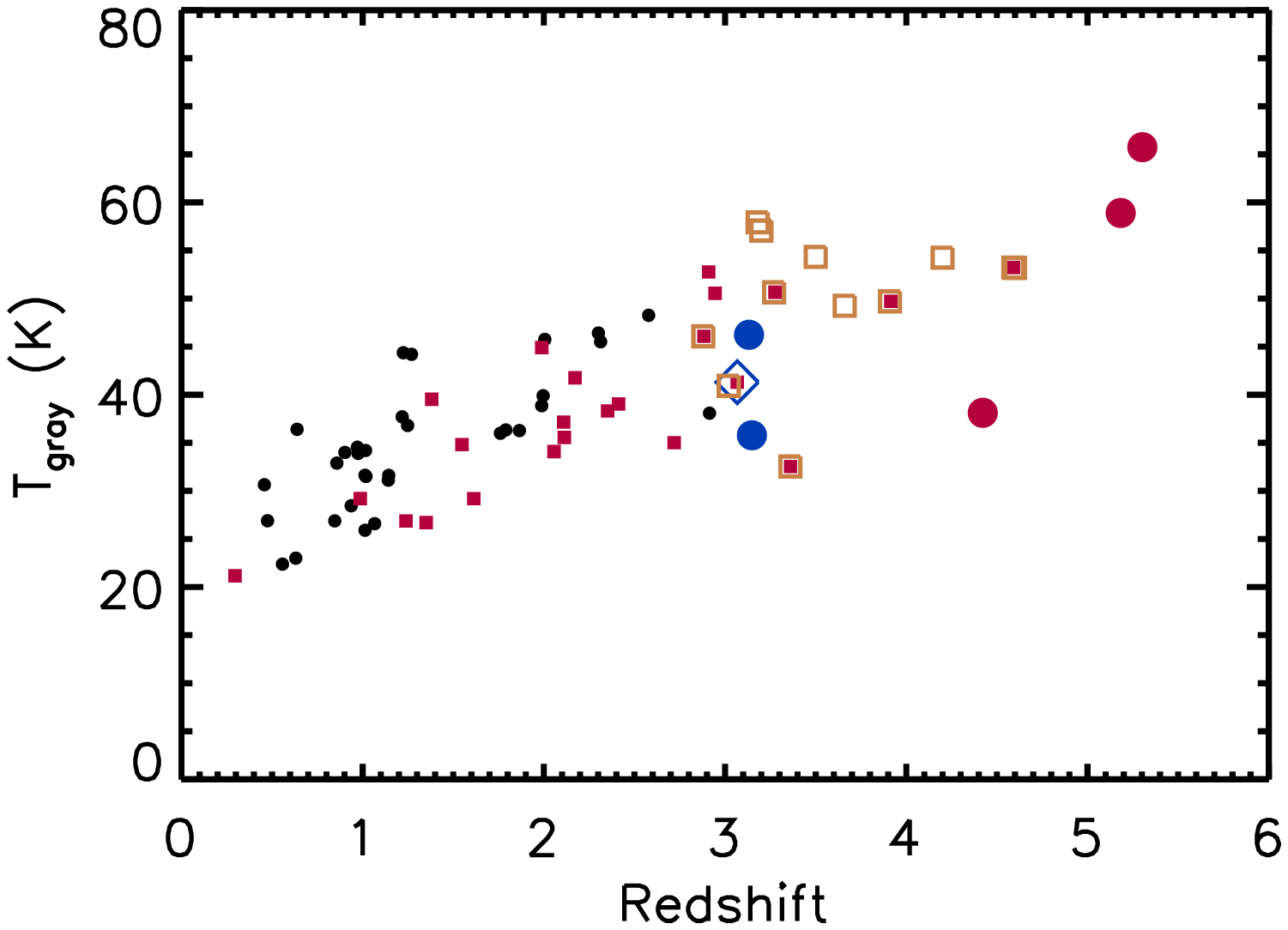}
\includegraphics[width=9.5cm,angle=0]{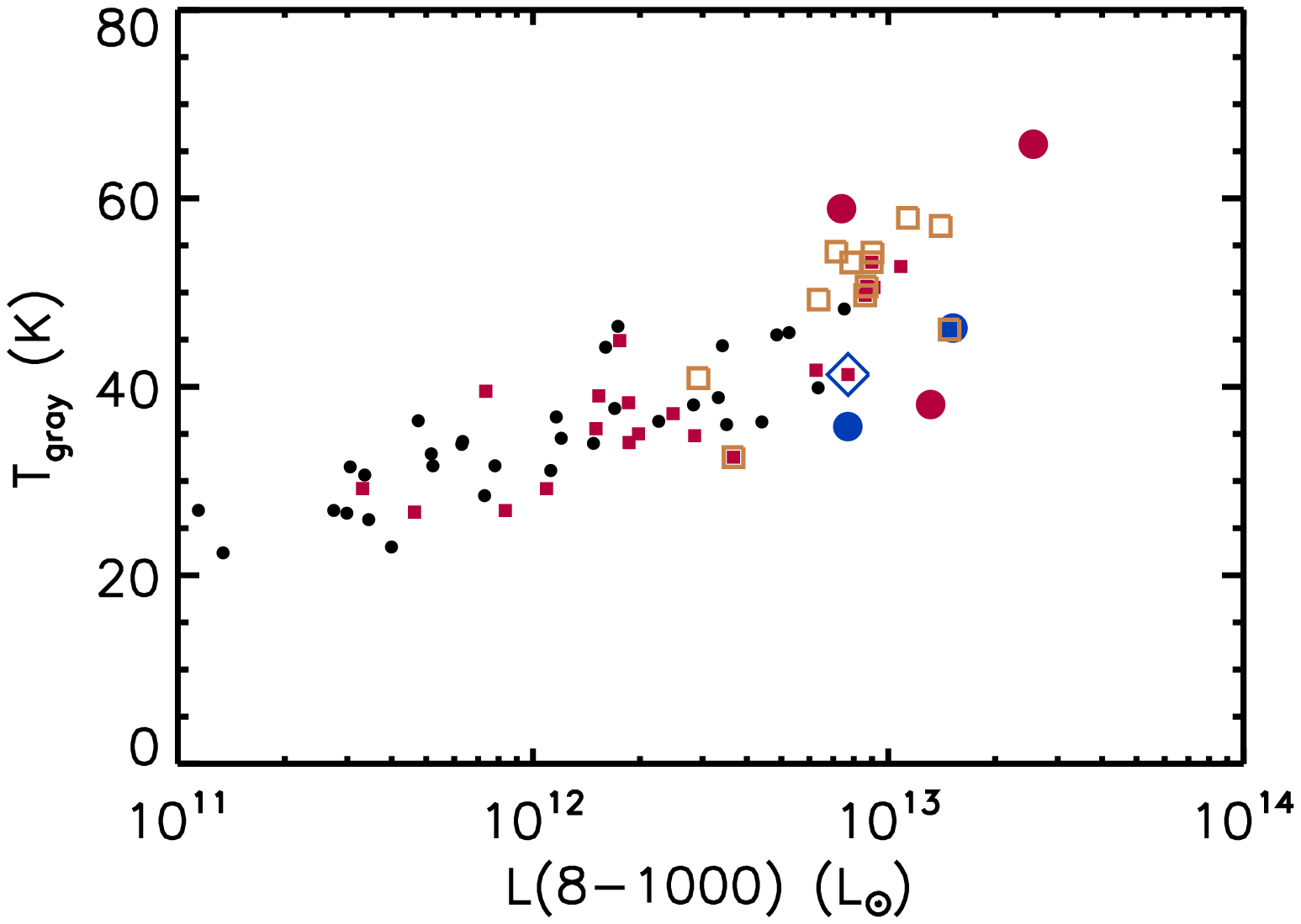}}
\caption{
Gray body dust temperature vs. (a) redshift and (b) \lfir\
for our \afluxa selected GOODS-N sample with gray body temperature measurements.
Sources with $z_{spec}<3$ are shown as black small circles. Sources 
with $z_{phot}$ are shown as red small squares.
Sources with $3\le z_{spec}<4$
are denoted by blue large circles, while those with $z_{spec}\ge4$ are denoted by red
large circles. 
The high-redshift candidates from Table~\ref{highztab} are denoted by gold open squares
(we use the photzs for the five that have a photz).
A source with $3\le z_{phot}<4$ that was not also identified as a high-redshift candidate
from the FIR is enclosed in a blue large open diamond.
\label{greyTdL}
}
\end{figure*}

We show the evolution in dust temperature
for the present \afluxa selected GOODS-N
sample in Figures~\ref{greyTdL}(a) and \ref{greyTdL}(b), respectively.
In both panels, we code the data points by redshift. 
From Figure~\ref{greyTdL}(a), we see that there is a strong increase
in dust temperature with redshift.
From Figure~\ref{greyTdL}(b), we see that there is a strong increase
in dust temperature with FIR luminosity.

%-----------------------------------------------------------------------------
% FIGURE 13 ; gray_temp_z.ps, gray_temp_lum.ps
%-----------------------------------------------------------------------------
%-----------------------------------------------------------------------------

%-----------------------------------------------------------------------------
% FIGURE 14 ; lfir_z_GN.ps, grey_temp_z_GN.ps
%-----------------------------------------------------------------------------
\begin{figure*}
\hskip -0.5cm
\centerline{\includegraphics[width=9.5cm,angle=0]{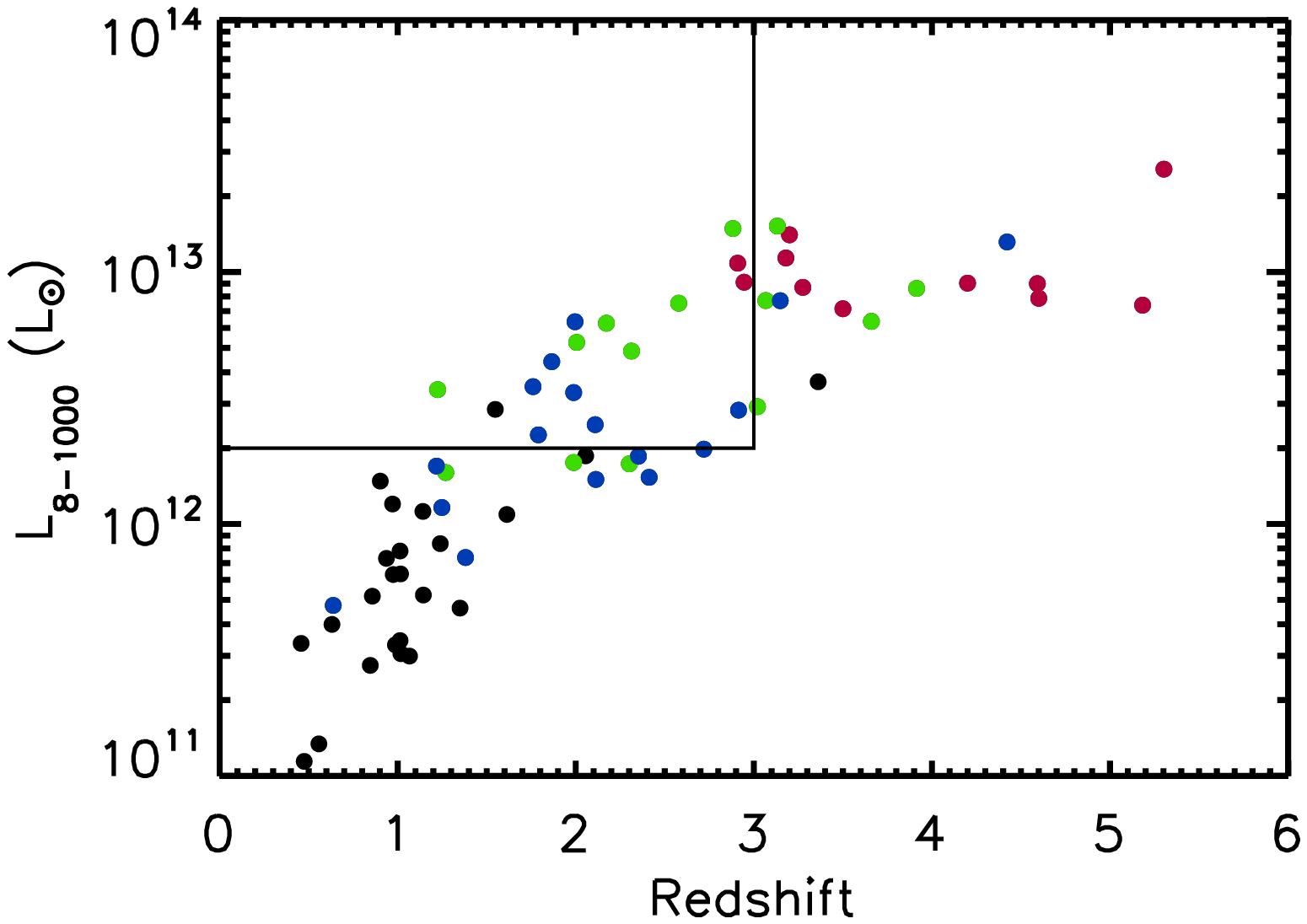}
\includegraphics[width=9.5cm,angle=0]{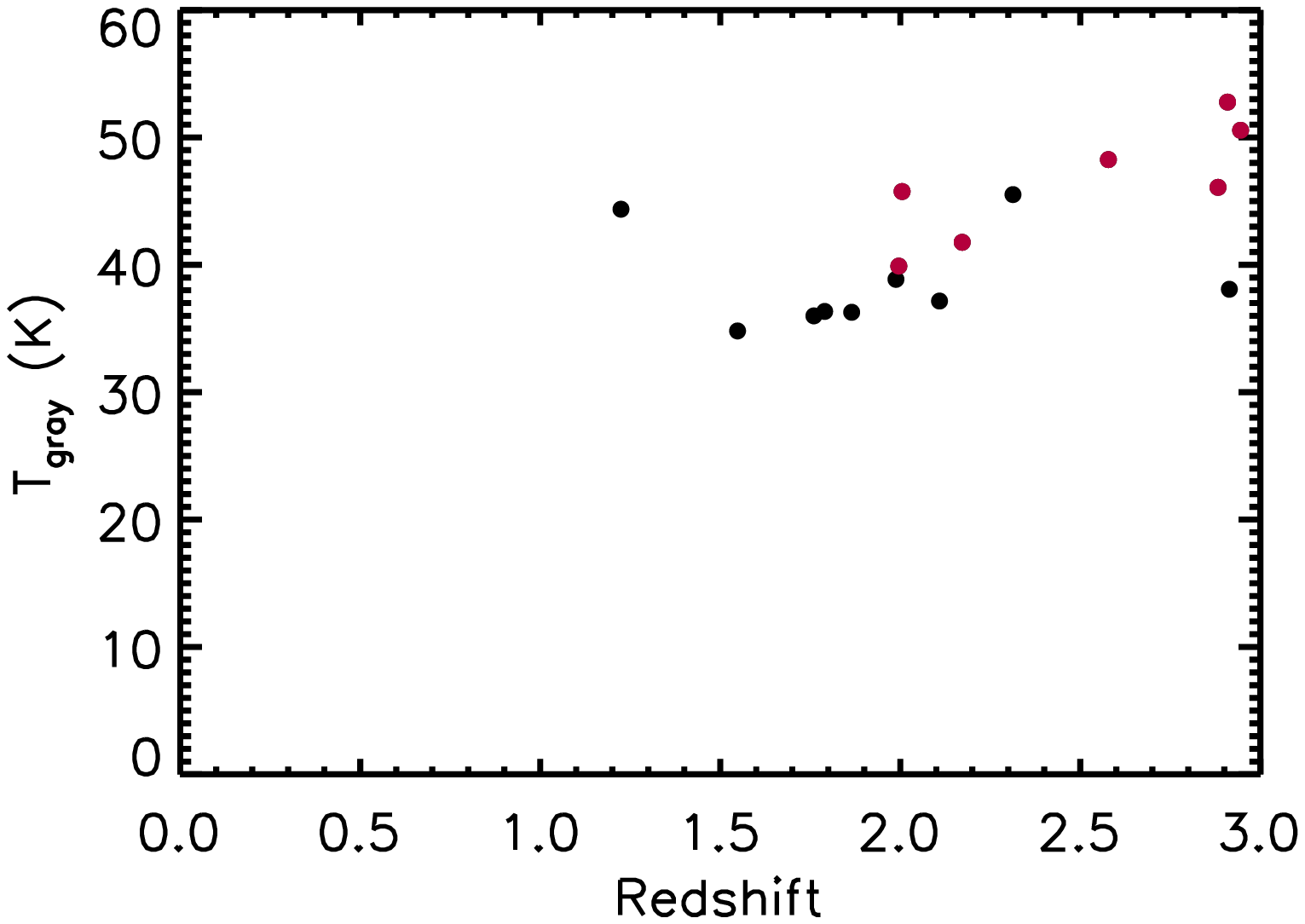}}
\caption{
(a) \lfir\ and (b) gray body dust temperature (from selected region in (a)) vs. redshift
for our \afluxa selected GOODS-N sample with gray body temperature measurements.
In (a), the symbols are color-coded by temperature 
(black$<35$~K; blue=35--40~K; green=40--50~K; red$>50$~K).
The temperatures are strongly correlated with \lfir.
The black lines show a restricted region with 
redshifts $z<3$ and \lfir\ $>2\times10^{12}~L_\odot$ where our
sample should be substantially complete.
In (b), we show sources from only that restricted region.
The symbols are color-coded by \lfir\ (black$>2\times 10^{12}~L_\odot$;
red$>5\times 10^{12}~L_\odot$).
\label{uniformLFIR}
}
\end{figure*}
%-----------------------------------------------------------------------------

However, there is controversy over whether the evolution
of the dust temperature with redshift is a real effect or simply
a consequence of the selection bias to higher
FIR luminosity as we move to higher redshifts. 
Most recently, Lim et al.\ (2020) and Dudzevi\u{c}i\={u}t\.{e} et al.\ (2020) 
did not find any evidence for redshift evolution in a uniform FIR luminosity
subsample of their data. Using a spectroscopically complete 1.4~mm 
selected 
South Pole Telescope high-redshift, strongly gravitationally lensed sample, 
Reuter et al.\ (2020) also found that their data could
be consistent with no evolution.

Our data agree with these recent results. 
From Figure~\ref{greyTdL}(b), we can see at once that at 
the high FIR luminosity end for this relatively small sample, the range 
of dust temperatures is similar for both
high ($z>3$) and more moderate ($z<3$) redshift sources.
We look at this in more detail in Figure~\ref{uniformLFIR}.
Here we restrict to sources with redshifts $z<3$ and \lfir\
$>2\times10^{12}~L_\odot$, where we expect our
sample to be substantially complete.
We find no statistically significant evidence of evolution with 
redshift with a Pearson coefficient of $R=0.3433$.

%-----------------------------------------------------------------------------
\subsection{Do \afluxa Samples Select Warmer Sources Than an \afluxb Sample?}
\label{secdiscsample}
\begin{figure} 
\centerline{\includegraphics[width=9cm]{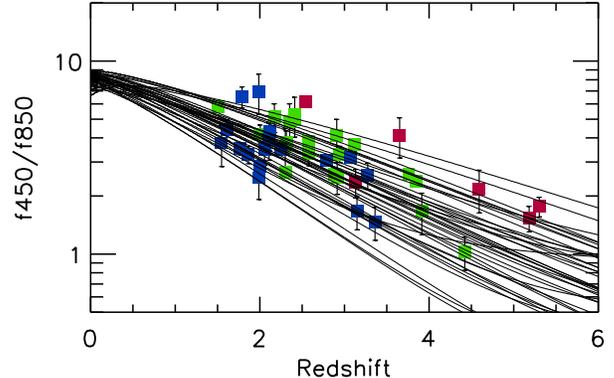}
}
\caption{
Predicted \flipsmmfluxratio\ vs. redshift (black curves)
for all of the \afluxa selected sources in the
GOODS-N with $z<1.5$ based on the
fitted MAGPHYS SEDs.
Squares show the measured values for all the sources
with $z>1.5$, color-coded by the MAGPHYS dust
temperature (blue $<40$~K; green=40--50~K; red$>50$~K).
\label{f4_f8_extra}
}
\end{figure}

%-----------------------------------------------------------------------------
As we discussed in Section~\ref{secdisczdist}, the
\afluxa selection only adds sources that would not be found in an 
\afluxb sample below $z = 1.5$.
We characterize the evolution in
\flipsmmfluxratio\ with redshift by using the fitted MAGPHYS SEDs 
for the \afluxa selected GOODS-N sample sources
and determining what the \flipsmmfluxratio\ values would be
if the lower-redshift sources were moved up in redshift.
In Figure~\ref{f4_f8_extra}, we show predicted \flipsmmfluxratio\ 
versus redshift (black curves) for all of the sources with $z<1.5$. 
We compare these with the measured values for all the sources
with $z>1.5$ (squares; color-coded by the MAGPHYS dust 
temperature: blue $<40$~K; green=40--50~K; red$>50$~K). 
The higher-redshift sources are consistent
with having the same SEDs as the lower-redshift sources 
(i.e., the squares lie in the range of the extrapolated lower-redshift values).
Thus, there is no evidence for the selection of warmer sources
with a \afluxa sample than with an \afluxb sample.

%-----------------------------------------------------------------------------
\section{Summary}
\label{secsummary}
%-----------------------------------------------------------------------------
In this fifth paper in our SUPER GOODS series, we presented SCUBA-2 \afluxa
selected samples in the two GOODS fields.  Our main results are as follows:

$\bullet$ 
We constructed the differential number counts at \afluxa and found
excellent agreement with results in the literature.

$\bullet$ 
We used the extensive
redshift information available on the GOODS fields to see how well
redshifts could be estimated from simple flux ratios (20~cm/\afluxb
and \afluxar/\afluxbr). We found tight correlations for both ratios with redshift. 

$\bullet$
We provided a table of 12 high-redshift candidates based on these correlations.
Most of the candidates are very faint in the optical/NIR and do not have reliable 
photzs, but for the five that have photzs, the photzs are consistent 
with the sources being at high redshift. 

$\bullet$
We found a strong correlation of redshift with 4.5~$\mu$m flux. This alternate high-redshift 
diagnostic strongly confirms our high-redshift candidate selection.
Without FIR methods to identify high-redshift candidates that cannot be found with photzs,
the redshift distribution for \afluxa selected samples is not  
complete and is biased to lower redshifts.

$\bullet$
We found a wide distribution of redshifts for the \afluxa selected sample, making the 
median redshift a poor characterization of this distribution.

$\bullet$
We constructed full SEDs and used the publicly available
SED-fitting code MAGPHYS (da Cunha et al.\ 2015) to obtain best fits at our adopted redshifts.
These SED fits gave SFRs and dust properties for our \afluxa selected samples.

$\bullet$
At low redshifts, 
MAGPHYS may insert hot components that raise the temperatures and FIR luminosities of
the sources.
We checked these components by also constructing gray body fits for our \afluxa selected
GOODS-N sample. Through comparisons with observed-frame 70~$\mu$m fluxes that were
not included in either type of fit, we concluded that the gray body fits are preferred at $z<1$.

$\bullet$
We found that the \afluxa selected sample
introduces a number of $z<1.5$ sources,
but beyond this, there is no difference
in the redshift distributions for \afluxa and \afluxb samples.

$\bullet$ 
We found that the observed evolution of dust temperature with redshift is 
primarily driven by a selection bias of higher FIR luminosities at higher redshifts.

$\bullet$
We did not find evidence that warmer sources
are selected in a \afluxa sample than in an \afluxb sample.

%-----------------------------------------------------------------------------
% FIGURE 15 ; plot_f4_f8_extra.ps
%-----------------------------------------------------------------------------
%-----------------------------------------------------------------------------

%-----------------------------------------------------------------------------
\begin{acknowledgments}
%-----------------------------------------------------------------------------
We thank the anonymous referee for a very constructive report that helped us to improve 
the manuscript.
We gratefully acknowledge support for this research from the William F. Vilas Estate
and a Kellett Mid-Career Award and a WARF Named Professorship from the 
University of Wisconsin-Madison Office of the 
Vice Chancellor for Research and Graduate Education with funding from the 
Wisconsin Alumni Research Foundation (A.J.B.), NASA grants NNX17AF45G
and 80NSSC22K0483 (L.L.C.), 
a Wisconsin Space Grant Consortium (WSGC) Undergraduate Scholarship (A.H.B.), 
and a WSGC Graduate and Professional Research Fellowship (L.H.J.).

The James Clerk Maxwell Telescope is operated by the East Asian Observatory on 
behalf of The National Astronomical Observatory of Japan, Academia Sinica Institute 
of Astronomy and Astrophysics, the Korea Astronomy and Space Science Institute, 
the National Astronomical Observatories of China and the Chinese Academy of 
Sciences (grant No. XDB09000000), with additional funding support from the Science 
and Technology Facilities Council of the United Kingdom and participating universities 
in the United Kingdom and Canada. The W.~M.~Keck Observatory is operated as 
a scientific partnership among the California Institute of Technology, the 
University of California, and NASA, and was made possible by the generous financial 
support of the W.~M.~Keck Foundation. The authors wish to recognize and acknowledge 
the very significant cultural role and reverence that the summit of Maunakea has always 
had within the indigenous Hawaiian community. We are most fortunate to have the 
opportunity to conduct observations from this mountain.
\end{acknowledgments}

%======================================================================
%   References
%======================================================================

%\clearpage
%-----------------------------------------------------------------------------
% CDF-N TABLE
%-----------------------------------------------------------------------------
%\movetableright=-1.0cm
\startlongtable
\begin{deluxetable*}{lccrrccrrcl}
\centerwidetable
\setcounter{table}{1}
\renewcommand\baselinestretch{1.0}
\tablewidth{0pt}
\tablecaption{CDF-N SCUBA-2 \afluxa Sample ($>4\sigma$) \label{cdfntab}}
\scriptsize
\tablehead{No. and Name & R.A.$^{\rm 450}$ & Decl. & \afluxa & SNR  & R.A.$^{\rm VLA}$ & Decl. &  20~cm & \afluxb & $z$ & offset \\
& \multicolumn{2}{c}{(J2000.0)} & (mJy) & & \multicolumn{2}{c}{(J2000.0)} & ($\mu$Jy) & (mJy) & & ($''$) \\
(1) & (2) & (3) & (4) & (5) & (6) & (7) & (8) & (9) & (10) & (11)}
\startdata
1 SMM 123730+621300 &       12       37 30.62 &       62       13 0.09 &  32.3(2.20) &  14.64 &       12       37 30.80 &       62       12 58.7 &   123(6.1) &    12.60(0.26) &   2.88 &    1.82\cr
2 SMM 123537+622239 &       12       35 37.82 &       62       22 39.2 &  28.6(4.38) &  6.52 &       12       35 38.15 &       62       22 41.0 &   154(8.6) &    3.93(0.53) &   1.23 &    3.01\cr
3 SMM 123546+622012 &       12       35 46.80 &       62       20 12.5 &  25.5(3.91) &  6.53 &       12       35 46.66 &       62       20 13.4 &   46(6.8) &    10.79(0.50) &   3.132 &    1.32\cr
4 SMM 123707+621408 &       12       37 7.309 &       62       14 8.29 &  24.4(1.69) &  14.36 &       12       37 7.210 &       62       14 8.19 &   59(8.7) &    7.95(0.20) &   3.06 &    1.46\cr
5 SMM 123721+620709 &       12       37 21.09 &       62 07 9.20 &  25.1(4.02) &  6.24 &       12       37 21.40 &       62 07 8.30 &   293(9.3) &    4.85(0.52) &   2.17 &    2.36\cr
6 SMM 123711+621330 &       12       37 11.31 &       62       13 30.2 &  25.5(1.71) &  14.90 &       12       37 11.34 &       62       13 30.9 &   126(6.3) &    8.71(0.20) &   1.995 &   0.73\cr
7 SMM 123618+621549 &       12       36 18.48 &       62       15 49.2 &  23.4(2.40) &  9.76 &       12       36 18.35 &       62       15 50.4 &   169(7.7) &    7.53(0.28) &   1.865 &    1.60\cr
8 SMM 123701+621145 &       12       37 1.442 &       62       11 45.4 &  22.6(1.51) &  14.92 &       12       37 1.578 &       62       11 46.4 &   95(5.7) &    6.55(0.19) &   1.760 &    1.42\cr
9 SMM 123527+622218 &       12       35 27.34 &       62       22 18.7 &  22.7(5.02) &  4.51 & \nodata & \nodata & off radio &    2.49(0.64) &   2.71 & \nodata\cr
10 SMM 123610+622042 &       12       36 10.05 &       62       20 42.9 &  21.9(4.25) &  5.14 &       12       36 9.880 &       62       20 45.5 &   149(20.5) &    5.20(0.53) &  0.6309 &    2.81\cr
11 SMM 123623+622009 &       12       36 23.01 &       62       20 9.19 &  21.2(4.66) &  4.55 &       12       36 23.05 &       62       20 7.90 &   51(7.6) &    3.09(0.56) &   1.98 &    1.34\cr
12 SMM 123622+621628 &       12       36 22.78 &       62       16 28.2 &  19.9(2.36) &  8.42 &       12       36 22.67 &       62       16 29.7 &   81(7.8) &    3.03(0.27) &   1.790 &    1.74\cr
13 SMM 123556+622236 &       12       35 56.21 &       62       22 36.7 &  19.7(3.56) &  5.54 &       12       35 55.90 &       62       22 39.0 &   56(7.9) &    11.74(0.44) &   3.148 &    3.20\cr
14 SMM 123558+621353 &       12       35 58.21 &       62       13 53.7 &  20.2(4.47) &  4.51 &       12       35 58.19 &       62       13 53.9 &   75(16.1) &    1.10(0.45) & \nodata &    1.62\cr
15 SMM 123551+622144 &       12       35 51.65 &       62       21 44.5 &  19.1(3.39) &  5.65 &       12       35 51.40 &       62       21 47.2 &   51(7.3) &    18.84(0.43) &   4.422 &    3.17\cr
16 SMM 123629+621048 &       12       36 29.28 &       62       10 48.3 &  17.6(1.99) &  8.87 &       12       36 29.03 &       62       10 45.5 &   91(7.1) &    1.60(0.26) &   1.013 &    3.25\cr
17 SMM 123645+621448 &       12       36 45.99 &       62       14 48.4 &  18.4(1.71) &  10.74 &       12       36 46.08 &       62       14 48.6 &   103(3.7) &    5.63(0.20) &   2.94 &   0.65\cr
18 SMM 123616+621515 &       12       36 16.21 &       62       15 15.0 &  19.4(2.38) &  8.17 &       12       36 16.10 &       62       15 13.7 &   38(2.9) &    5.60(0.28) &   2.578 &    1.54\cr
19 SMM 123635+621423 &       12       36 35.84 &       62       14 23.2 &  18.0(1.88) &  9.53 &       12       36 35.59 &       62       14 24.0 &   78(5.1) &    4.35(0.22) &   2.005 &    1.86\cr
20 SMM 123634+621242 &       12       36 34.84 &       62       12 42.3 &  17.7(1.68) &  10.55 &       12       36 34.51 &       62       12 40.9 &   188(7.7) &    1.48(0.21) &   1.224 &    2.65\cr
21 SMM 123735+621057 &       12       37 35.43 &       62       10 57.9 &  16.5(2.58) &  6.39 &       12       37 35.55 &       62       10 55.9 &   32(5.8) &    2.51(0.31) &   1.141 &    2.15\cr
22 SMM 123713+621824 &       12       37 13.53 &       62       18 24.2 &  16.7(3.05) &  5.47 &       12       37 13.89 &       62       18 26.2 &   623(19.4) &    5.45(0.33) & \nodata &    3.21\cr
23 SMM 123709+620841 &       12       37 9.700 &       62 08 41.3 &  17.1(2.17) &  7.87 &       12       37 9.748 &       62 08 41.2 &   156(6.9) &    3.34(0.28) &  0.9021 &   0.37\cr
24 SMM 123544+622241 &       12       35 44.85 &       62       22 41.4 &  16.9(3.81) &  4.45 &       12       35 44.82 &       62       22 42.2 &   61(8.2) &    4.61(0.47) &   1.54 &   0.90\cr
25 SMM 123603+621113 &       12       36 3.142 &       62       11 13.9 &  17.2(3.33) &  5.17 &       12       36 3.262 &       62       11 10.9 &   142(6.6) &    1.11(0.41) &  0.6380 &    3.12\cr
26 SMM 123633+621407 &       12       36 33.39 &       62       14 7.29 &  17.2(1.89) &  9.14 &       12       36 33.42 &       62       14 8.50 &   33(5.4) &    9.56(0.23) &   5.302 &    1.19\cr
27 SMM 123635+621922 &       12       36 35.07 &       62       19 22.2 &  17.5(3.95) &  4.42 &       12       36 34.92 &       62       19 23.5 &   81(4.5) &    6.00(0.44) & \nodata &    1.59\cr
28 SMM 123646+620834 &       12       36 46.72 &       62 08 34.4 &  17.0(2.02) &  8.43 &       12       36 46.68 &       62 08 33.2 &   95(5.8) &    2.31(0.28) &  0.9710 &    1.15\cr
29 SMM 123700+620909 &       12       37 0.428 &       62 09 9.41 &  16.5(1.93) &  8.56 &       12       37 0.270 &       62 09 9.70 &   297(10.1) &    3.68(0.26) &   1.61 &    1.14\cr
30 SMM 123725+620858 &       12       37 25.10 &       62 08 58.0 &  15.0(2.55) &  5.90 &       12       37 25.00 &       62 08 56.5 &   84(7.3) &    2.94(0.33) &  0.9367 &    1.78\cr
31 SMM 123631+620958 &       12       36 31.28 &       62 09 58.3 &  15.2(2.06) &  7.40 &       12       36 31.26 &       62 09 57.6 &   140(3.8) &    3.97(0.28) &   2.313 &   0.64\cr
32 SMM 123741+621220 &       12       37 41.61 &       62       12 20.7 &  14.6(2.96) &  4.96 &       12       37 41.70 &       62       12 23.7 &   18(3.0) &    3.61(0.34) &   2.90 &    2.95\cr
33 SMM 123628+620713 &       12       36 28.05 &       62 07 13.3 &  13.7(3.38) &  4.07 & \nodata & \nodata & \nodata &   0.870(0.43) &   1.35 & \nodata\cr
34 SMM 123726+620824 &       12       37 26.39 &       62 08 24.0 &  14.1(2.93) &  4.83 &       12       37 26.66 &       62 08 23.2 &   51(5.4) &    3.61(0.38) &   2.10 &    2.08\cr
35 SMM 123658+620931 &       12       36 58.42 &       62 09 31.4 &  12.8(1.88) &  6.83 &       12       36 58.55 &       62 09 31.4 &   28(2.7) &    4.46(0.25) & contam. &   0.92\cr
36 SMM 123716+621642 &       12       37 16.36 &       62       16 42.2 &  13.4(2.31) &  5.79 &       12       37 16.62 &       62       16 43.4 &   80(3.9) &    2.77(0.25) &  0.5573 &    2.14\cr
37 SMM 123618+621408 &       12       36 18.09 &       62       14 8.08 &  13.3(2.13) &  6.23 &       12       36 17.83 &       62       14 7.91 &   14(2.5) &    1.99(0.27) &  0.8460 &    1.83\cr
38 SMM 123706+620723 &       12       37 6.536 &       62 07 23.2 &  12.4(2.75) &  4.50 &       12       37 6.807 &       62 07 22.2 &   86(6.6) &   0.914(0.37) &   1.218 &    2.10\cr
39 SMM 123730+621617 &       12       37 30.97 &       62       16 17.0 &  12.6(2.76) &  4.57 &       12       37 31.20 &       62       16 20.2 &   65(3.9) &    1.69(0.30) &  0.9750 &    3.55\cr
40 SMM 123719+621022 &       12       37 19.42 &       62       10 22.1 &  12.8(2.15) &  5.94 &       12       37 19.55 &       62       10 21.2 &   22(2.7) &    3.94(0.27) & \nodata &    1.35\cr
41 SMM 123637+620853 &       12       36 37.02 &       62 08 53.3 &  13.1(2.06) &  6.34 &       12       36 37.03 &       62 08 52.4 &   90(5.4) &    3.70(0.29) &   2.05 &    1.00\cr
42 SMM 123712+621035 &       12       37 12.42 &       62       10 35.2 &  11.1(2.04) &  5.43 &       12       37 12.48 &       62       10 35.6 &   21(2.7) &    5.40(0.26) & \nodata &   0.48\cr
43 SMM 123622+621615 &       12       36 22.21 &       62       16 15.2 &  12.7(2.35) &  5.41 &       12       36 22.10 &       62       16 15.9 &   20(2.8) &    3.08(0.27) & \nodata &   0.99\cr
44 SMM 123631+621716 &       12       36 31.92 &       62       17 16.2 &  11.7(2.33) &  5.02 &       12       36 31.94 &       62       17 14.7 &   22(2.8) &    8.54(0.26) &   3.36 &    1.59\cr
45 SMM 123633+621006 &       12       36 33.59 &       62       10 6.29 &  12.2(2.03) &  6.01 &       12       36 33.69 &       62       10 5.79 &   50(6.0) &   0.927(0.28) &   1.016 &   0.88\cr
46 SMM 123621+621710 &       12       36 21.32 &       62       17 10.2 &  12.2(2.63) &  4.64 &       12       36 21.28 &       62       17 8.40 &   148(4.1) &    4.92(0.29) &   1.988 &    1.81\cr
47 SMM 123717+620803 &       12       37 17.39 &       62 08 3.20 &  11.0(2.65) &  4.15 &       12       37 17.46 &       62 08 4.40 &   32(5.7) &    4.26(0.35) &   1.42 &    1.28\cr
48 SMM 123611+621033 &       12       36 11.72 &       62       10 33.0 &  11.1(2.53) &  4.37 &       12       36 11.52 &       62       10 33.5 &   21(2.7) &    2.38(0.33) &   2.35 &    1.55\cr
49 SMM 123619+621006 &       12       36 19.44 &       62       10 6.19 &  10.4(2.26) &  4.63 &       12       36 19.11 &       62       10 4.30 &   29(2.8) &    2.68(0.30) &   2.11 &    2.99\cr
50 SMM 123637+621156 &       12       36 37.13 &       62       11 56.4 &  10.4(1.67) &  6.25 & \nodata & \nodata & \nodata &    4.43(0.22) &   3.27 & \nodata\cr
51 SMM 123702+621301 &       12       37 2.307 &       62       13 1.40 &  11.6(1.42) &  8.13 &       12       37 2.570 &       62       13 2.40 &   29(2.7) &    2.56(0.18) & \nodata &    2.09\cr
52 SMM 123612+621146 &       12       36 12.55 &       62       11 46.0 &  11.2(2.29) &  4.89 & \nodata & \nodata & \nodata &    1.95(0.30) &  0.29 & \nodata\cr
53 SMM 123634+620943 &       12       36 34.87 &       62 09 43.2 &  12.4(2.03) &  6.10 & \nodata & \nodata & \nodata &    3.69(0.28) &   2.10 & \nodata\cr
54 SMM 123608+621249 &       12       36 8.961 &       62       12 49.0 &  10.3(2.49) &  4.15 &       12       36 8.671 &       62       12 51.0 &   42(6.1) &    2.51(0.31) & \nodata &    2.83\cr
55 SMM 123730+621254 &       12       37 30.47 &       62       12 54.0 &  10.1(2.19) &  4.63 & \nodata & \nodata & \nodata & blend(1) & \nodata & \nodata\cr
56 SMM 123702+621401 &       12       37 2.600 &       62       14 1.40 &  9.74(1.54) &  6.32 &       12       37 2.761 &       62       14 1.70 &   17(2.6) &    1.96(0.19) &   1.248 &    1.18\cr
57 SMM 123712+621212 &       12       37 12.16 &       62       12 12.2 &  8.88(1.78) &  4.97 &       12       37 12.05 &       62       12 11.9 &   32(2.7) &    3.94(0.22) &   2.914 &   0.86\cr
58 SMM 123655+620808 &       12       36 55.00 &       62 08 8.41 &  8.53(2.14) &  3.98 & \nodata & \nodata & \nodata &    2.53(0.29) & \nodata & \nodata\cr
59 SMM 123713+621543 &       12       37 13.34 &       62       15 43.2 &  9.56(2.20) &  4.33 &       12       37 13.69 &       62       15 44.4 &   29(7.3) &    2.61(0.24) &   2.300 &    2.72\cr
60 SMM 123629+621515 &       12       36 29.67 &       62       15 15.2 &  8.98(2.13) &  4.21 &       12       36 29.45 &       62       15 13.1 &   14(2.5) &    2.28(0.25) &   3.652 &    2.59\cr
61 SMM 123652+621227 &       12       36 52.01 &       62       12 27.3 &  8.73(1.35) &  6.45 &       12       36 52.03 &       62       12 25.9 &   12(2.4) &    6.10(0.17) &   5.183 &    1.51\cr
62 SMM 123718+621135 &       12       37 18.28 &       62       11 35.2 &  8.18(2.03) &  4.02 &       12       37 18.24 &       62       11 33.0 &   23(2.6) &    1.99(0.26) &   1.013 &    2.11\cr
63 SMM 123642+621544 &       12       36 42.40 &       62       15 44.4 &  8.78(1.97) &  4.45 &       12       36 42.22 &       62       15 45.4 &   185(7.2) &    1.98(0.22) &  0.8575 &    1.66\cr
64 SMM 123720+621247 &       12       37 20.60 &       62       12 47.2 &  8.95(1.95) &  4.58 & \nodata & \nodata & \nodata &    1.25(0.24) & contam. & \nodata\cr
65 SMM 123653+620848 &       12       36 53.71 &       62 08 48.4 &  8.63(1.94) &  4.43 &       12       36 53.60 &       62 08 49.9 &   20(2.5) &    1.65(0.27) &   2.41 &    1.69\cr
66 SMM 123659+621449 &       12       36 59.02 &       62       14 49.4 &  8.28(1.70) &  4.85 & \nodata & \nodata & \nodata &    3.51(0.20) & \nodata & \nodata\cr
67 SMM 123711+621324 &       12       37 11.75 &       62       13 24.2 &  8.51(1.72) &  4.94 &       12       37 12.00 &       62       13 25.7 &   55(5.1) & blend(6) &   1.996 &    2.31\cr
68 SMM 123636+621436 &       12       36 36.11 &       62       14 36.2 &  8.98(1.92) &  4.67 &       12       36 35.89 &       62       14 36.0 &   32(2.7) &    1.18(0.23) &   1.018 &    1.57\cr
69 SMM 123638+621112 &       12       36 38.01 &       62       11 12.4 &  8.40(1.79) &  4.68 & \nodata & \nodata & \nodata &    1.41(0.24) &  0.98 & \nodata\cr
70 SMM 123648+621217 &       12       36 48.71 &       62       12 17.4 &  8.40(1.38) &  6.06 &       12       36 48.65 &       62       12 15.7 &   22(2.7) &    1.80(0.18) &   1.066 &    1.67\cr
71 SMM 123634+621212 &       12       36 34.55 &       62       12 12.2 &  8.17(1.71) &  4.77 &       12       36 34.47 &       62       12 12.9 &   221(8.5) &   0.142(0.22) &  0.4575 &   0.90\cr
72 SMM 123713+621157 &       12       37 13.87 &       62       11 57.2 &  7.64(1.87) &  4.07 &       12       37 14.05 &       62       11 56.5 &   21(2.6) &    3.62(0.24) &   4.59 &    1.41\cr
73 SMM 123634+621359 &       12       36 34.54 &       62       13 59.3 &  7.91(1.83) &  4.30 &       12       36 34.28 &       62       14 0.59 &   36(5.3) &    1.77(0.22) &   1.38 &    2.20\cr
74 SMM 123635+621155 &       12       36 35.99 &       62       11 55.2 &  7.49(1.70) &  4.38 &       12       36 36.13 &       62       11 54.4 &   21(5.7) &    4.43(0.22) &   3.91 &    1.22\cr
75 SMM 123645+621146 &       12       36 45.85 &       62       11 46.4 &  7.30(1.48) &  4.90 & \nodata & \nodata & \nodata &   0.177(0.19) & \nodata & \nodata\cr
76 SMM 123653+621137 &       12       36 53.14 &       62       11 37.4 &  6.43(1.46) &  4.39 &       12       36 53.37 &       62       11 39.5 &   74(7.6) &    1.31(0.19) &   1.269 &    2.72\cr
77 SMM 123639+621255 &       12       36 39.42 &       62       12 55.4 &  7.12(1.54) &  4.61 &       12       36 38.89 &       62       12 56.9 &   27(6.9) &   0.861(0.20) &   1.144 &    3.97\cr
78 SMM 123656+621204 &       12       36 56.28 &       62       12 4.41 &  6.49(1.40) &  4.61 &       12       36 56.59 &       62       12 7.40 &   38(5.2) &    3.86(0.18) &  0.29 &    3.64\cr
79 SMM 123649+621314 &       12       36 49.71 &       62       13 14.4 &  5.59(1.37) &  4.06 &       12       36 49.73 &       62       13 12.9 &   57(5.0) &   0.834(0.17) &  0.4750 &    1.41\cr
\enddata
\tablecomments{The columns are 
(1) the SCUBA-2 \afluxa source number and name,
(2 and 3) the SCUBA-2 \afluxa R.A. and Decl., 
(4) the \afluxa flux and, in parentheses, the error; these were
measured from the SCUBA-2 matched-filter image, 
(5) the SNR from the \afluxa measurements,
(6 and 7) the accurate R.A. and Decl. of the corresponding VLA 20~cm source, when available,
(8) the VLA 20~cm flux and, in parentheses, the error,
(9) the SCUBA-2 \afluxb flux and, in parentheses, error; these were measured at the \afluxa position, 
though note that two of the \afluxa sources are
blended at \afluxbr, so these are labeled ``blend" with the number in parentheses giving
the \afluxa source number of the brighter 
source in the pair to which we assigned all of the measured \afluxb flux,
(10) the specz (three or four significant figures after the decimal point; see Section~\ref{subsecz} for 
references), 
or the photz (two significant figures after the decimal point); some sources have
neither, and two sources are
marked ``contam.", because the photometry is contaminated by a neighboring star or galaxy,
and (11) the offset between the SCUBA-2 \afluxa and VLA source positions.
}
\end{deluxetable*}

%-----------------------------------------------------------------------------
% CDF-S TABLE
%-----------------------------------------------------------------------------
\begin{deluxetable*}{lcccrccrrrcl}
\centerwidetable
\setcounter{table}{2}
\renewcommand\baselinestretch{1.0}
\tablewidth{0pt}
\tablecaption{CDF-S SCUBA-2 \afluxa Sample ($>4\sigma$) \label{cdfstab}}
\scriptsize
\tablehead{No. and Name & R.A.$^{\rm 450}$ & Decl.& \afluxa & SNR  & R.A.$^{\rm ALMA}$ & Decl. & 870~$\mu$m & 20~cm & \afluxb & $z$ & offset
\\  & & & & & & & (ALMA) & (VLA) & (SCUBA-2) & & 
\\ & \multicolumn{2}{c}{(J2000.0)} & (mJy) & & \multicolumn{2}{c}{(J2000.0)} & (mJy) & ($\mu$Jy) & (mJy) & & ($''$)
\\ (1) & (2) & (3) & (4) & (5) & (6) & (7) & (8) & (9) & (10) & (11) & (12)}
\startdata
1 SMM033204-274647 &        3       32 4.950 &      -27       46 47.2 &  27.0(4.6) &   5.8 &        3       32 4.889 &      -27       46 47.7 &  6.45 &   130 &     7.7(0.38) &   2.252 &   0.94\cr
2 SMM033248-274934 &        3       32 48.52 &      -27       49 34.2 &  19.7(4.4) &   4.4 &        3       32 48.55 &      -27       49 34.7 & \nodata &   135 &     2.6(0.38) &   1.120 &   0.64\cr
3 SMM033249-274917 &        3       32 49.11 &      -27       49 17.2 &  19.5(4.5) &   4.3 & \nodata & \nodata & \nodata & \nodata &   blend(2) & \nodata & \nodata\cr
4 SMM033207-275119 &        3       32 7.349 &      -27       51 19.2 &  19.2(4.7) &   4.0 &        3       32 7.289 &      -27       51 20.8 &  8.93 &   120 &     3.3(0.36) & contam. &    1.87\cr
5 SMM033219-274603 &        3       32 19.57 &      -27       46 3.29 &  18.7(3.2) &   5.8 &        3       32 19.70 &      -27       46 2.20 &  4.90 &   51 &     3.7(0.30) &   2.41 &    2.03\cr
6 SMM033235-274917 &        3       32 35.70 &      -27       49 17.2 &  18.0(3.1) &   5.7 &        3       32 35.73 &      -27       49 16.2 &  5.09 &   80 &     4.7(0.27) &   2.576 &    1.16\cr
7 SMM033215-275037 &        3       32 15.49 &      -27       50 37.2 &  17.2(3.4) &   4.9 &        3       32 15.33 &      -27       50 37.6 &  6.61 &   50 &     4.6(0.28) &   3.12 &    2.26\cr
8 SMM033222-274935 &        3       32 22.42 &      -27       49 35.2 &  15.3(2.6) &   5.8 &        3       32 22.47 &      -27       49 35.2 & 0.934 &   77 &     3.8(0.24) &  0.7323 &   0.67\cr
9 SMM033222-274807 &        3       32 22.28 &      -27       48 7.29 &  14.6(2.5) &   5.7 &        3       32 22.28 &      -27       48 4.79 &  5.21 &   42 &     6.1(0.23) &   3.847 &    2.49\cr
10 SMM033232-274545 &        3       32 32.53 &      -27       45 45.2 &  14.6(3.3) &   4.3 & \nodata & \nodata & near edge & \nodata &     2.9(0.33) & \nodata & \nodata\cr
11 SMM033243-274639 &        3       32 43.53 &      -27       46 39.2 &  14.4(3.4) &   4.2 &        3       32 43.53 &      -27       46 39.2 &  3.18 & \nodata &     4.7(0.34) &   2.794 &    0.0\cr
12 SMM033238-274633 &        3       32 38.41 &      -27       46 33.2 &  13.9(3.0) &   4.5 &        3       32 38.55 &      -27       46 34.5 &  2.04 & \nodata &     2.2(0.31) &   2.543 &    2.32\cr
13 SMM033228-275041 &        3       32 28.16 &      -27       50 41.2 &  12.9(3.2) &   4.0 & \nodata & \nodata & \nodata & \nodata &    0.76(0.28) & \nodata & \nodata\cr
14 SMM033228-274659 &        3       32 28.46 &      -27       46 59.2 &  12.9(2.6) &   4.8 &        3       32 28.50 &      -27       46 58.3 &  6.39 &   103 &     4.8(0.26) &   2.309 &    1.11\cr
15 SMM033234-274940 &        3       32 34.33 &      -27       49 40.2 &  12.5(3.0) &   4.0 &        3       32 34.27 &      -27       49 40.3 &  4.73 & \nodata &     4.8(0.27) &   3.76 &   0.92\cr
16 SMM033228-274827 &        3       32 28.75 &      -27       48 27.2 &  9.65(2.3) &   4.0 &        3       32 28.80 &      -27       48 29.7 &  1.57 & \nodata &     2.0(0.23) &   1.83 &    2.58\cr
\enddata
\tablecomments{The columns are 
(1) the SCUBA-2 \afluxa source number and name,
(2 and 3) the SCUBA-2 \afluxa R.A. and Decl., 
(4) the \afluxa flux and, in parentheses, the error; these were
measured from the SCUBA-2 matched-filter image, 
(5) the SNR from the \afluxa measurements,
(6 and 7) the accurate R.A. and Decl. of the corresponding ALMA 870~$\mu$m source, when available
(for source~2, it is the VLA 20~cm source position),
(8) the ALMA 870~$\mu$m flux, 
(9) the VLA 20~cm flux,
(10) the SCUBA-2 \afluxb flux and, in parentheses, error; these were measured at the position of the 
\afluxa source when there is no ALMA 870~$\mu$m position, 
though note that one of the \afluxa sources is blended at \afluxbr, so it is labeled “blend” 
with the number in parentheses giving the \afluxa source number of the brighter source 
in the pair to whom we assigned all of the measured \afluxb flux, 
(11) the specz (three or four significant figures after the decimal point; see Section~\ref{subsecz} 
for references) or the photz (two significant figures after the decimal point);
some sources have neither, and one source is marked ``contam.", because the ALMA 
position places the source at the edge of another source (see bottom-left thumbnail of
Figure~\ref{magphysSEDCDFS}),
and (12) the offset between the SCUBA-2 \afluxa and ALMA source positions, 
except for source~2,
where the offset is between the SCUBA-2 \afluxa and VLA source positions.
}
\label{tab3}
\end{deluxetable*}

%-----------------------------------------------------------------------------
% HIGH-Z CANDIDATES TABLE 4
%-----------------------------------------------------------------------------
\begin{deluxetable*}{lcccccccc}
\setcounter{table}{3}
\renewcommand\baselinestretch{1.0}
\tablewidth{0pt}
\tablecaption{High-Redshift Submillimeter Galaxies}
\scriptsize
\tablehead{No. and Name & \afluxa & \afluxb & 20~cm & 20~cm/\afluxbr & \afluxar/\afluxb 
& $z$ & $z_{20/850}$ & $z_{450/850}$ \\
& (mJy) & (mJy) & (mJy) & $(\times10^{-3})$ & & (from Table~2) & & \\
(1) & (2) & (3) & (4) & (5) & (6) & (7) & (8) & (9)}
\startdata
\hline\hline
\multicolumn{9}{c}{High-Redshift Candidates Based on the Criteria of Figure~\ref{fluxratios} \label{highztab}} \cr
\hline
GN-1    SMM 123730+621300  & 32.3 & 12.60 & 123 & 9.78   &   2.58 & 2.88 & \nodata  & 3.38   \cr
GN-27  SMM 123635+621922  & 17.5 & 6.00 & 81 & 13.63 &  2.78 & \nodata & \nodata & 3.18 \cr
GN-42  SMM 123712+621035  & 11.1 & 5.40 & 21 & 4.05    &  2.33 &  \nodata & 4.31  & 3.66  \cr
GN-44  SMM 123631+621716  & 11.7 & 8.54 & 22 & 2.57     & 1.46 &  3.36 & 5.03 & 4.91   \cr
GN-47  SMM 123717+620803   & 11.0 & 4.26 & 32 & 7.59    & 2.76 & \nodata & \nodata & 3.20 \cr
GN-50  SMM 123637+621156  & 10.4 & 4.43 & $<11$ & $<2.48$  &  2.56  & 3.27 & $>5.09$ & 3.40  \cr
GN-53  SMM 123634+620943  & 12.4 & 3.69 & $<11$ & $<2.97$  &  2.95  & \nodata & $>4.80$ & 3.02 \cr
GN-58 SMM 123655+620808  & 8.53 & 2.53 & $<11$ & $<4.33$ & 3.82 & \nodata & $>4.20$ & \nodata \cr
GN-66  SMM 123659+621449  & 8.28 & 3.51 & $<11$ & $<3.13$  &  2.47 & \nodata & $>4.72$  & 3.50 \cr
GN-72  SMM 123713+621157  & 7.64 & 3.62 & 21 & 5.96 & 2.18 & 4.59 & \nodata & 3.84 \cr
GN-74  SMM 123635+621155  & 7.49 & 4.43 & 21 & 4.80 & 1.66 & 3.91 & 4.04 & 4.57 \cr
GN-78  SMM 123656+621204 & 6.49 & 3.86 & 38 & 10.02  & 1.64 & \nodata & \nodata & 4.60 \cr
\hline\hline
\multicolumn{9}{c}{Additional Photometric High-Redshift Candidates} \cr
\hline
GN-4 SMM 123707+621408 & 24.4 & 7.95 & 59 & 7.47 & 3.18 & 3.06 & & \cr
GS-7 SMM 033215-275037 & 17.2 & 6.61 & 50 &  10.77 & 3.69 & 3.12 & & \cr %non-IDL 86 currently ALMA flux
GS-15 SMM 033234-274940 & 12.5 & 4.73 & $<29$ & $<6.02$ & 2.60 & 3.76 & & \cr %non-IDL 94 currently ALMA flux
\hline\hline
\multicolumn{9}{c}{Spectroscopic High-Redshift Galaxies} \cr
\hline
GN-3 SMM 123546+622012 & 25.5 & 10.79 & 46     & 4.28   & 2.35    & 3.132 & & \cr
GN-13 SMM 123556+622236 & 19.7 & 11.74 & 56  &  4.81   & 1.67  &  3.148 & & \cr
GN-15 SMM 123551+622144 & 19.1 & 18.84 & 51 & 2.75    & 1.02   &  4.422 & & \cr
GN-26 SMM 123633+621407 &17.2 &  9.56  & 33  & 3.45   & 16.87   & 5.302 & & \cr
GN-60 SMM 123629+621515 & 8.98 & 2.28 & 14 & 6.2 & 9.42 & 3.652 & & \cr
GN-61 SMM 123652+621227 & 8.73 & 6.10 & 12 &  1.96   & 1.54   &  5.183 & & \cr
GS-9 SMM 033222-274807 & 14.6 & 5.21 & 42 & 6.8 & 14.69 & 3.847 & &  \cr
%88 currently ALMA flux
\hline
 \enddata
 \end{deluxetable*}
 %-----------------------------------------------------------------------------

\end{document}